\documentclass[twocolumn]{aastex7}

\shorttitle{AB Aur b Spectroscopic Detection with VLT/MUSE}
\shortauthors{Currie et al.}

\graphicspath{{./}{figures_new/}}

\usepackage{appendix}

\begin{document}

\title{VLT/MUSE Detection of the AB Aurigae b Protoplanet with $H _{\rm \alpha}$ Spectroscopy}

\author[0000-0002-7405-3119]{Thayne Currie}
\affil{Department of Physics and Astronomy, University of Texas-San Antonio, San Antonio, TX, USA}
\affil{National Astronomical Observatory of Japan, Subaru Telescope, 650 N. Aohoku Pl., Hilo, HI, USA}
\email[show]{thayne.currie@utsa.edu}

\author[0000-0002-3053-3575]{Jun Hashimoto}
\affil{Astrobiology Center, National Institutes of Natural Sciences, 2-21-1 Osawa, Mitaka, Tokyo 181-8588, Japan}
\affil{Academia Sinica Institute of Astronomy \& Astrophysics (ASIAA), 11F of Astronomy-Mathematics Building, AS/NTU, No.1, Sec. 4, Roosevelt Rd., Taipei 106319, Taiwan}
\email[]{thayne.currie@utsa.edu}  

\author[0000-0003-0568-9225]{Yuhiko Aoyama}
\affil{Kavli Institute for Astronomy and Astrophysics, Peking University, Beijing 100871, People’s Republic of China}
\email[]{thayne.currie@utsa.edu}

\author[0000-0001-9290-7846]{Ruobing Dong}
\affil{Kavli Institute for Astronomy and Astrophysics, Peking University, Beijing 100871, People’s Republic of China}
\affil{Department of Physics \& Astronomy, University of Victoria, Victoria, BC, V8P 5C2, Canada}
\email[]{thayne.currie@utsa.edu}  

\author{Misato Fukagawa}
\affil{ALMA Project, National Astronomical Observatory of Japan, 2-21-1 Osawa, Mitaka, Tokyo 181-8588, Japan}
\email[]{thayne.currie@utsa.edu}  

\author{Takayuki Muto}
\affil{Division of Liberal Arts, Kogakuin University, 1-24-2, Nishi-Shinjuku, Shinjuku-ku, Tokyo 163-8677, Japan}
\affil{Leiden Observatory, Leiden University, P.O. Box 9513, NL-2300 RA Leiden, The Netherlands}
\affil{Department of Earth and Planetary Sciences, Tokyo Institute of Technology, 2-12-1 Oh-okayama, Meguro-ku, Tokyo 152-8551, Japan}
\email[]{thayne.currie@utsa.edu}  

\author[]{Erica Dykes}
\affil{Department of Physics and Astronomy, University of Texas-San Antonio, San Antonio, TX USA}
\email[]{erica.dykes@my.utsa.edu}

\author[]{Mona El Morsy}
\affil{Department of Physics and Astronomy, University of Texas-San Antonio, San Antonio, TX USA}
\email[]{mona.elmorsy@.utsa.edu}

\author{Motohide Tamura}
\affil{Astrobiology Center, National Institutes of Natural Sciences, 2-21-1 Osawa, Mitaka, Tokyo 181-8588, Japan}
\affil{Subaru Telescope, National Astronomical Observatory of Japan, Mitaka, Tokyo 181-8588, Japan}
\affil{Graduate School of Science, The University of Tokyo, 7-3-1 Hongo, Tokyo 113-0033, Japan}
\email[]{thayne.currie@utsa.edu}  

\begin{abstract}
We analyze high-contrast, medium-spectral-resolution $H_{\rm \alpha}$ observations of the star AB Aurigae using the Very Large Telescope's Multi Unit Spectroscopic Explorer (MUSE).   
In multiple epochs, MUSE detects the AB Aur b protoplanet discovered from Subaru/SCExAO data
in emission at wavelengths slightly blue-shifted from the $H_{\rm \alpha}$ line center (i.e. at 6558.88--6560.13 \AA; $\sim$ -100 km s$^{-1}$) and in absorption at redshifted wavelengths (6562.8--6565.1 \AA; $\sim$ 75 km s$^{-1}$). 
AB Aur b's $H_{\rm \alpha}$ spectrum is inconsistent with that of the host star or the average residual disk spectrum and is dissimilar to that of PDS 70 b and c.    Instead, the spectrum's shape resembles that of an inverse P Cygni profile seen in some accreting T Tauri stars and interpreted as evidence of infalling cold gas from accretion, although we cannot formally rule out all other nonaccretion origins for AB Aur b's MUSE detection.  
AB Aurigae  hosts only the second protoplanetary system detected in $H_{\rm \alpha}$ thus far and the first with a source showing a spectrum resembling an inverse P Cygni profile.  Future modeling and new optical data will be needed to assess how much AB Aur b's emission source(s) originates from protoplanet accretion reprocessed by the disk, a localized scattered-light feature with a unique $H_{\rm \alpha}$ profile, or another mechanism.
\end{abstract}


\section{Introduction} \label{sec:intro}

\textit{Protoplanets} directly imaged within their natal protoplanetary disks shed light on where, when, and how gas giant planet formation can occur \citep[e.g.,][]{Benisty2023,Currie2023PPVII}.   In young star-forming regions, many gas-rich protoplanetary disks show evidence for cavities in scattered light and/or the submillimeter indicating that protoplanets may be accreting disk gas \citep[e.g.][]{Ginski2016,Andrews2020,fran2020a}.   Massive protoplanets could also be responsible for spiral density waves launched in disks and visible in scattered light in some disks \citep{Muto2012,Paardekooper2023PPVII}.

\begin{deluxetable*}{cccccccc}[htbp]
\tablewidth{0pt} 
\tablecaption{MUSE $H_{\rm \alpha}$ observations\label{tab:obs}}
\tablehead{
\colhead{OB (1)}      &\colhead{UT Date} & \colhead{Airmass} & \colhead{DIMM seeing (\arcsec{})}& \colhead{$\tau_{\rm o}$ (ms)$^a$ } & \colhead{DIT (s) $\times$ NDIT} & FWHM$_{H_{\rm \alpha}}$ (mas) & Aperture Correction$^b$
}
\startdata
1 & 2022 Oct 22  & 1.75--1.79 & 0.6--0.7 & 3--5 & 1~$\times$~2, 20~$\times$~30 & 78.0 & 10.9\\
2 & 2022 Nov 19 & 1.75--1.79 & 0.3--0.4 & 9--11 & 1~$\times$~2, 20~$\times$~30 & 88.0 & 10.1\\
3 & 2022 Nov 19 & 1.81--2.05 & 0.3--0.6 & 4-10 & 1~$\times$~2, 20~$\times$~30 & $>$300 & \nodata\\
4 & 2022 Nov 21 & 1.84--1.75 & 0.4--0.5 & 11-15 & 1~$\times$~2, 20~$\times$~30 & 94.0 & 7.0\\
\enddata
\tablecomments{
$^a$Coherence time as reported in the ESO Data Achive.
$^b$Correction factor for photometry between small (FWHM$_{H_{\rm \alpha}}$ in diameter from column~6) and large (2\farcs{}5 in diameter) apertures. 
}
\end{deluxetable*}

After PDS 70 bc \citep{Keppler2018,Haffert2019}, AB Aurigae b presents the most comprehensive case for being a directly-imaged protoplanet \citep{Currie2022ABAurb}\footnote{While the literature, especially in light of this study, favors AB Aur b as identified as a site of active jovian planet formation (i.e. not a brown dwarf), its exact emission sources are less well constrained.  For example, the near-IR signal from AB Aur b has been modeled as thermal emission from an embedded protoplanet atmosphere but alternative models (e.g. a circumplanetary disk/sphere, shock emission from a region surrounding an unseen protoplanet) may be plausible.}.   However, its true emission.  Discovered from near-infrared (IR) high-contrast imaging with \textit{Subaru Coronagraphic Extreme Adaptive Optics} Project (SCExAO), AB Aur b appears as a spatially extended, concentrated source nearly 0\farcs{}6 ($\sim$100 au) due south from the star at a location consistent with the predicted position of a protoplanet driving CO gas spirals seen in submillimeter data and within the submillimeter-imaged dust cavity \citep{Tang2012,Tang2017}.   The SCExAO data, obtained over 4 yr and combined with archival \textit{Hubble Space Telescope} (HST) imaging taken over a decade prior (2007), provide evidence for counterclockwise orbital motion.   The CHARIS spectra and HST Space Telescope Imaging Spectrograph (STIS) optical photometry combined together appear inconsistent with a simple scattered light model \citep[][]{Currie2022ABAurb}.   Complementary near-IR polarimetry 
likewise favor a non-scattered-light origin for AB Aur b's emission \citep{Dykes2024}.

Furthermore, \citet{Currie2022ABAurb} reported the detection of an $H_{\rm \alpha}$ signal at the position of AB Aur b with SCExAO using the VAMPIRES instrument that was also much later detected with HST \citep{Zhou2022ABAurb,Bowler2025}.  For PDS 70 bc, an $H_{\rm \alpha}$ imaging detection was interpreted as an accretion signature. However, as the \citet{Currie2022ABAurb} discovery paper clearly noted, AB Aur b's $H_{\rm \alpha}$ emission results from narrowband photometry and could either originate from scattered light at the disk's surface or could be due to intrinsic $H_{\rm \alpha}$ emission from accretion.

Medium-to-high-resolution $H_{\rm \alpha}$ spectroscopy presents a clear next step to search for accretion onto AB Aur b \citep[e.g.][]{Haffert2019,Hashimoto2020MUSE,Currie2024}.  For stars, circumstellar gas accretes at high latitudes along magnetic field lines near freefall velocities, producing a shock when the hot gas (T $\gtrsim$ 10,000 K) reaches the stellar photosphere \citep[e.g.][]{Muzerolle1998Model,Hartmann2016} and results in $H_{\rm \alpha}$ line emission.  Some models for protoplanet growth extend the magnetospheric accretion framework to smaller radii and masses \citep{Thanathibodee2019PDS70}, while others attribute $H_{\rm \alpha}$ emission to heating of postshock gas \citep{Aoyama2018}.  Both scenarios result in  $H_{\rm \alpha}$ emission consistent with that detected from the PDS 70 bc protoplanets as well as wide-separation planet-mass companions.  Other lines considered to be diagnostic of accretion for young stars (e.g. $Pa_{\rm \beta}$, $H_{\rm \beta}$) have not been consistently detected from PDS 70 bc or other planet-mass objects \citep{Hashimoto2020MUSE,Uyama2021,Demars2023}.  

In this paper, we present medium-spectral-resolution observations of the AB Aurigae protoplanetary system enclosing $H_{\rm \alpha}$ ($\lambda=6562.8$~\AA) and obtained with the Multi Unit Spectroscopic Explorer \citep[MUSE;][]{Bacon2010MUSE} on the Very Large Telescope (VLT). 
Compared to previous $H_{\rm \alpha}$ imaging data, 
MUSE decisively detects AB Aur b; we compare AB Aur b's extracted spectrum with spectra expected for scattered light and contrast its features with those for other protoplanets and young accreting objects.

\section{Observations and Basic Data Reduction} \label{sec:obs}

We observed AB Aurigae using MUSE in Narrow Field Mode (NFM) covering the $H_{\rm \alpha}$ and $H_{\rm \beta}$ lines in four observing blocks (OB) executed between 22 October 2022 and 21 November 2022, all taken in excellent seeing conditions.   This paper focuses solely on the $H_{\rm \alpha}$ data (Table~\ref{tab:obs}) over a wavelength range of 6440.13 \AA to 6688.88 \AA.  A separate work will analyze the $H_{\rm \beta}$ data and compare AB Aur b's MUSE data in detail to various models (J. Hashimoto et al. 2025, in preparation).  


For each OB, we obtained 600 s of exposure time of high-contrast observations in $H_{\rm \alpha}$ with the AB Aurigae primary saturated in some channels.    
We bracketed these deep observations with short, 1-second exposures to obtain unsaturated spectra for the AB Aur primary star.  Except for OB3, which had an airmass of more than 1.8 and whose sequence was impacted by clouds and generally poor conditions, the Adaptive Optics Facility \citep{Arsenault2008VLT-AO,Strobele2012MUSE-AO} delivered an AO correction with a PSF full width at half maximum (FWHM) of $\approx$0.1\arcsec{} in H$\alpha$. 
We calculated a correction factor for photometry between small, FWHM-sized apertures and larger apertures (2\farcs{}5 in diameter). 

Inspection of both the saturated and unsaturated data show that OB4 has the highest-quality data: it has the sharpest PSF, smallest aperture correction, and faintest halo at $\rho$ $\approx$1\arcsec{}. At AB Aur b's separation the stellar halo is $\approx$ 2--2.5 times fainter than for OB1 and OB2 in the $H_{\rm \beta}$ science sequences. OB2 is the second-highest-quality observing block as the halo at AB Aur b's separation is $\approx$10-15\% fainter than OB1 at AB Aur b's separation.  These differences in image quality -- despite comparable seeing values -- are likely due to varying atmospheric coherence times ($\tau_{o}$), where OB4 had by far the longest coherence time.   Thus, we focus our analysis on OB4 and consider OB2 and OB1 as supplementary data sets.  We do not analyze OB3 further due to its poor quality.


The MUSE line-spread function (LSF) is asymmetrical and exhibits slight variations from one spectral slice to another, as well as from one spaxel (representing a spectrum in a datacube) to another \citep{Husser2016a}. When emission lines are particularly strong, certain regions in proximity to these emission lines within a datacube can exhibit a stripe pattern \citep[e.g. Figure 6 in][]{Xie2020MUSE}. To minimize the impact of this stripe pattern on the detectability of AB~Aur~b, we fixed the detector direction to ensure that the stripe pattern remained oriented horizontally during observations (i.e. not overlapping with AB Aur b's position). To mitigate bad and hot pixels, we applied dithering to our deep sequences.

We calibrated raw frames with the MUSE pipeline version 2.8.7
\citep{Weilbacher2020MUSEpipeline}.   Basic calibrations included bias and dark subtraction, flat-fielding, wavelength calibration, measurement of line-spread function, geometric calibration, and illumination correction.  For wavelength calibration, in addition to the calibration with the arc frames taken during the morning, the MUSE pipeline calibrates the wavelength shift by sky lines at 5577.339 and 6300.304~\AA. However, when the sky lines are faint due to insufficient integration time, the wavelength calibration by sky lines fails \citep[refer to Figure A1 in][]{Hashimoto2020MUSE}. Thus we do not apply the wavelength shift by sky lines. 

The typical wavelength solution precision (i.e. scatter per spaxel) from the MUSE pipeline is $\lesssim$0.1 \AA\ or less than 10\% of the MUSE channel width of 1.25~\AA\ \citep[][S. Haffert, pvt. comm.]{Xie2020MUSE}. For wavelengths covering $H_{\rm \alpha}$, the characteristic precision is $\sim$ 0.065~\AA\ (3 km s$^{-1}$), or $\sim$5\% of the MUSE channel width\footnote{See page 46 of \url{https://www.eso.org/sci/facilities/paranal/instruments/muse/doc/ESO-261650_MUSE_User_Manual_P116.pdf} \footnote{\textbf{We directly investigated the possible wavelength solution scatter by computing the Pearson’s correlation coefficient between the spectrum at the H-alpha line (6555-6570~\AA) centered on AB Aur b and those in an annulus between 0\farcs{}3 and 2\farcs{}5 from the star. }
These spaxels probe a region where the AB Aur A spectrum is in a very high-count regime.   We then repeated this measurement by forcing a $\pm$0.05 bandpass offset in the wavelength solution.  Different offsets in the wavelength solution result in different correlation coefficients: the one maximizing the Pearson r value best aligns two spectra.  We find no evidence for a significant wavelength solution drift. 
In our highest-quality data, the cross-correlation of AB Aur A’s scattered light spectrum assuming zero offset is systematically higher than offsets of 0.05 channels widths or greater (0.065 \AA\ or greater).
 }
 }.

The maximum wavelength offset during one night is 0.78~\AA, or $\sim2/3$ of the channel width at wavelengths close to H$\alpha$ and H$\beta$\footnote{Same link as footnote 14: page 42, row 4.} 

Subsequent to the wavelength calibration, we performed flux calibration, sky subtraction, and distortion corrections.  We calibrated the absolute flux from an atmospheric extinction curve at Cerro Paranal and a spectro-photometric standard star observation obtained after our AB Aur sequence (e.g. for OB4 this was the CD-32 5613 white dwarf) and as MUSE master calibrations. The instrumental distortion is corrected by a multi-pinhole mask; astrometry is calibrated and monitored by observing stellar cluster fields with high astrometric quality from HST observations.

Finally, we windowed our data to cubes with spatial dimensions of 200~$\times$~200 pixels (5\farcs0~$\times$~5\farcs0) encompassing the central star.   To register the data, we applied subpixel shifts by calculating the approximate centroid positions using a center-of-light estimate and applying a least-squares fit to the centroid vs. wavelength measurements, masking saturated channels near $H_{\rm \alpha}$.
MUSE's atmospheric dispersion corrector enables a stable centroid position around $H_{\rm \alpha}$: the measured centroid deviation across all wavelength channels is $\lesssim$0.2 pixels.   At the star's $H_{\rm \alpha}$ line center ($\sim$6563 \AA), the inner 0\farcs{}3 are affected by a combination of saturation, detector nonlinearity, and residual striping.  Exterior to these separations, the halo is unsaturated and in the linear response regime for all channels.  At the separation of AB Aur b ($\approx$0\farcs{}6)), the halo is a factor of $\approx$6 fainter than at $\rho$ $\sim$ 0\farcs{}3 (see Appendix A).

\vspace{-0.1in}
\section{PSF subtraction}\label{sec:psfsub}

For PSF subtraction, we followed two separate approaches in this Letter\footnote{We also considered the pyKLIP \citep{Wang2015pyKLIP} package for a more aggressive least-squares PSF subtraction to identify any additional emission lines nominally below the subtraction residuals from methods considered in this work.  The results of this reduction will be presented in our companion paper including $H_{\rm \beta}$ (Hashimoto et al. 2025 in prep.).  pyKLIP also yields a detection of AB Aur b.}.   First, we follow recent MUSE studies \citep{Xie2020MUSE} by fitting and subtracting a reference spectrum to each pixel, correcting for field-dependent variations.  Second, we consider a classical \textit{spectral differential imaging} (SDI) approach to identify bright excess signals near the $H_{\rm \alpha}$ and $H_{\rm \beta}$ lines without algorithm biasing.

\subsection{Reference Spectrum Subtraction from High-Resolution SDI}

This approach follows the high-resolution SDI (HRSDI) approach informed by \citet{Haffert2019} and \citet{Xie2020MUSE}.   First, we normalize the signal of each spaxel by the total flux along the wavelength dimension at each location in the image plane spatial location, $S_{\rm total}$, yielding a normalized spectrum ($S_{norm}$).  We then divide the image plane into annuli, compute the median normalized spectrum within this annulus ($S_{\rm \star ref}$), and divide each spaxel by this normalized spectrum.   

The residuals consistent of field-dependent low and higher-order deviations from the median spectrum.   To remove residuals due to a field-dependent stellar spectrum, we then fit a third-degree polynomial to the residuals smoothed by a Savitzky-Golay filter with a window size of 12.6 \AA\  \citep{Haffert2019}, masking the wavelength channels bracketing $H_{\rm \alpha}$ ($\lambda$ =6550--6570 \AA) \footnote{In practice, we found negligible differences in our results between third and second-order polynomial fit or with our nominal mask vs. one expanded to 6550--6575 \AA (see Appendix B).}.  Finally, we subtracted the normalized reference spectrum (reweighted by the polynomial fit) from each flux-normalized spaxel and multiplied by the original total flux to recover the true residual signal.  

Thus, the reference star subtraction for each spaxel $j$ in an annulus $i$ has the form:
\begin{equation}
    S_{residual,j} = [S_{norm,j}-S_{\rm \star, ref, i}(a+b\lambda+c\lambda^{2}+d\lambda^{3})]S_{total, j}
\end{equation}

The width and geometry of the annuli within which we compute and subtract a reference spectrum are key parameters in our fit.  In addition to instrumental effects \citep[see][]{Xie2020MUSE}, astrophysical effects may explain reference spectrum variations for accreting stars surrounded by protoplanetary disks, like AB Aurigae.  AB Aur's $H_{\rm \alpha}$ flux and line profile vary on the scale of hours \citep{Catala1999,Costigan2014}.  Our MUSE data confirm a variable $H_{\rm \alpha}$ line intensity and shape over the course of both days and over $\sim$a few hours timescale at the $\gtrsim$10\% level in our MUSE data (see Sect. 5.1), and HST observations of AB Aurigae likewise confirm this trend (T. Currie et al. 2025 in preparation, see also \citealt{Bowler2025}).   Due to finite light travel time, light emitted from the accretion shock/disk wind region is reflected off of inner disk regions earlier than outer disk regions: scattered light from the AB Aur primary and innermost disk region reaches the location of AB Aur b $\approx$ 12-13 hr later and thus has a different spectrum than light received earlier or later.   Thus, the reference spectrum at AB Aur b's location -- and the $H_{\rm \alpha}$ line-to-continuum ratio -- could differ from locations in the disk closer to and further away from the star.

To guard against nonphysical residuals due to a reference spectrum mismatch from variability, we compute reference spectra in 1-pixel wide annuli with an elliptical geometry consistent with the position angle and inclination angle of the AB Aur disk \citep[$PA$ = 234$^{o}$, $i$ = 30$^{o}$][]{Perrin2009,Tang2017,Betti2022,Dykes2024}\footnote{Note, the position angle refers to the angle of the major axis counterclockwise from north.  The plane of the major axis is equivalent to that from the 324$^{o}$ quoted in \citet{Betti2022}. }. 
We verified the robustness of this approach by applying it to MUSE PDS 70 data first presented in \citet{Haffert2019} and yielding a clear detection of both PDS 70 b and c and a spectrum consistent with published results (see Appendix C).

\subsection{Classical SDI PSF Subtraction}
In classical SDI, the reference PSF for a given target wavelength channel is built from a median/mean combination of other channels and weighted to minimize the residuals when subtracted from the target channel.  
Here, we applied a radial-dependent SDI-based PSF subtraction\footnote{We also explored subtractions where each reference slice was magnified by the ratio of its wavelength to the target slice prior to constructing a reference PSF, as with SDI typically performed with extreme AO systems coupled with near-IR integral field spectrographs \citep[e.g. SCExAO/CHARIS, SPHERE/IFS][]{Currie2023a,Zurlo2016}.  Aligning the speckles resulted in slightly better suppression of the stellar halo for the bluest and reddest MUSE channels but otherwise offered no significant advantage for channels near the $H_{\rm \alpha}$ and $H_{\rm \beta}$ central wavelengths, especially within 1\arcsec{} of the star.  }. 
We perform SDI-based PSF subtraction in annuli extending from 1 $\lambda$/D to the edge of the visible stellar halo.  To avoid under/oversubtraction of astrophysical signals, we 
scaled the reference PSF for each wavelength slice by comparing the total signal of the reference and target within a given annulus\footnote{We obtain the same results if instead we scaled the reference PSF by the median pixel value.}.


As with the reference spectrum subtraction, we varied the annular width used in constructing the reference PSF, adopting 1 pixel wide annuli.  For annular widths larger than$\approx$ 5 pixels, the residuals revealed some over/undersubtraction of the PSF between 6550--6570 \AA near the optical axis.  However, our detection/non-detections of astrophysical features described later in this Letter are robust for all annular widths considered.

\section{Detection of AB Aurigae b at \texorpdfstring{$H_{\rm \alpha}$}{H-alpha}} \label{sec:results}
\begin{figure}[!h]
    \includegraphics[width=1.0\linewidth,clip]{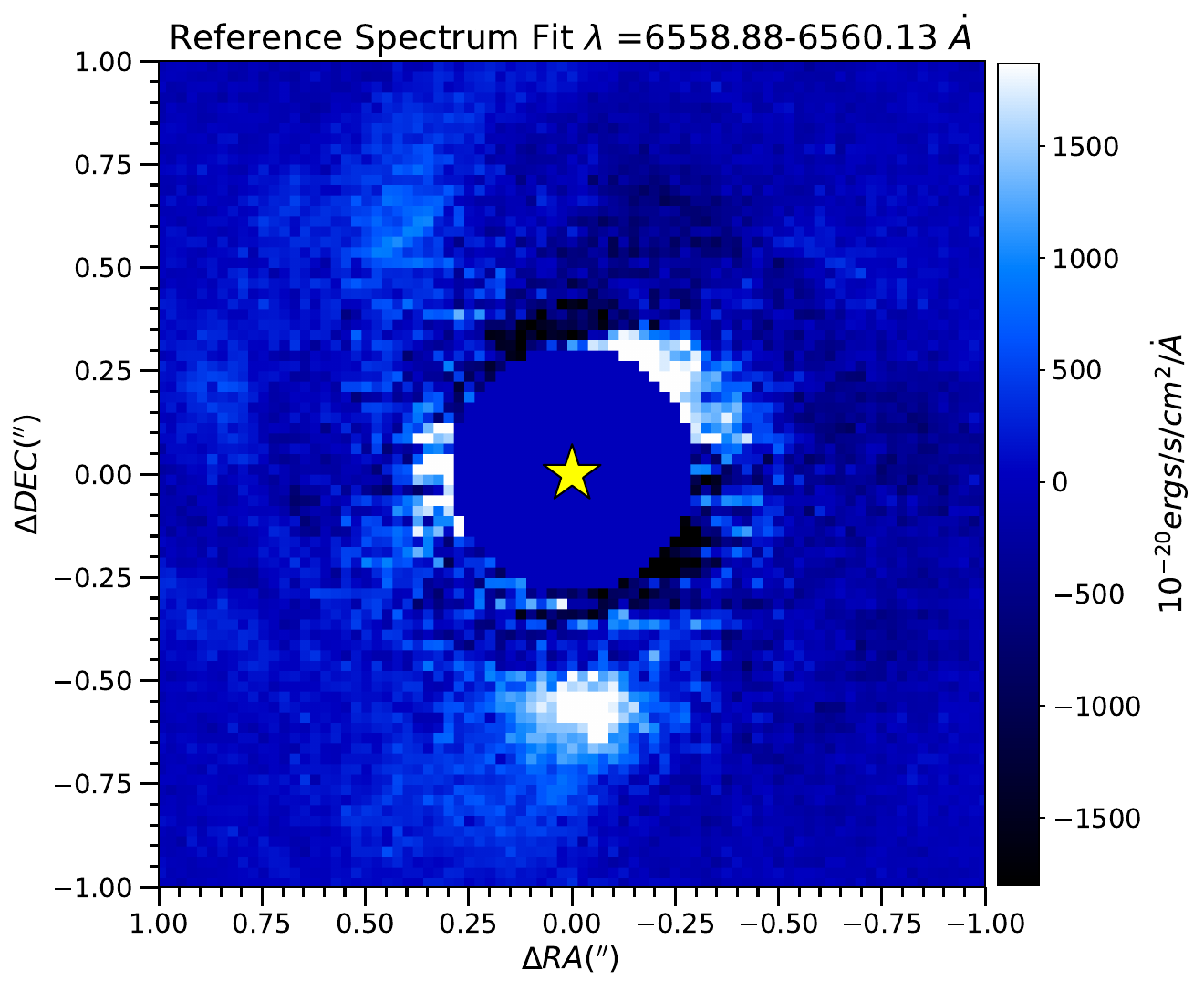}
    \caption{PSF-subtracted MUSE image of AB Aurigae constructed from the average of the 6558.88 \AA and 6560.13 \AA slices (i.e. slightly blue-shifted from $H_{\rm \alpha}$) from OB4, showing a clear detection of AB Aurigae b at $\approx$ 0\farcs{}6 slightly counterclockwise of due-south.  The region within 0\farcs{}3 is masked for clarity because it is heavily affected by saturated/non-linear pixels, especially in some channels centered on the $H_{\rm \alpha}$ line.
    }\label{fig:abaur_blueshift1} 
\end{figure}

\begin{figure*}[htbp]
\includegraphics[width=0.329\textwidth]{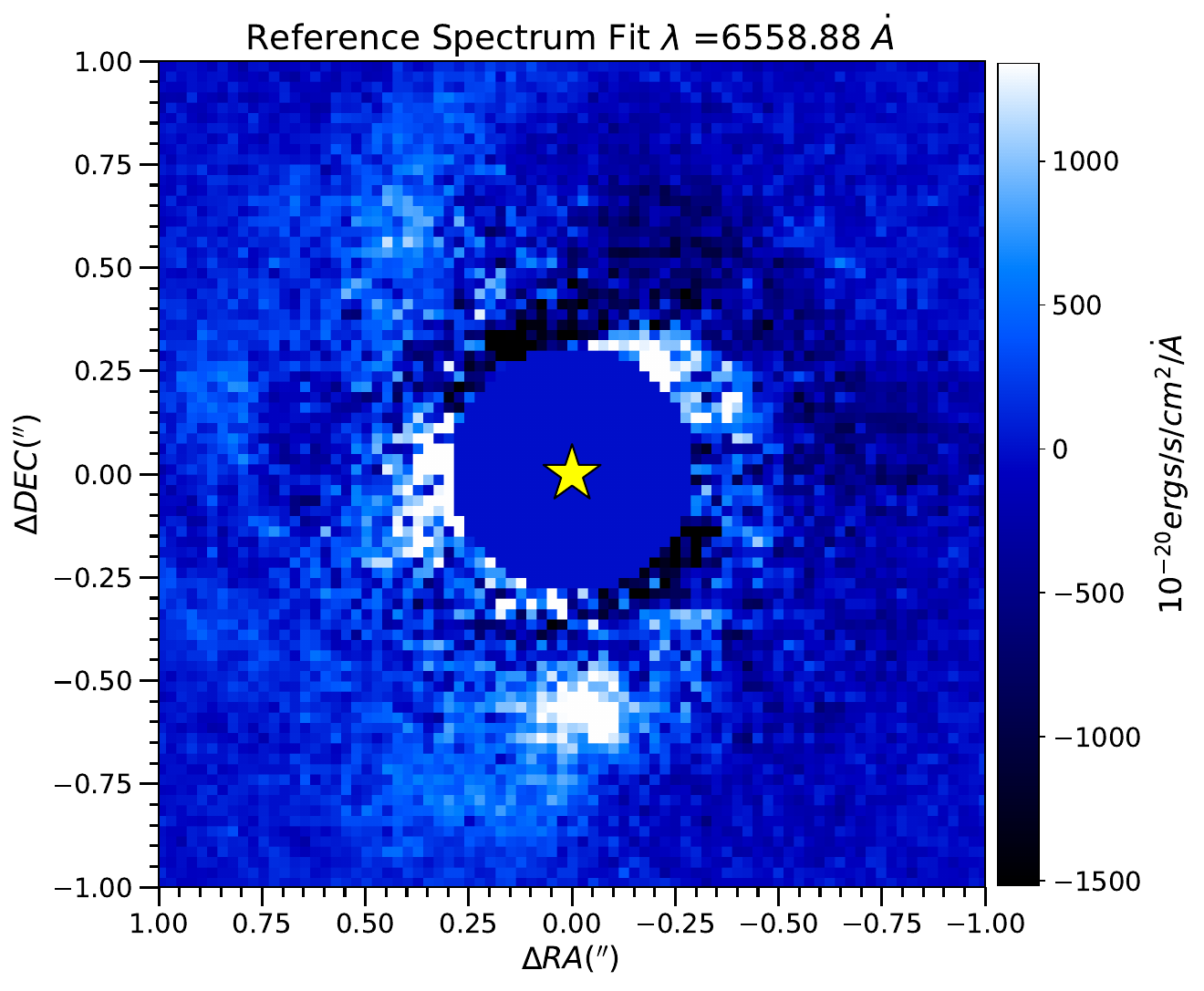}
\includegraphics[width=0.329\textwidth]{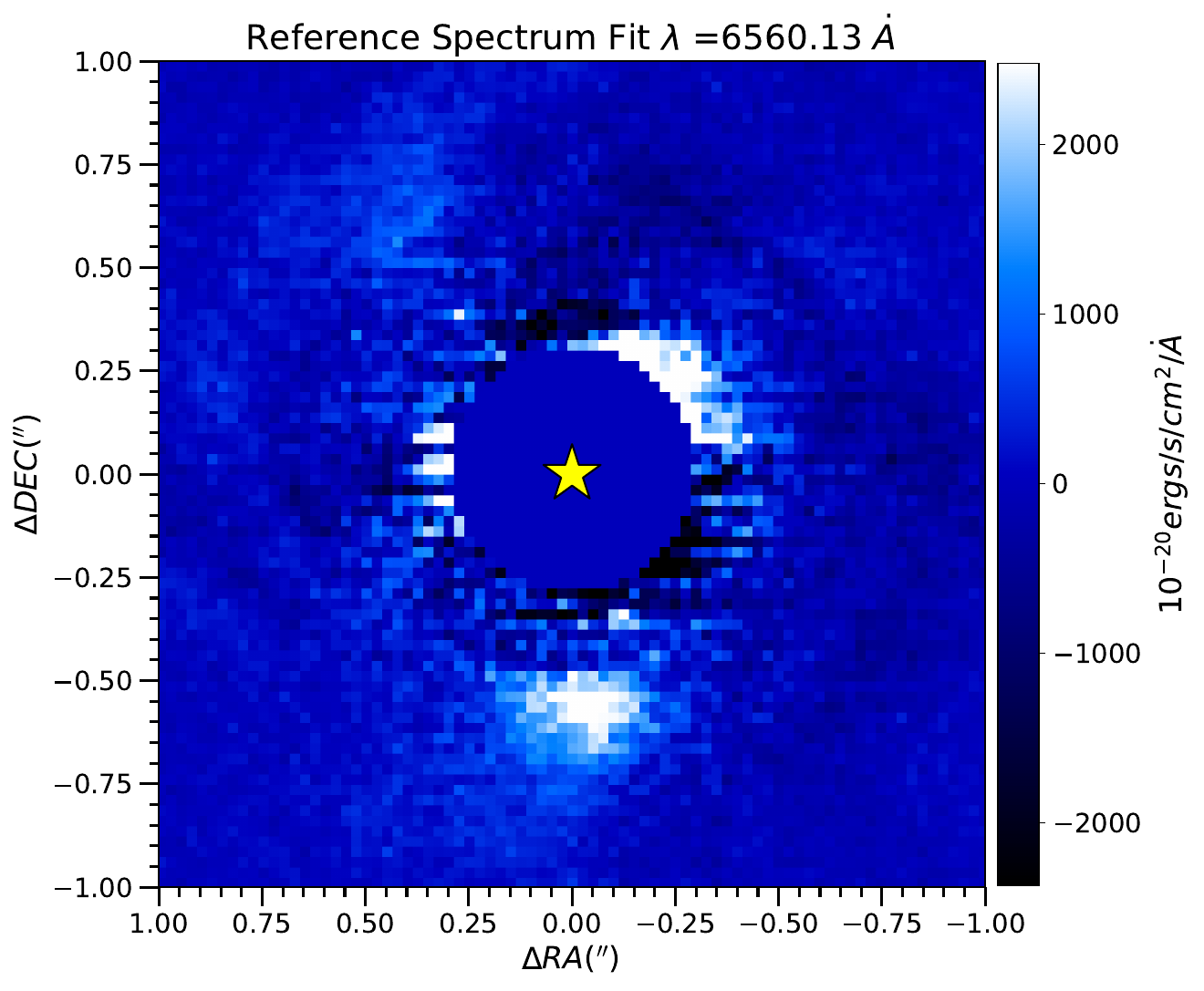}
\includegraphics[width=0.329\textwidth]{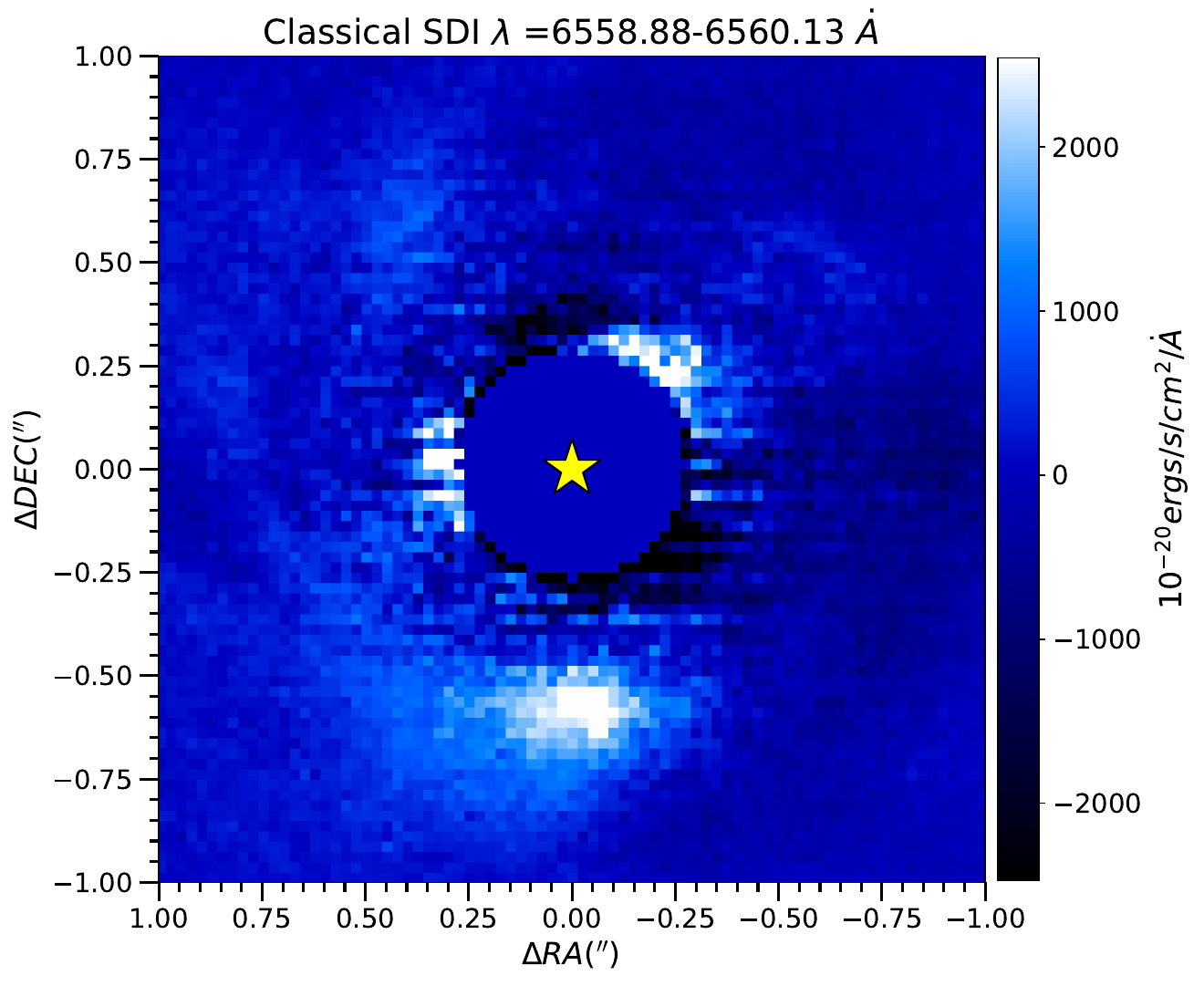}\\

   \vspace{-0.25in}
    \caption{Other MUSE reductions showing the detection of AB Aur b from OB4.  (left, middle) 6558.88 \AA and 6560.13 \AA from the reduction shown in Figure \ref{fig:abaur_blueshift1}.  
    Average of the 6558.88 \AA and 6560.13 \AA slices from a classical SDI reduction (right).  In each panel, we mask the field out to a separation affected by horizontal striping and saturation.
    }\label{fig:abaur_blueshift2} 
\end{figure*}

As shown by Figures \ref{fig:abaur_blueshift1}--\ref{fig:abaur_blueshift2}, after PSF subtraction we easily detect AB Aur b in blue-shifted $H_{\rm \alpha}$ 
in our highest-quality data set (OB4).  The emission is most clearly visible in the individual 6558.88 \AA\ and 6560.13 \AA\ channels and is detectable but fainter and more weakly visible at slightly shorter wavelength (6557.63 \AA) and longer wavelength (6561.14 \AA) channels.  
The detection is robust against slight differences in PSF subtraction approaches.   
The detections with different reduction approach settings are consistent: e.g. AB Aur b is detected regardless of the annular width and geometry adopted for reference spectrum subtraction.

\begin{figure*}
\includegraphics[width=0.328\textwidth]{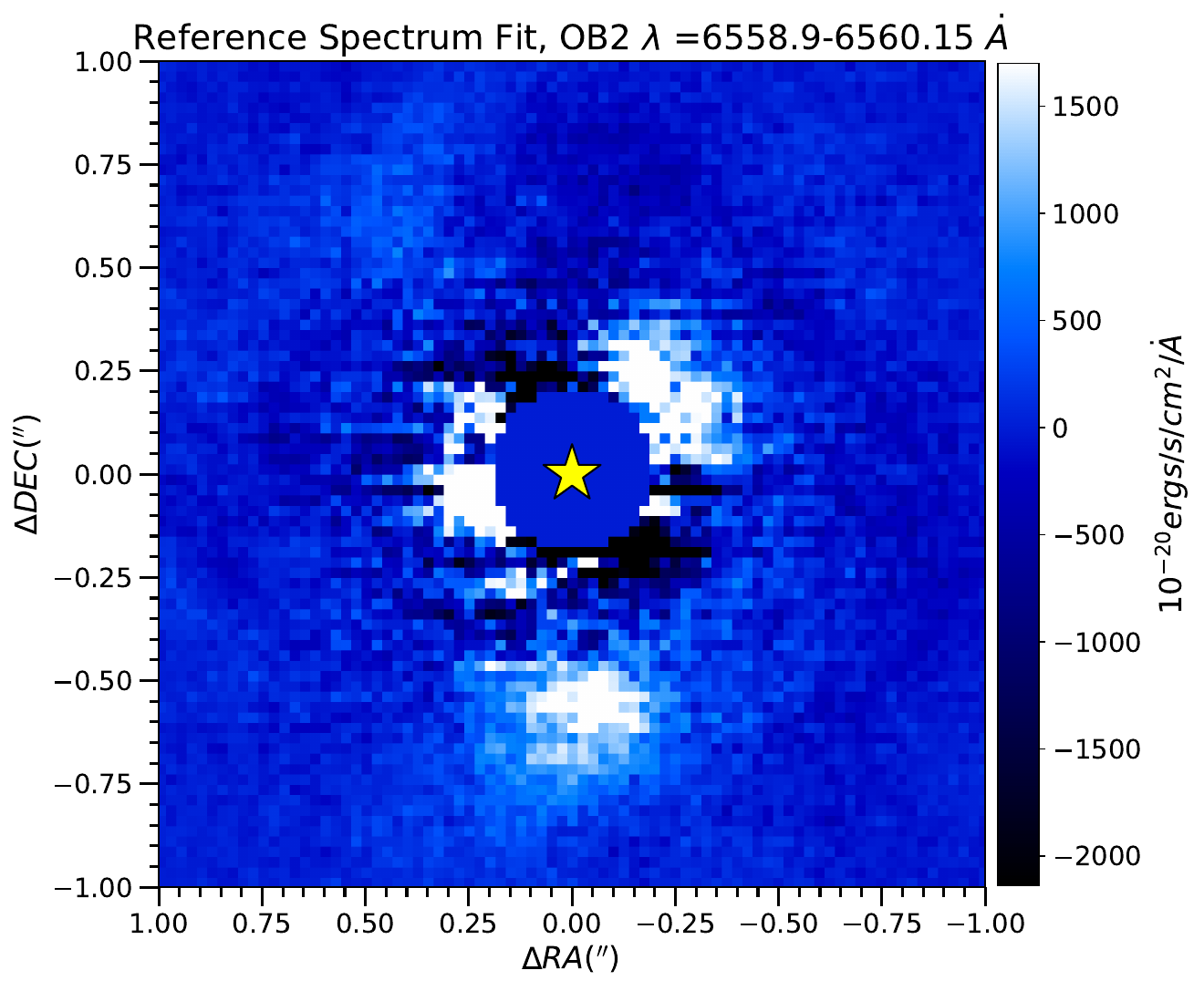}
\includegraphics[width=0.328\textwidth]
{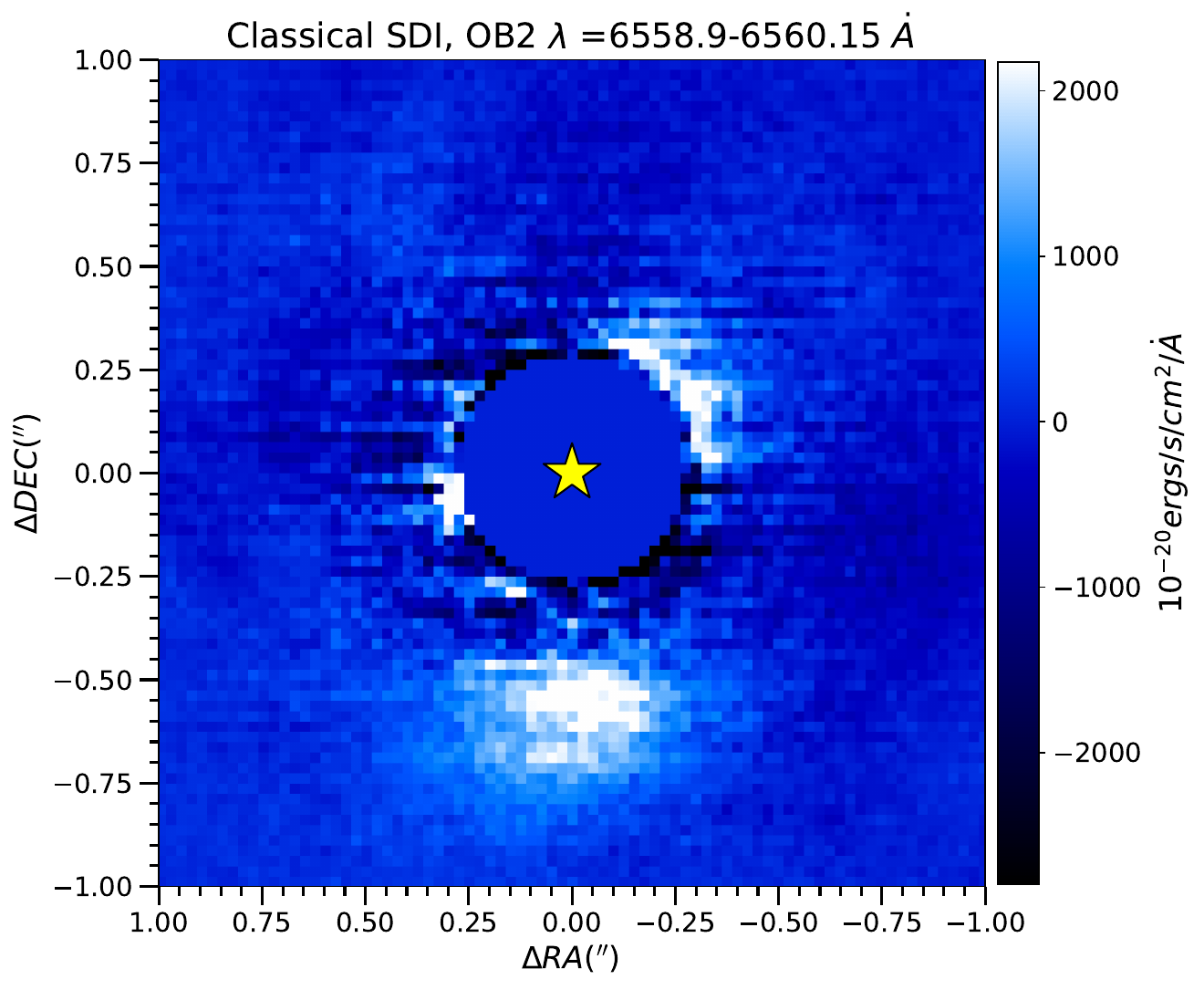}
\includegraphics[width=0.328\textwidth]{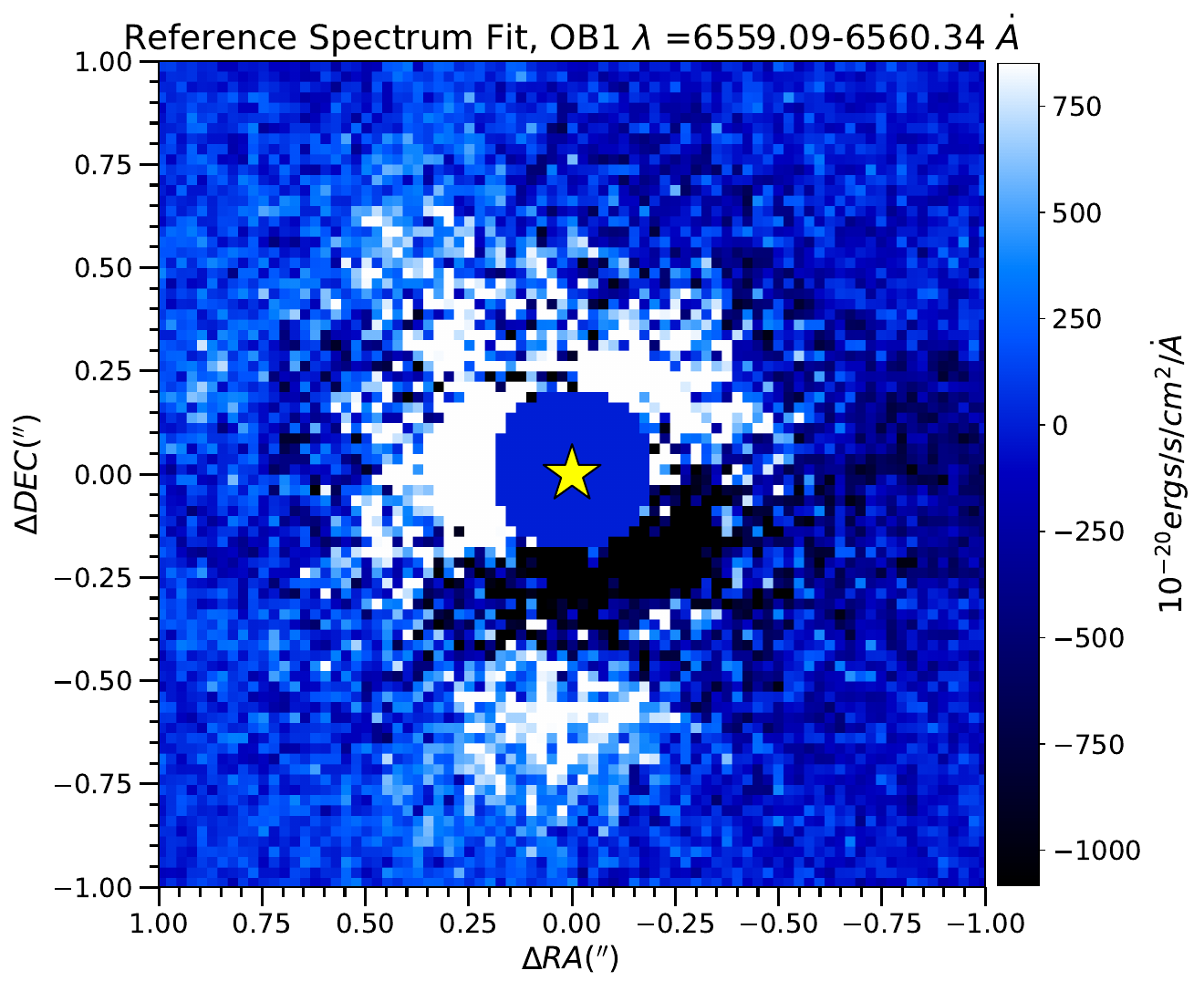}
\caption{Detection of AB Aur b from OB2 (left, middle panels) and weaker detection from OB1 in blue-shifted $H_{\rm \alpha}$ (right).}
\label{fig:abaur_blueshiftob12}
\end{figure*}

In addition to detecting AB Aur b, inspection of our reference spectrum and classical SDI reductions for OB4 reveals far fainter signal from the centers of some spiral arms in the AB Aur disk, which have been imaged previously in scattered light \citep{Fukagawa2004ABAur,Boccaletti2020ABAur,Currie2022ABAurb,Dykes2024}, where the spiral labeled S1 in \citet{Hashimoto2011ABAur} is the most visible. 
Relative to AB Aur b, the spiral signals are typically fainter in the reference spectrum subtraction by at least a factor of 5 in the combined 6558.88 \AA\ and 6560.13 \AA\ channels (i.e. in Figure \ref{fig:abaur_blueshift1}) and by a factor of 10 in the 6560.13 \AA channel alone.   

To calculate the signal-to-noise ratio (SNR) of AB Aur b, we first replace each pixel in each channel by the sum within a FWHM-sized aperture.  We compare AB Aur b's signal in individual channels covering $H_{\rm \alpha}$ to the standard deviation of the spectrum at AB Aur b's spatial position from this aperture-summed cube, masking the channels covering 6550--6570 \AA.    To compute the SNR from the two blueshifted $H_{\rm \alpha}$ channels combined, we convolved the spectrum with a moving-box filter of length two channels.  

For the reduction shown in Figure \ref{fig:abaur_blueshift1}, AB Aur b is detected at SNR $\sim$ 23.3 (13.8 and 22.8 in the individual channels centered on 6558.88 \AA\ and 6560.13 \AA); the detection significance with classical SDI is $\sim$ 17.9$\sigma$.  
As a second approach, we adopt the standard high-contrast imaging approach of calculating the robust standard deviation of the residual noise in each channel as a function of angular separation, correcting for finite sample sizes \citep{Currie2011,Mawet2014}.  This method yields SNR $\sim$ 10 for the reference spectrum subtraction and $\ge$6 for all images shown in Figure \ref{fig:abaur_blueshift2}.   

We also recover AB Aur b in OB2 data with both reference spectrum subtraction and classical SDI (Figure \ref{fig:abaur_blueshiftob12}).  Compared to the OB4 data, the AB Aur b detection in OB2 is slightly more extended, likely due to poorer AO performance.  In the OB1 data, AB Aur b is detected, albeit more weakly (8$\sigma$).

\begin{figure}[!htbp]
\centering
        \includegraphics[width=1\linewidth]{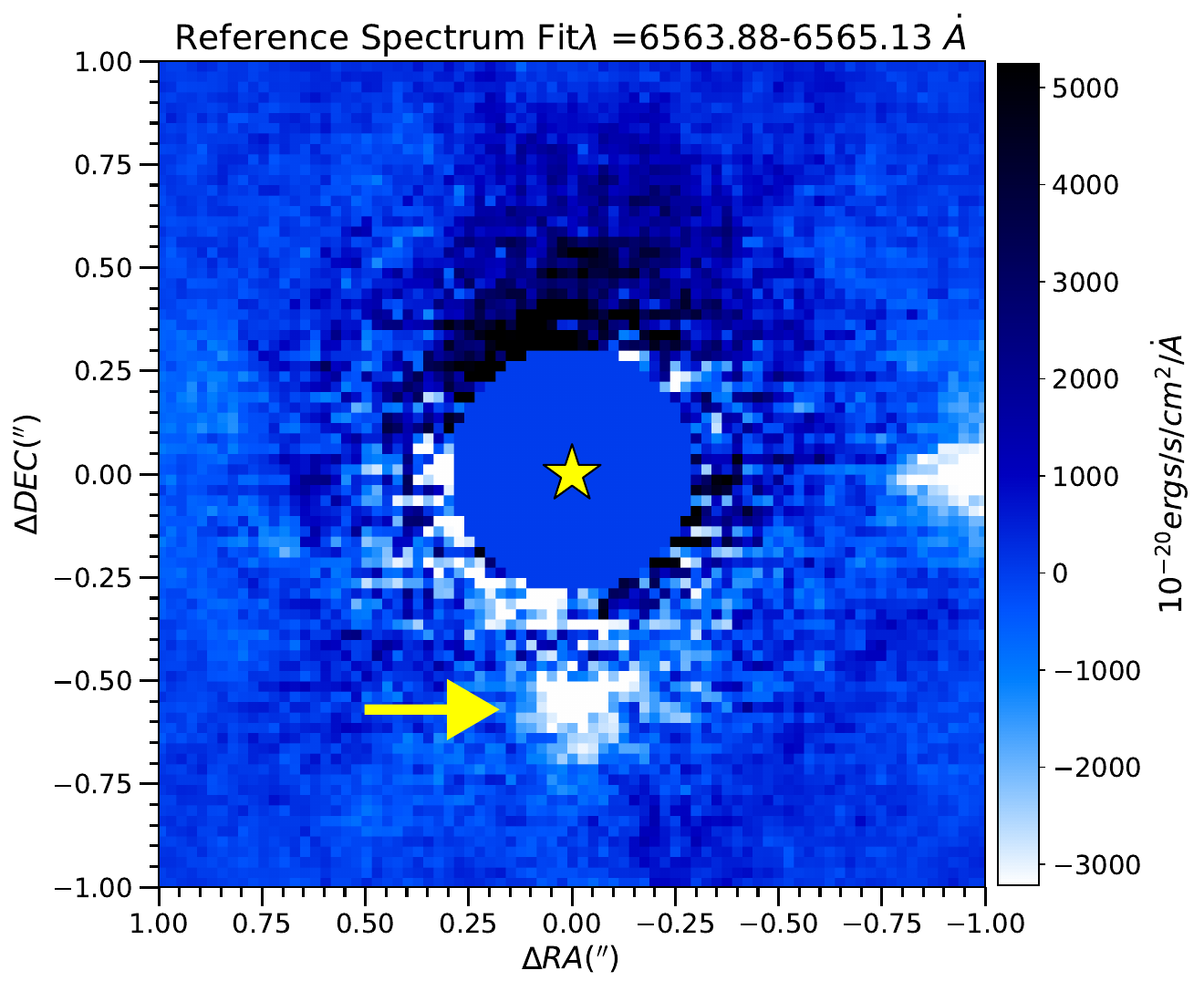}
      \includegraphics[width=1\linewidth]{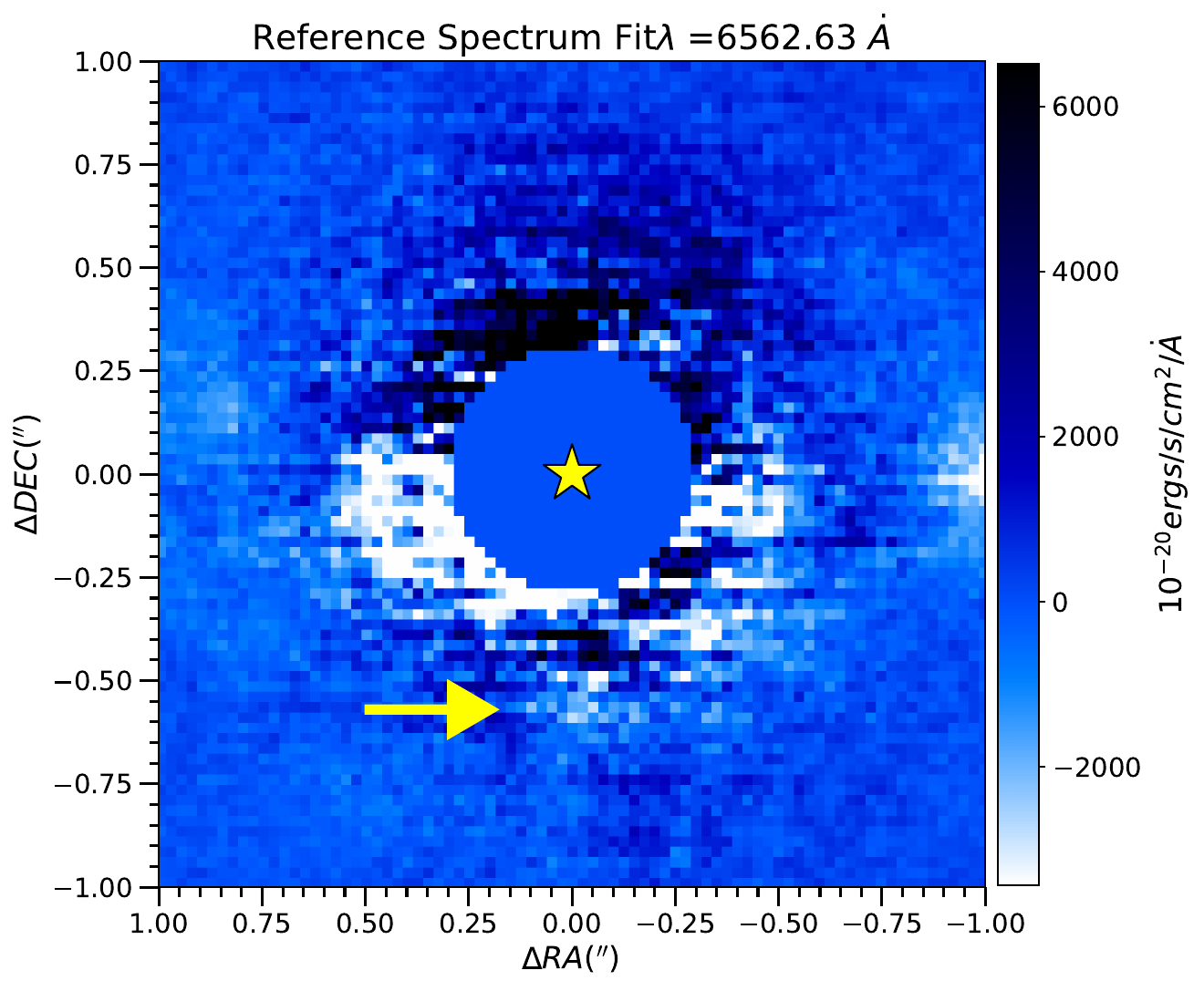}  
    \caption{Reference spectrum subtraction of the $H_{\rm \alpha}$ data slightly redshifted from the $H_{\rm \alpha}$ line center (6563.88-6565.13 \AA) where AB Aur b's signal is clearly visible and at the $H_{\rm \alpha}$ line center (6562.63 \AA; bottom) where it is marginally detectable.  Note that the color intensity scaling is \textit{reversed} compared to prior figures.  For clarity, the yellow arrow points to the location of AB Aur b's emission in the blueshifted channels.  The bright region along the image's center right is a residual detector striping artifact caused by saturation.
    }\label{fig:abaur_redshift1} 
\end{figure}

The $H_{\rm \alpha}$ line core and redshifted channels also reveal AB Aur b
in absorption (Figure \ref{fig:abaur_redshift1}).   At the channels corresponding to deepest absorption level (6563.88-6565.13 \AA), AB Aur b's signal is $\sim$45$\sigma$ significant (top panel).  The signal fades at the $H_{\rm \alpha}$ line center (6562.63 \AA, bottom panel) and at a flanking channel redward of the absorption minimum (6566.38 \AA; not shown) before becoming undetectable at redder wavelengths.  
Like the detection of AB Aur b in blue-shifted emission, its signal in red-shifted absorption is robust against PSF subtraction methods.

Collectively considering all channels covering $H_{\rm \alpha}$, AB Aur b appears to have emission peaking at about $\sim$3 \AA\ blueward of the $H_{\rm \alpha}$ rest wavelength of $\sim$ 6562.8 \AA\ ($\sim$6569-6560 \AA), which weakens at the line center and then transitions into absorption at $\sim$6563-6565 \AA.  The behavior of any residual disk signal is slightly different.  In contrast to the blue-shifted channels, the red-shifted channels display weaker evidence for residual signal at the location of the spirals.  

We will more quantitatively explore these trends in the next section.

\begin{figure}[!htbp]
\centering
    \includegraphics[width=1.\linewidth]{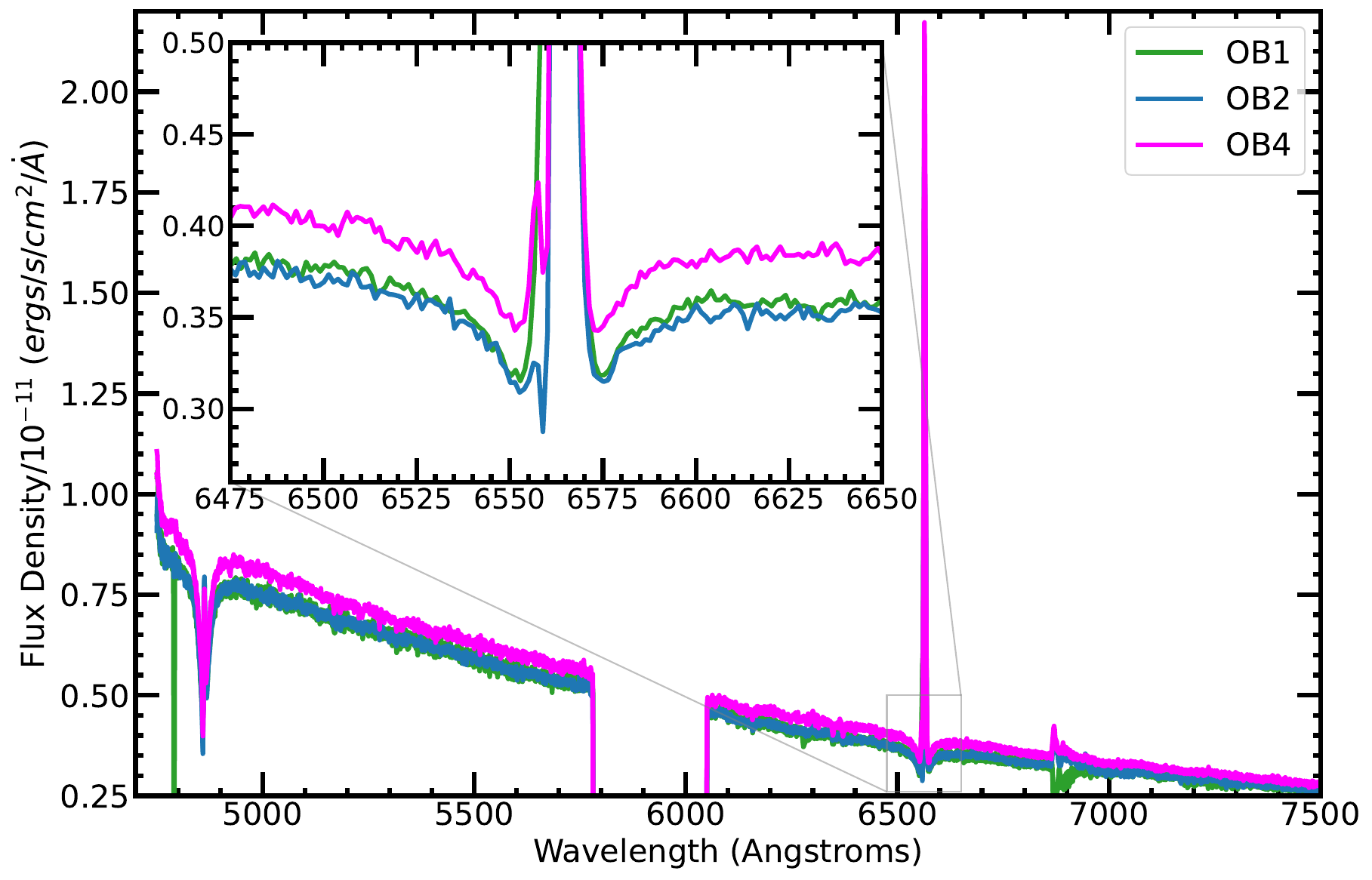}\\
    \includegraphics[width=1.\linewidth]{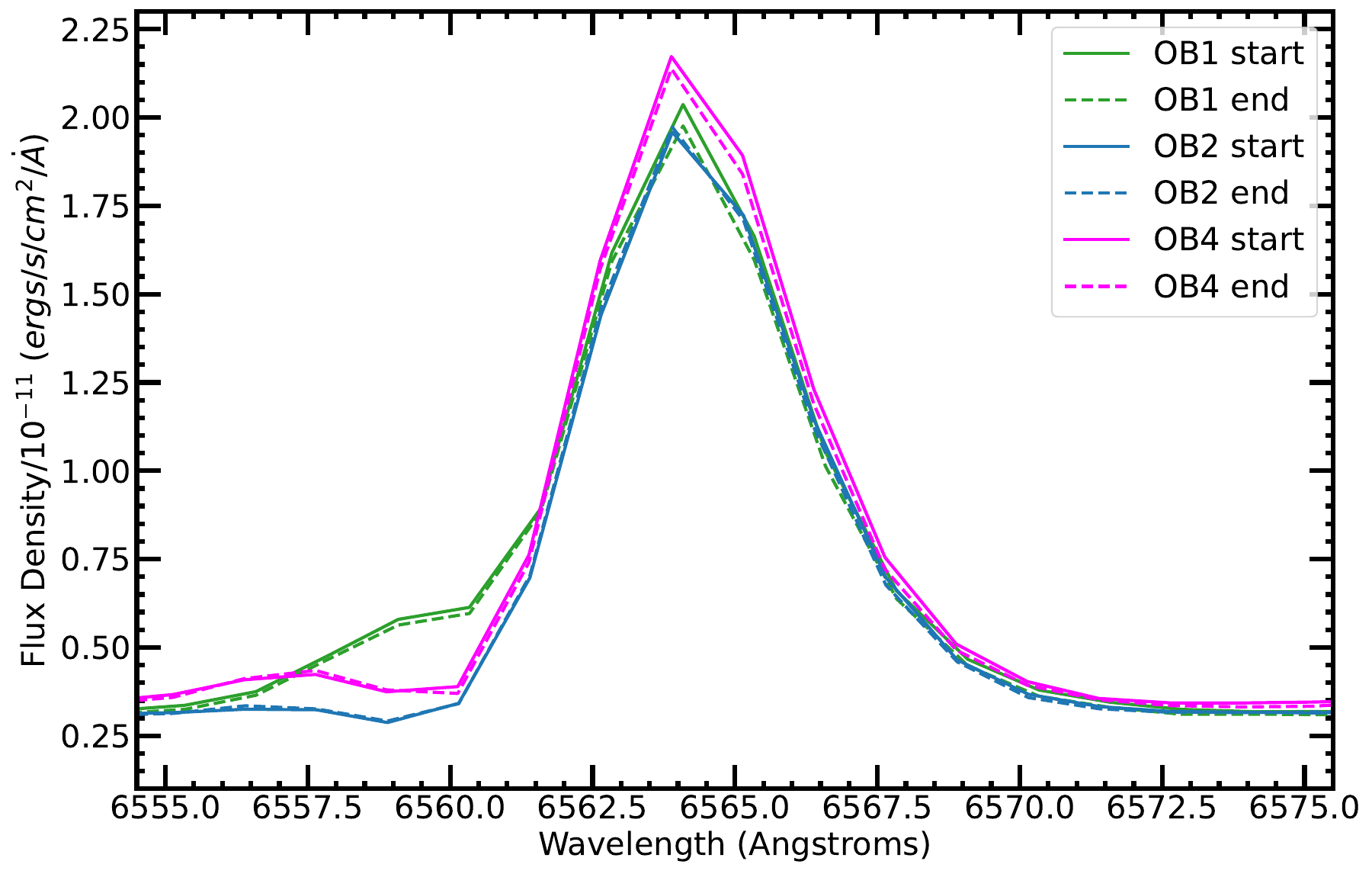}
    \caption{The AB Aurigae primary's spectrum over the entire MUSE wavelength range, with an inset showing the full $H_{\rm \alpha}$ region, including the absorption component (top) and focused on the $H_{\rm \alpha}$ emission line.  
    }\label{fig:abaur_balmer} 
            \vspace{-0.1in}
\end{figure}

\section{Analysis \label{sec:analysis}}
\subsection{\texorpdfstring{$H_{\rm \alpha}$}{H-alpha} Spectrum of the AB Aur primary star}
To provide a reference point for the MUSE detection and spectrum of AB Aur b, we first analyze spectra for the primary star.
Figure \ref{fig:abaur_balmer} shows the AB Aur primary star spectrum from unsaturated exposures for different OBs.  
At $H_{\rm \alpha}$, the star shows strong emission, peaking at a flux density of $F_{\lambda}$ $\sim$ 2.25$\times$10$^{-11}$ $ergs$ $s^{-1}$ $cm^{-2}$ $\dot{A}^{-1}$: roughly 4 times higher than the continuum.   The star shows a weak secondary peak at $\sim$ 6556--6557 \AA\ but no emission line feature at 6559--6560 \AA\ where AB Aur b is detected in excess emission and no absorption at $\sim$ 6563--6565 \AA where AB Aur b is seen in absorption.

The AB Aur $H_{\rm \alpha}$ line peak varies by $\approx$ 20\% between OB4 and other epochs; the continuum also varies by $\sim$10\%.   The line profile at 6555--6560 \AA\ also changes in morphology and varies in brightness by 10-50\%.  Thus, we detect AB Aur b at wavelengths where the primary's $H_{\rm \alpha}$ line profile and continuum both show variability.

\subsection{The Location and Morphology of AB Aur b in \texorpdfstring{$H_{\rm \alpha}$}{H-alpha}\label{peakmcmc}}
Because AB Aur b is a spatially extended additive signal to the disk background \citep{Currie2022ABAurb}, one must jointly model its properties as well as the disk background in order to derive posterior distributions for its position, morphology, and brightness.  For this purpose, we use the affine-invariant Markov chain Monte Carlo (MCMC) ensemble sampler \texttt{emcee} \citep{foreman-mackey+2013}.   In this section we first model its peak emission at blue wavelengths (6559--6560 \AA) and absorption at red wavelengths (6563--6565 \AA).   Guided by these results, we then repeat our MCMC analysis for individual channels in order to extract AB Aur b's spectrum and derive more conservative uncertainties on its morphology and position in Section \ref{abaurbspectrum}.

Building upon previous work \citep[][]{Currie2022ABAurb,Currie2024}, we model AB Aur b's signal as a gaussian intensity distribution with six variable parameters of:
\begin{equation}
I(x,y) = Ae^{-[(x-x_{o})^{2}/2\sigma_{x}^{2}+ (y-y_{o})^{2}/2\sigma_{y}^{2}] } + B, 
\end{equation}. 

$A$ is the peak intensity, $B$ is the background level, ($x_{o}$, $y_{o}$) is the centroid location, and the FWHM in x and y are 2$\sqrt{2 ln(2)}$[$\sigma_{x}$,$\sigma_{y}$].  For simplicity, we assume that the gaussian is aligned along the cardinal directions (i.e. $\theta$ = 0).

To set starting MCMC parameter positions, we perform a maximum-likelihood (ML) fit to AB Aur b's signal using the same gaussian functional form (Table \ref{tab:mcmcprior}, second column).  Each MCMC walker parameter combination is drawn from a gaussian centered on the ML best fit.
The MCMC exploration uses 100 walkers and 10$^{4}$ steps.  The first 100 steps are treated as burn-in\footnote{We explored other values for the number of steps and burn-in (e.g. 10$^{5}$ steps) and found no difference in the derived posterior distributions).}.  We set the \textit{thin} parameter to 35 steps, comparable to the autocorrelation time for each parameter.   Table \ref{tab:mcmcprior} (third column) lists our priors.  


The fourth and fifth columns of Table \ref{tab:mcmcprior} summarize our results; the Appendix D displays the full corner plots.    
Modeled as a gaussian intensity distribution, AB Aur b is spatially extended and azimuthally elongated with a $FWHM_{E,N}$ of $\sim$0\farcs{}255,0\farcs{}173.   
The mean positional values for our posterior distributions -- [E,N] $\sim$ [-0.013, -0.589] -- are broadly consistent with positions measured from \citet{Currie2022ABAurb} (see Appendix E).  Formally, the 68\% confidence interval on the posterior distribution is about 3--4 mas for each coordinate; however, the MCMC simulation does not consider the intrinsic SNR of the detection, the true morphology may slightly depart from a gaussian distribution oriented to $\theta$ = 0, any small gradient in the background may slightly bias the centroid determination and channel-to-channel scatter.  Section \ref{abaurbspectrum} further investigates AB Aur b's positional uncertainty.

\begin{deluxetable*}{ccccc}[htbp]
\tablewidth{0pt} 
\tablecaption{MCMC Priors and Posterior Distributions from Peak Emission and Absorption Minimum\label{tab:mcmcprior}}
\tablehead{
\colhead{Parameter}   &\colhead{ML Best Fit} & \colhead{Prior Range$^{a}$} & \colhead{Posterior (Median)} & \colhead{Posterior [0.16, 0.84]}}

\startdata
\hline
\textit{6558.88--6560.13 \AA\ (Peak Emission)}\\
\hline
Gaussian Amplitude$^{b}$ & 1875.78& [1000, 3000]& 1862.2 &[1773.8,1953.8]\\
E Position (mas) & -13.3 & [-125, 25] & -13.0 & [-8.8, -17.2]\\
N Position (mas) & -589.2 & [-500, -650]  & -589.2 & [-586.1, -592.4]\\
X FWHM (mas) & 253.1 &[100, 375]&255.9&[242.6, 270.6]\\
Y FWHM (mas) & 171.3 &[100, 375]&173.1&[162.6, 184.6]\\
Background Intensity$^{b}$ & 321.3 & [0, 1000]&315.7&[262.3,361.2]\\
\hline\\
\textit{6563.88--6565.13 \AA\ (Peak Absorption)}\\
\hline
Gaussian Amplitude$^{b}$ & -3807.66& [-6000, -1000]& -3800.8 &[-3706.7,-3896.3]\\
E Position (mas) & -12.7 & [-125, 25] & -15.4 & [-8.8, -17.2]\\
N Position (mas) & -567 & [-475, -625]  & -566.4 & [-564.3, -568.6]\\
X FWHM (mas) & 214 &[100, 375]&214.4&[207.8,221.5]\\
Y FWHM (mas) & 219 &[100, 375]&218.7&[209.9,228.0]\\
Background Intensity$^{b}$ & -41.0 & [-1000, 1000]&-36.2&[-94.7,27.5]\\
\enddata
\tablecomments{a) - We adopt a flat prior between the limits listed. b) The intensity units are 10$^{-20}$ $ergs/s/cm^{2}/\dot{A}$.
}
\end{deluxetable*}

\begin{figure}[!htbp]
\centering
    \includegraphics[width=1.\linewidth]{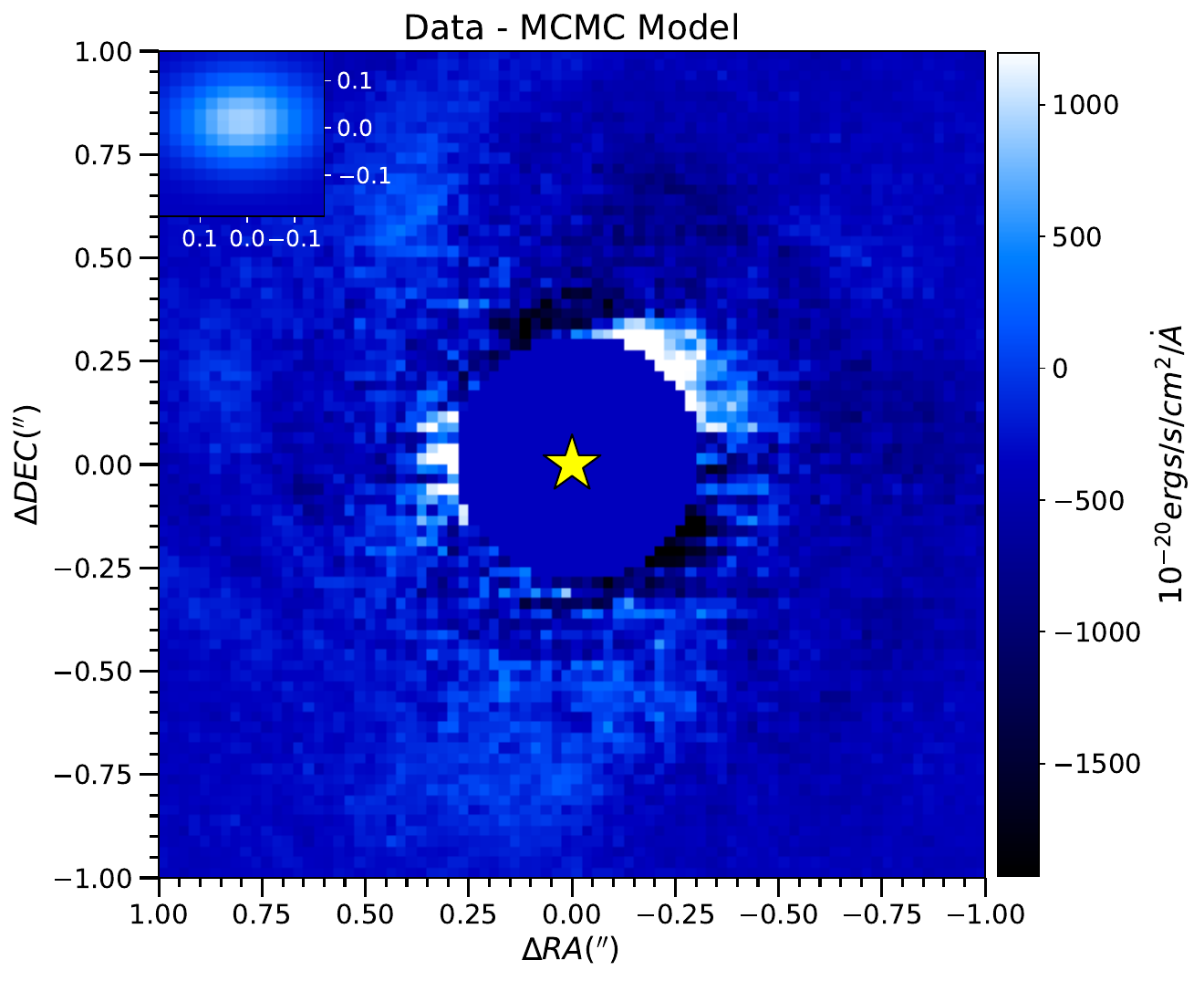}
    \caption{Subtraction of the blue-shifted $H_{\rm \alpha}$ data from Figure \ref{fig:abaur_blueshift1} by the MCMC model corresponding to the median of the posterior distributions.  The inset shows the MCMC model with the same color intensity scaling. 
    }\label{fig:abaur_modelsubtracted_halphablue} 
\end{figure}

The source model derived from MCMC leaves a flat residual background: the residuals are comparable to the characteristic 1$\sigma$ noise at AB Aur b's location (Figure \ref{fig:abaur_modelsubtracted_halphablue}).  The background count values are not zero, although they are low ($\sim$316 ergs s$^{-1}$ cm$^{-2}$ $\dot{A}^{-1}$): the peak signal from AB Aur b is $\approx$6 times brighter than the residual background disk emission after subtracting the best-fit model from the data.  Thus assuming a gaussian intensity distribution for AB Aur b, our modeling is consistent with at least some additional extended emission detected with MUSE at blue-shifted $H_{\rm \alpha}$. 
In STIS broadband optical imaging, which better probes the contribution of scattered starlight, AB Aur b is only 50\% as bright as the disk background \citep{Currie2022ABAurb}.

\begin{figure}[!htbp]
\centering
    \includegraphics[width=1.\linewidth]{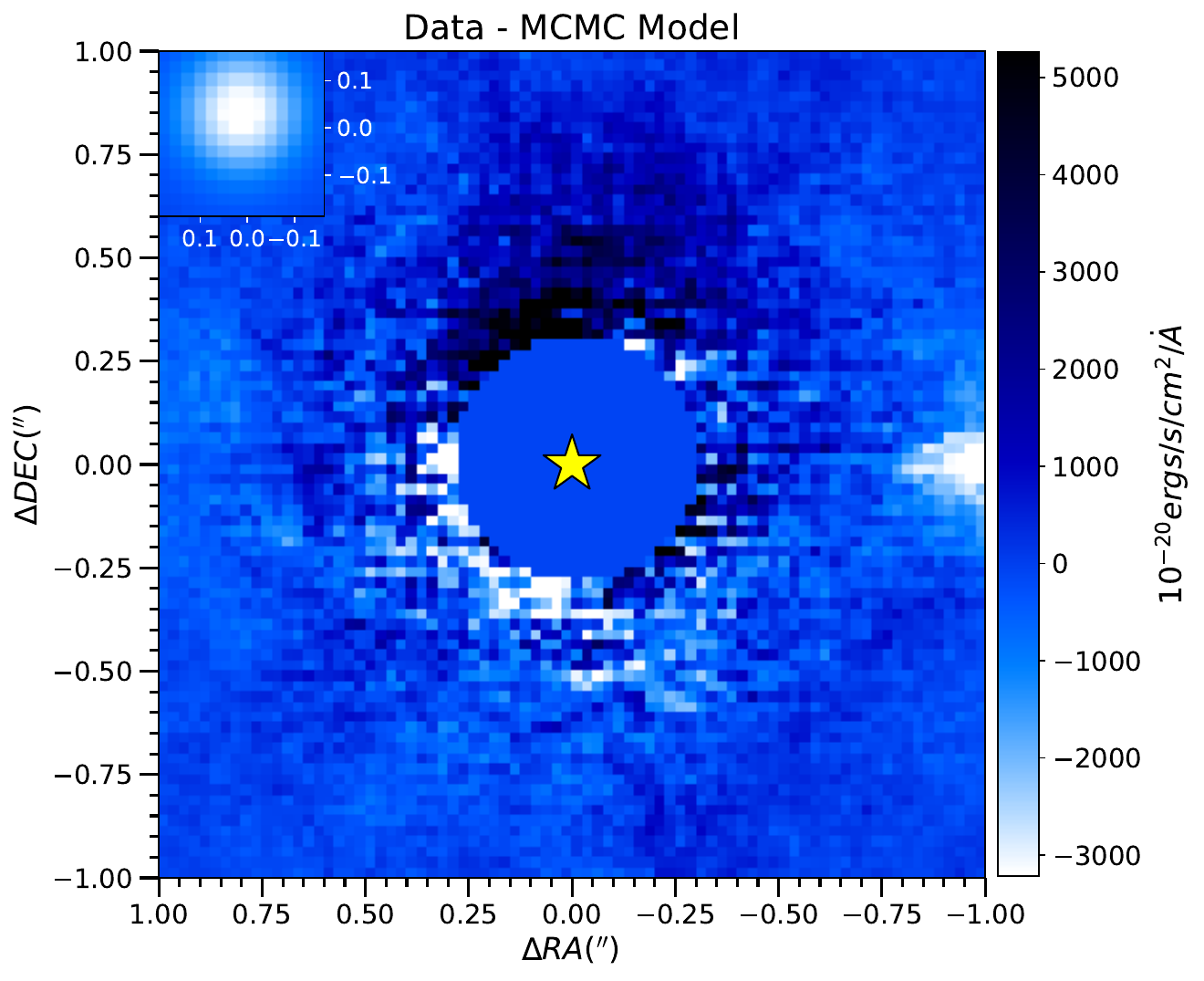}
    \caption{Subtraction of the red-shifted $H_{\rm \alpha}$ data from Figure \ref{fig:abaur_redshift1} (bottom panel) by the MCMC model corresponding to the median of the posterior distributions.  The inset shows the MCMC model with the same color intensity scaling. 
    }\label{fig:abaur_modelsubtracted_halphared} 
\end{figure}

Table \ref{tab:mcmcprior} and Figure \ref{fig:abaur_modelsubtracted_halphared} show results for redshifted $H_{\rm \alpha}$.   The position and morphology for AB Aur b as seen in absorption is similar to that found for blueshifted $H_{\rm \alpha}$, although AB Aur b's centroid position is offset $\sim$3/4th of a pixel closer to the star and it may be more azimuthally symmetric in appearance.  However, we find no evidence for residual disk emission or absorption at AB Aur b's location.

\subsection{\texorpdfstring{$H_{\rm \alpha}$}{H-alpha} Spectrum of AB Aur b\label{abaurbspectrum}}
To extract the spectrum for AB Aur b and estimate more conservative errors on its position and morphology, we jointly estimated the peak AB Aur b signal and background for each channel from MCMC modeling using \texttt{emcee}.  
We performed two separate simulations.  In the first one,
we fixed the location and morphology of AB Aur b to the mean posterior values from our blue-shifted $H_{\rm \alpha}$ and red-shifted $H_{\rm \alpha}$ simulations described in the previous section.  In the second simulation, we allow AB Aur b's position and morphology to freely vary in each channel (i.e. perform the analysis described in Sect. \ref{peakmcmc} independently for each channel).  This free fit also allows us to estimate a more conservative uncertainty on the source position and morphology.
From the fitted peak signal and background and the adopted location and morphology, we estimate AB Aur b's flux density in each channel.  The total signal's 68\% confidence interval in each channel provides a 1$\sigma$ uncertainty.  

\begin{figure*}[!htbp]
\centering
    \includegraphics[width=0.45\linewidth]{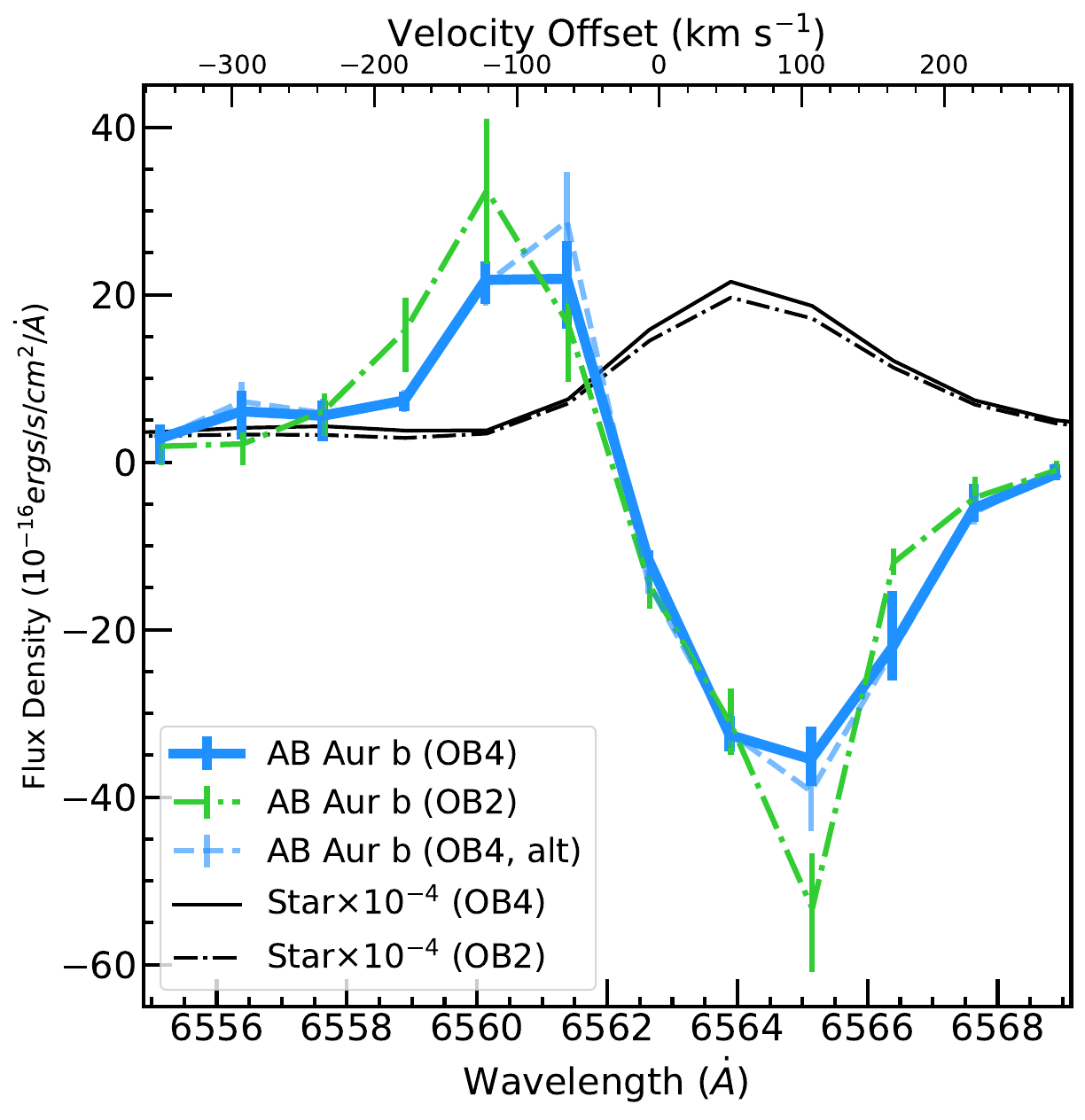}
    \includegraphics[width=0.45\linewidth]{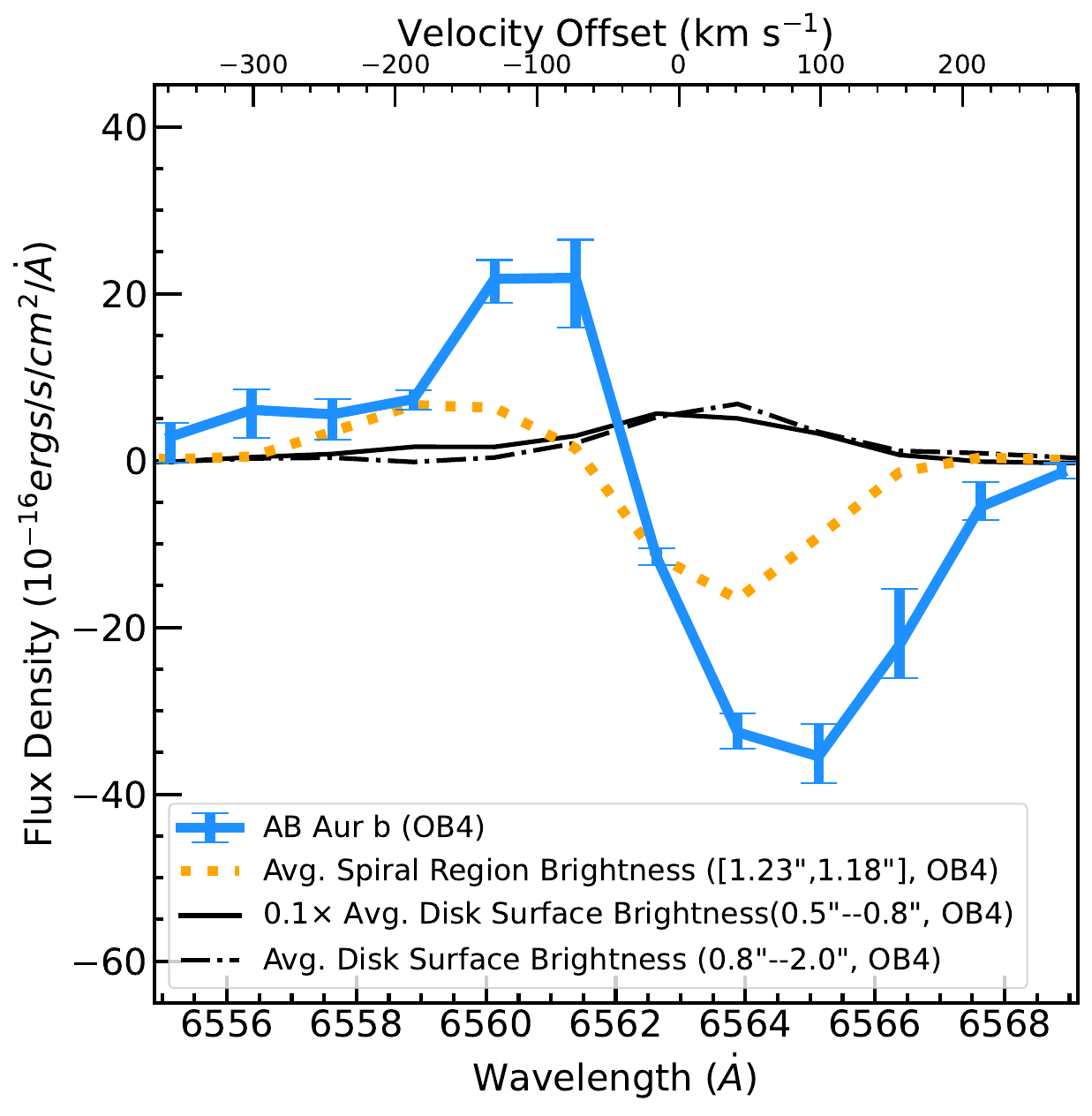}
    \caption{(Left) Spectrum of AB Aur b compared to scaled stellar spectra for OB4 and OB2.  The corresponding velocities are in the system rest frame and are corrected for the system's radial velocity \citep{gaia2022}.  We obtain consistent AB Aur b spectra from an alternate extraction, treating the position and spatial extent of AB Aur b as free parameters for each channel instead of fixing them to results obtained in Sect \ref{sec:analysis}.1.  (Right) Spectrum of AB Aur b compared to the average spectrum of the disk at two angular separations and within one spiral arm.  
    }\label{fig:abaur_musespec} 
\end{figure*}

Figure \ref{fig:abaur_musespec} shows our results.  In both OB4 and OB2, AB Aur b's $H_{\rm \alpha}$ spectrum is strikingly dissimilar to that of its host star (left panel).   The AB Aur primary's $H_{\rm \alpha}$ line is single-peaked at $\sim$6562.8 \AA\ and narrow.  In contrast, AB Aur b's $H_{\rm \alpha}$ line is broad with an emission peak blue-shifted by $\sim$ 2--3 \AA\ ($\sim$ -100 km s$^{-1}$) from the star's $H_{\rm \alpha}$ line center and absorption reaching a maximum depth at 1--2 \AA\ redward ($\sim$ 75 km s$^{-1}$).      

AB Aur b's spectrum is also dissimilar to the average residual signal from the disk (right panel). 
The average flux density per pixel of the disk at 6558.88--6560.13 \AA\ in an annulus enclosing AB Aur b (0\farcs{}5--0\farcs{}8) is $\approx$18 times lower than the background-subtracted peak emission from AB Aur b and comparable to the rms noise; similarly, the disk's typical signal at 6563.88--6565.13 \AA\ is $\approx$15 times smaller than AB Aur b's maximum absorption.   At wider separations (0\farcs{}8--2\farcs{}0), the disk's average residual signal is $\approx$ an additional factor of 10 smaller.  
However, some regions of the disk's spiral arms
may vaguely share AB Aur b's blue-shifted $H_{\rm \alpha}$ emission and red-shifted absorption, albeit at signals far 
lower than AB Aur b's peak signal (dashed orange line)\footnote{We explored this possibility further by computing Pearson's correlation coefficient between the AB Aur b spectrum and the per-spaxel spectrum over the entire MUSE field of view.  Only the region identifying AB Aur b itself had $\rho$ $\gtrsim$ 0.95 over a $\gtrsim$FWHM-sized footprint.  Compared to the map of spiral structure seen with ALMA \citep{Speedie2024}, only part of one spiral (S1) could be identified from the correlation map and only at $\rho$ $\sim$ 0.8.   Thus, while some disk regions may have spectra with apparent positive/negative wavelength ranges like AB Aur b, all are very low SNR, and none are correlated at the $\gtrsim$ 2-$\sigma$ level.} .  

Table \ref{tab:bprop} summarizes our results from fitting individual MUSE channels.   Independently fitting the channels results in a position of [E,N]\farcs{} = [-0\farcs{}021, -0\farcs{}578] $\pm$ [0\farcs{}012, 0\farcs{}015] and an apparent spatial extent of [0\farcs{}236, 0\farcs{}188] $\pm$ [0\farcs{}028, 0\farcs{}35].  These positions agree to within 0.2--2$\sigma$ of the latest CHARIS and STIS astrometric measurements from \citet{Currie2022ABAurb}.   Additional astrometric errors -- e.g. pixel scale and north position angle uncertainties -- likely further limit our measurements' precisions but must be computed for MUSE: considering them is beyond the scope of this work.   

For the blue-shifted wavelengths with greater than 5-$\sigma$ detections, we derive a line flux of 8.19 $\pm$ 0.93 $\times$$10^{-15}$ ergs s$^{-1}$ cm$^{-2}$.   For the red-shifted absorption, the line flux is -13.59 $\pm$ 0.90 $\times$$10^{-15}$ ergs s$^{-1}$ cm$^{-2}$.  AB Aur b's spectrum is insensitive to the choice between fixing its morphology and position to those determined from the peak emission/absorption channels in Sect. \ref{peakmcmc} and allowing \texttt{emcee} to independently model each channel (labeled as `alt' in Figure \ref{fig:abaur_musespec}).   

The companion's $H_{\rm \alpha}$ line shape appears to resemble that of an inverse p Cygni profile seen in a small subset of young accreting objects \citep[e.g.][]{Reipurth1996}. We explore this similarity and its possible implications further in Section \ref{sect:empcomp}.


\begin{deluxetable*}{ll}[htbp]
\tablewidth{0pt} 
\tablecaption{AB Aur b $H_{\rm \alpha}$ Line Properties Derived from MCMC Modeling of Individual MUSE Channels \label{tab:bprop}}
\tablehead{
\colhead{Parameter} & \colhead{Value}}
\startdata
Mean Channel Position, Blue-Shifted Emission [E,N] (mas) & [-17, -583] $\pm$ [6, 2]\\
Mean Channel Position, Red-Shifted Absorption [E,N] (mas)  & [-24, -575] $\pm$ [14, 19]\\
Mean Channel Position, Combined [E,N] (mas)  & [-21, -578] $\pm$ [12, 15]\\\\
Mean Channel FWHM, Blue-Shifted Emission (mas) & [241, 169] $\pm$ [32, 16]\\
Mean Channel FWHM, Red-Shifted Absorption (mas) & [232, 201] $\pm$ [23, 38] \\
Mean Channel FWHM, Combined (mas) & [236, 188] $\pm$ [28, 35]\\\\
Line Flux, Emission ($\times$10$^{-15}$ ergs s$^{-1}$ cm$^{-2}$) & 8.19 $\pm$ 0.93 \\
Line Flux, Absorption ($\times$10$^{-15}$ ergs s$^{-1}$ cm$^{-2}$) & -13.59 $\pm$ 0.90\\
\enddata
\end{deluxetable*}




\begin{figure}[!htbp]
\centering
    \includegraphics[width=0.995\linewidth]{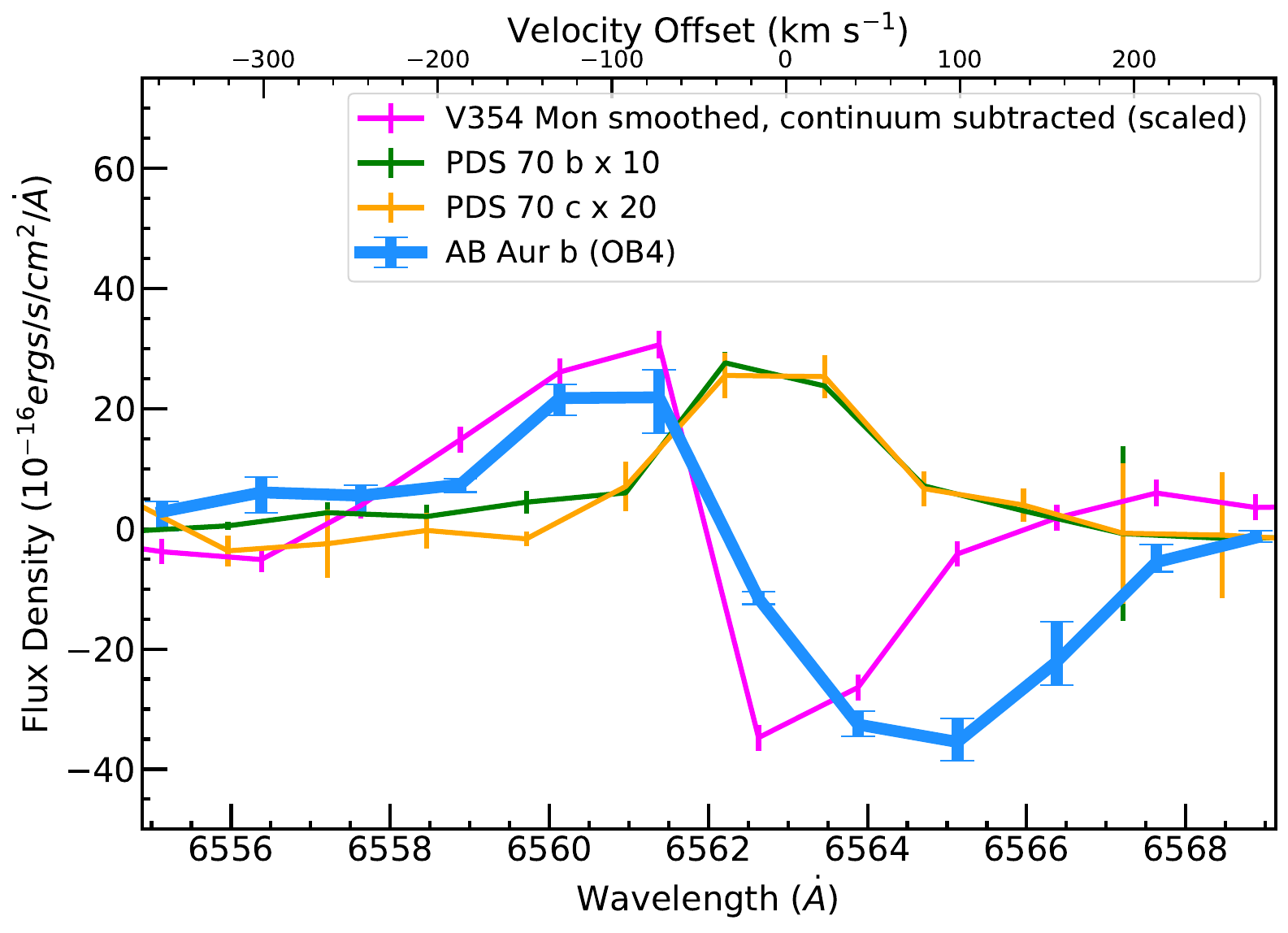}
    \caption{Spectrum of AB Aur b compared to that of the protoplanets PDS 70 b and c and the accreting T Tauri star V345 Mon.  The VAMPIRES $H_{\rm \alpha}$ bandpass used for a detection of AB Aur b in \citet{Currie2022ABAurb} covers $\approx$6558--6568 \AA; the HST/WFC3 $H_{\rm \alpha}$ bandpass (F656N) used for AB Aur b's detection \citep{Zhou2022ABAurb,Bowler2025} covers $\approx$6552.5--6570.4 \AA. }
    \label{fig:abaur_musespecemp} 
\end{figure}

\section{Discussion} \label{sec:discuss}
\subsection{Summary of Results and Basic Interpretation\label{sect:resultssummary}}
VLT/MUSE NFM high contrast medium-resolution spectroscopy reveals AB Aurigae b at $H_{\rm \alpha}$ in multiple epochs, verified using two separate data reduction methods to remove stellar halo light. 
In contrast to the single-peaked AB Aur primary's $H_{\rm \alpha}$ line and the average residual disk spectrum, AB Aur b appears in emission at blue-shifted wavelengths and in absorption at the $H_{\rm \alpha}$ line center and red-shifted wavelengths.  AB Aur b's location and morphology are consistent with previously published astrometry and morphologies for AB Aur b \citep{Currie2022ABAurb}.  

Variability does not provide a clear explanation for AB Aur b's MUSE detection.   Our PSF subtraction approach mitigates against a spurious detection induced by light travel time delays from the star \citep[e.g. see][]{Bowler2025}.  AB Aur b is detected in three different epochs where the $H_{\rm \alpha}$ line has different intensities, different H-alpha peak ($\sim$6563 \AA) to continuum (6555-6560 \AA, 6570-6575 \AA) ratios, and different shapes (especially see the 6555-6560 \AA\ region) (see Figure \ref{fig:abaur_balmer}).   In the two epochs with detections suitable for subsequent analysis (OB4 and OB2), the AB Aur b spectrum is consistent. 

Considered holistically, our findings indicate that AB Aur b's $H_{\rm \alpha}$ signal is highly unlikely to simply be reprocessed starlight from protoplanetary dust with a spectrum and scattering properties characteristic of the disk as a whole.   If it were otherwise, the AB Aur b signal would subtracted out, leaving no residual signal, as seen in many MUSE observations of structured protoplanetary disks \citep[e.g. see Figs. 6 and 12 in][]{Xie2020MUSE}. AB Aur b's $H_{\rm \alpha}$ signal must have a different explanation.

\subsection{Comparisons with Other Objects and Models of Accretion \label{sect:empcomp}}

To further explore the empirical context for AB Aur b's $H_{\rm \alpha}$ spectrum, we compare it with $H_{\rm \alpha}$ spectra from the PDS 70 protoplanets (Figure \ref{fig:abaur_musespecemp}).  To achieve the most direct comparisons, we downloaded and reduced the MUSE PDS 70 data published in \citet{Haffert2019} with the same reference spectrum approach used for AB Aur.  To these data, we add $H_{\rm \alpha}$ spectra for the young, accreting K star V354 Mon from 2010 Jan 18 \citep{Schneider2018} extracted from X-Shooter data downloaded from the ESO archive\footnote{\url{https://doi.org/10.18727/archive/71}}.  Appendix C details the data reduction procedures for PDS 70 and V345 Mon.

AB Aur b's broad blue-shifted emission and red-shifted absorption distinguish it from the spectra of protoplanets lying in disk cavities evacuated of gas and dust and isolated wide-separation planet-mass companions.  The PDS 70 b and c protoplanets (green and yellow curves) have narrow single-peaked $H_{\rm \alpha}$ lines \citep{Haffert2019} with a 50\% line width of $\approx$110 km s$^{-1}$ centered within $\approx$18-26 km s$^{-1}$ of the $H_{\rm \alpha}$ rest wavelength \citep{Hashimoto2020MUSE}.  The spectra for other very wide separation planet-mass companions studied with MUSE like Delorme 1 (AB)b have similar profiles \citep[e.g.,][]{Eriksson2020}.  

However, the shape of AB Aur b's spectrum is broadly reproduced by V345 Mon's blue-shifted emission and red-shifted absorption (cyan curve).   During the epoch at which V345 Mon's spectrum was obtained, its profile exhibits characteristics of a Reipurth class IV R object\footnote{We note that V345 Mon's spectrum is variable, where other epochs retain an inverse p Cygni profile but bear less resemblance to AB Aur b's spectrum \citep{Schneider2018}.}, representing a subset of young accreting (sub)stellar objects with inverse p Cygni profiles.   Other  objects with broad inverse p Cygni profiles qualitatively like AB Aur b's include the young K8-M0 T Tauri star V409 Tau \citep{Akimoto2019}, the older (5 Myr) solar-mass T Tauri star CVSO 1335 \citep{Thanathibodee2019acc}, and a large number of dipper stars such as TIC 434229695 \citep{Kasagi2022}.  

The physical mechanisms invoked to explain these objects' inverse p Cygni profiles provide a reference point for AB Aur b's $H_{\rm \alpha}$ spectrum.  For T Tauri stars, inverse P Cygni profiles are interpreted as evidence for infalling gas along the observer line of sight onto a central object (e.g. accretion), while their P Cygni profile counterparts identify $H_{\rm \alpha}$ detections from outflowing material (e.g. stellar winds) \citep[e.g.][]{Edwards1994Accretion,Hartmann1994Model}.   Recent models for Balmer emission from accretion and winds predict that P Cygni profiles are common only for high ratios of stellar wind mass loss rates vs accretion rates \citep{Wilson2022Model}.   Inverse P Cygni profiles from \citet{Kurosawa2006} and \citet{Wilson2022Model} qualitatively similar to V345 Mon's and AB Aur b's occur for a limited parameter phase space dominated by high inclinations (e.g. $i$ $\sim$ 80$^{o}$)\footnote{Variability seen in the line profiles for T Tauri stars with inverse P Cygni profiles is also in part attributed to high, edge-on inclinations where emission is modulated by circumstellar disk material along the line-of-sight \citep[e.g.][]{Schneider2018}.} where cold gas/dust lie in between the observer and the region where Balmer line emission is generated.   According to these models, inverse P Cygni profiles are also more often present for low accretion rates (e.g. 10$^{-9}$--10$^{-8}$ M$_{\odot}$ yr$^{-1}$) and maximum temperatures below 8500 K for the magnetospheric accretion funnel.   

Using the interpretive framework for T Tauri stars as a guide, AB Aur b's $H_{\rm \alpha}$ detection appears broadly consistent with the type of profile expected from infalling, accreting gas without a substantial disk wind at its source, albeit from an embedded source whose light is reprocessed by disk material.  The reason PDS 70 bc lack such a profile is then because it resides in a cleared region of the disk.

\subsection{Comparisons with Prior Work on AB Aur b's Optical Emission and Accretion Diagnostics \label{sect:context}}

In the \citet{Currie2022ABAurb} discovery paper, AB Aur b is detected with HST/STIS in broadband coronagraphic imaging and SCExAO/VAMPIRES $H_{\rm \alpha}$ narrowband imaging.   
The VAMPIRES $H_{\rm \alpha}$ bandpass used is centered on 6562.8 \AA\ with 
bandwidth of $\approx$10 \AA\ (T. Uyama, pvt. comm.).  
Later studies recover AB Aur b in HST/WFC3 $H_{\rm \alpha}$ narrowband imaging with a bandpass centered on 6561.5 \AA\ and a bandwidth of 17.9 \AA\ \citep{Zhou2022ABAurb,Bowler2025}.  
Despite these detections, the MUSE AB Aur b signal summed over either the VAMPIRES or WFC3 bandpasses would be negligible or even negative (net absorption).

Two scenarios allow us to reconcile the MUSE $H_{\rm \alpha}$ spectra with VAMPIRES and WFC3 $H_{\rm \alpha}$ imaging.   1) 
The MUSE data reduction is removing the $H_{\rm \alpha}$ continuum signal or a broader but single-peaked $H_{\rm \alpha}$ line due to overfitting the reference spectrum, while retaining the superimposed inverse P Cygni profile component. 
2) The VAMPIRES and WFC3 $H_{\rm \alpha}$ detections include multiple emission sources. For example, they result from some combination of signal coming from scattered light, Balmer continuum, and Balmer line emission, while MUSE only reveals the smaller contribution of the latter.  
Based on the details of our data reduction approach, scenario 2) is preferred
\footnote{Our MUSE data reduction masks the $H_{\rm \alpha}$ line region when constructing a per-spaxel reference spectrum to guard against overfitting; we reproduce line fluxes from PDS 70 b and c without the need for throughput corrections (see Appendix C).}.

Our results motivate a reconsideration of other recent optical imaging studies of AB Aur b.    
Based on F335W, F410N and F645N WFC3 photometry, \citet{Zhou2023ABAurb} claim that AB Aur b's UV-optical spectral energy distribution (SED) ``likely" originates from scattered light\footnote{In their previous work \citep{Zhou2022ABAurb}, the authors also claimed that AB Aur b's near-IR signal was consistent with a scattered light, since it purportedly matched a blackbody source of $\sim$3800 K.  This particular argument is incorrect.  Briefly, a 3800 K blackbody mismatches the AB Aur b SED: this is not apparent by inspection of \citet{Zhou2022ABAurb} because of the chosen y-axis limits and log scale.  Additionally, 3800 K does not correspond to the scattered light component of either the star or the reprocessed inner disk, and yet is consistent with the predicted shock temperatures possible during protoplanet growth \citep{Currie2022ABAurb}. 
\newline
\citet{Zhou2022ABAurb} also argue that Aur b does not incontrovertibly show near-IR molecular features, but neither do the PDS 70 bc protoplanets at the resolution of CHARIS ($R$ $\sim$20).  Thus, the lack of molecular features in AB Aur b's spectrum at this point is not a credible argument against it being an embedded protoplanet.  Similarly, they incorrectly draw skepticism from a supposed mismatch between the young planet spectrum used as a reference point in \citet{Currie2022ABAurb}.  The model matches the CHARIS spectrum to within 3-$\sigma$.  Furthermore, it was used to simply display the best-fitting model from one single, fiducial pre-computed grid to qualitatively guide the reader, not a determine a best-fit amongst a set of atmospheric grids with formal confidence limits.
}.  At minimum, our study complicates this statement, at least at $H_{\rm \alpha}$.

Furthermore, a reexamination of \citet{Zhou2023ABAurb}'s analyses identifies methodological errors that substantially undermine their conclusions more broadly.  AB Aur b is an concentrated, additive signal to the local disk intensity.  Thus, deriving its photometry requires an estimation of the signal properties (morphology, position, intensity) \textit{and} the background, where a proper estimate leaves a flat disk signal after removing the forward-modeled AB Aur b PSF, as in this work and in \citet{Currie2022ABAurb}\footnote{In particular, see Figures \ref{fig:abaur_modelsubtracted_halphablue} and \ref{fig:abaur_modelsubtracted_halphared} in this work and Supplemental Figure 9 in \citet{Currie2022ABAurb}.}.  However, the paper's model PSF clearly oversubtracts the disk background, leading to a miscalculation of AB Aur b's brightness: in middle/bottom-right panels Figure 3 of \citet{Zhou2023ABAurb}, the forward-modeled and scaled PSF should leave a flat, faintly red background instead of a blue background indicating a deficit in intensity shown.  Because the disk's optical signal can be at least as bright as AB Aur b's (or more) \citep[see][]{Currie2022ABAurb}, this error in photometry could be substantial.   As a result, it is premature to evaluate claims about AB Aur b's UV-optical SED and agreement with accretion models from these data at this time.

From multiepoch WFC3 $H_{\rm \alpha}$ imaging, \citet{Bowler2025} argued that AB Aur b shows variability up to 330\%, uncorrelated with the host star's variability, ruling out unobstructed scattered starlight as the only source of AB Aur b's emission.  Our data find no evidence of such large $H_{\rm \alpha}$ variability, though we analyze only two epochs and it is possible that only components besides the inverse P Cygni profile are variable\footnote{However, both CHARIS data presented in \citet{Currie2022ABAurb} and unpublished results from subsequent epochs (T. Currie, 2025 in preparation) fail to identify such large variability.}.  Like \citet{Zhou2023ABAurb}, they do not appear to consider the disk background intensity when computing AB Aur b's photometry.   However, their omission is far less impactful than for \citet{Zhou2023ABAurb} since they focus on relative changes in AB Aur b's $H_{\rm \alpha}$ brightness rather than a flux-calibrated SED.  On the other hand, scattered-light variability of the disk at AB Aur b's location could be mistaken for variability of AB Aur b.






\subsection{Evidence for Accretion onto a Protoplanet, Alternate Scenarios, and Future Directions}

Despite the disagreement between AB Aur b's spectrum and and the host star and its similarity with young accreting objects, we stop short of concluding that MUSE \textit{decisively} identifies accretion onto a protoplanet.  AB Aur b's $H_{\rm \alpha}$ profile is in strong tension with expectations for scattered starlight characteristic of the disk as a whole.  However, we cannot rule out scattered light for AB Aur b's $H_{\rm \alpha}$ line \textit{in some other form}.  

The most credible alternate hypothesis is that AB Aur b's MUSE detection is a complex radiative transfer effect.  For example, if the disk region at AB Aur b's projected position is compositionally unique, the $H_{\rm \alpha}$ line that it scatters may differ from that of the rest of the disk: subtracting the scaled average disk line profile from that of AB Aur b may induce a spurious residual resembling an inverse p Cygni profile.  The AB Aur primary's $H_{\rm \alpha}$ line includes contributions from both magnetospheric accretion and a disk wind.  Models for accretion and disk winds with different weighted contributions yield a diverse set of $H_{\rm \alpha}$ spectra \citep[][]{Kurosawa2006,Wilson2022Model}.  It may be possible to construct a model that reproduces the AB Aur b profile compromised purely of a disk wind and light scattered from the accreting primary, although the physical mechanism for doing so is as yet unclear.  Furthermore, jovian protoplanet accretion models may reproduce PDS 70 bc's relatively narrow $H_{\rm \alpha}$ profiles \citep[e.g.][]{Thanathibodee2019PDS70,Aoyama2018}.  Whether they could reproduce AB Aur b's significantly larger velocity offsets is unclear.

Residual signal from the AB Aur disk's spirals provides perhaps the best evidence at least complicating the picture of AB Aur b's signal originating from accretion.   At least some of this residual signal has a spectrum bearing a very coarse resemblance to AB Aur b's spectrum (i.e. blue-shifted emission and red-shifted absorption).   AB Aur b's $H_{\rm \alpha}$ location may bear some similarity to a broader spiral phenomenon.
However, the brightnesses of AB Aur b vs. the spirals limit their similarity.  In optical imaging (e.g. HST/STIS) and in near-IR total intensity (SCExAO/CHARIS) AB Aur b is $\approx$25-100\% brighter than the bulk disk (including the spirals), but the spirals are $\sim$ 5 times fainter in MUSE HRSDI: AB Aur b has excess signal at $H_{\rm \alpha}$.  Furthermore, as we noted in footnote 23, while some spiral regions may have spectra with apparent positive/negative wavelength ranges like AB Aur b, all are very low SNR, and none are correlated with AB Aur b’s spectrum at the $>$2$\sigma$ level. 

Emission from spirals may also have a non-scattered-light origin \citep[e.g. shocked thermal emission;][]{Hord2017}.  If the spirals are experiencing infall – as expected for a disk undergoing gravitational instability \citep{Speedie2024} or an accreting planet – it may be possible that emission from this process will not fully subtract out with MUSE HRSDI.

A major challenge for any nonaccretion scenario is how to explain why the scattering properties of the AB Aur b region happen to be different from the rest of AB Aurigae's protoplanetary disk without resorting to special pleading and why the colocation with an IR source showing evidence in favor of being a protoplanet is coincidental.  One option, that AB Aur b's $H_{\rm \alpha}$  line profile a by-product of scattering angle, is highly unlikely.  AB Aur b's position does not coincide with either the system's forward-scattering peak or back-scattering minimum \citep[e.g. see][]{Dykes2024,Currie2022ABAurb}: disk regions at similar angles are a null detection with MUSE.   

Fully considering the above analyses is well beyond the scope of this paper.  Exploring them in depth requires detailed models of accretion and radiative transfer.  Some of these topics will be further considered in a follow-up study (J. Hashimoto et al. 2025, in preparation).

AB Aurigae b joins the PDS 70 bc protoplanets as having high contrast, medium-spectral-resolution $H_{\rm \alpha}$ detections and is the first that appears to have a broad $H_{\rm \alpha}$ signal resembling an inverse P Cygni profile. 
New optical data could illuminate the nature of AB Aur b's emission sources\footnote{AB Aurigae is also included in a current JWST NIRCam coronagraphic imaging survey to image planets in protoplanetary disks with spiral structure (JWST PID 3947; PI B. Bowler).  It is highly unlikely that this program will provide clear new constraints on AB Aurigae b.  Relative to AB Aur b, the disk is particularly bright at 2--5 $\mu m$.  Convolving the native AB Aurigae scene with the JWST PSF (which has substantial off-axis power) causes severe contamination of AB Aur b signal by the highly structured disk, even considering transmission loss from the coronagraph.  Simulations of the NIRCam scene assuming even perfect PSF subtraction fail to yield a clear AB Aur b detection (K. Lawson, pvt. comm.).  However, this program could provide new detections of companions at much wider separations.}. 
\citet{Bowler2025} find no correlation between the star's $H_{\rm \alpha}$ variability and that of AB Aur b's: a lack of correlation between AB Aur b's variability (or lack thereof) and the \textit{disk} would disfavor the disk shadowing hypothesis considered by \citet{Bowler2025}.  STIS coronagraphic imaging would enable this measurement using RDI and classical PSF subtraction, bypassing forward-modeling challenges faced with WFC3 imaging processed with advanced, often-``aggressive" least-square algorithms like KLIP.  STIS coronagraphic spectroscopy would better measure the Balmer continuum from AB Aur b to fully explore differences with the disk's spectrum.  SCExAO/VAMPIRES could also probe the polarization of emission from AB Aur b at $H_{\rm \alpha}$.

\begin{acknowledgments}

The authors thank the anonymous reviewer for helpful comments that improved this Letter.  The authors also thank Jess Speedie for helpful conversations about ALMA data for AB Aurigae, Kellen Lawson for discussions about the (non)detectability of AB Aur b with JWST, and Mihoko~Konishi and Ko~Hosokawa for basic data reduction advice.   Finally, TC thanks Sebaastian Haffert for detailed discussions about the intricacies of VLT/MUSE NFM data reduction, possible pitfalls, and ways to assess the robustness of the AB Aur b detection, as well as helpful feedback on a preliminary version of these results.
This study is based on observations collected at the European Organisation for Astronomical Research in the Southern Hemisphere under ESO program(s) 0110.C-0259(A), 60.A-9100(K), and 084.C-1095(A).  M.T. is supported by JSPS KAKENHI grant No.24H00242.  T.C., M.E.M., and E.D. are supported by NASA-ROSES XRP grant 80NSSC23K0139.

\end{acknowledgments}

\software{
          astropy \citep{astropy:2013,astropy:2018,astropy:2022},
          Numpy \citep{VanDerWalt2011}, 
          emcee \citep{foreman-mackey+2013},
          corner \citep{foreman-mackey+2013}
          }

\facilities{VLT}

\bibliography{bibliography_abaur}{}

\begin{thebibliography}{}
\expandafter\ifx\csname natexlab\endcsname\relax\def\natexlab#1{#1}\fi
\providecommand{\url}[1]{\href{#1}{#1}}
\providecommand{\dodoi}[1]{doi:~\href{http://doi.org/#1}{\nolinkurl{#1}}}
\providecommand{\doeprint}[1]{\href{http://ascl.net/#1}{\nolinkurl{http://ascl.net/#1}}}
\providecommand{\doarXiv}[1]{\href{https://arxiv.org/abs/#1}{\nolinkurl{https://arxiv.org/abs/#1}}}

\bibitem[{H. {Akimoto} \& Y. {Itoh}(2019){Akimoto} \& {Itoh}}]{Akimoto2019}
{Akimoto}, H., \& {Itoh}, Y. 2019, \bibinfo{title}{{Optical Spectroscopic
  Monitoring Observations of a T Tauri Star V409 Tau},} International Journal
  of Astronomy and Astrophysics, 9, 321, \dodoi{10.4236/ijaa.2019.93023}

\bibitem[{S.~M. {Andrews}(2020){Andrews}}]{Andrews2020}
{Andrews}, S.~M. 2020, \bibinfo{title}{{Observations of Protoplanetary Disk
  Structures},} \araa, 58, 483, \dodoi{10.1146/annurev-astro-031220-010302}

\bibitem[{Y. {Aoyama} {et~al.}(2018){Aoyama}, {Ikoma}, \&
  {Tanigawa}}]{Aoyama2018}
{Aoyama}, Y., {Ikoma}, M., \& {Tanigawa}, T. 2018, \bibinfo{title}{{Theoretical
  Model of Hydrogen Line Emission from Accreting Gas Giants},} \apj, 866, 84,
  \dodoi{10.3847/1538-4357/aadc11}

\bibitem[{R. {Arsenault} {et~al.}(2008){Arsenault}, {Madec}, {Hubin},
  {Paufique}, {Stroebele}, {Soenke}, {Donaldson}, {Fedrigo}, {Oberti}, {Tordo},
  {Downing}, {Kiekebusch}, {Conzelmann}, {Duchateau}, {Jost}, {Hackenberg},
  {Bonaccini Calia}, {Delabre}, {Stuik}, {Biasi}, {Gallieni}, {Lazzarini},
  {Lelouarn}, \& {Glindeman}}]{Arsenault2008VLT-AO}
{Arsenault}, R., {Madec}, P.~Y., {Hubin}, N., {et~al.} 2008, in Society of
  Photo-Optical Instrumentation Engineers (SPIE) Conference Series, Vol. 7015,
  Adaptive Optics Systems, ed. N.~{Hubin}, C.~E. {Max}, \& P.~L. {Wizinowich},
  701524, \dodoi{10.1117/12.790359}

\bibitem[{ {Astropy Collaboration} {et~al.}(2013){Astropy Collaboration},
  {Robitaille}, {Tollerud}, {Greenfield}, {Droettboom}, {Bray}, {Aldcroft},
  {Davis}, {Ginsburg}, {Price-Whelan}, {Kerzendorf}, {Conley}, {Crighton},
  {Barbary}, {Muna}, {Ferguson}, {Grollier}, {Parikh}, {Nair}, {Unther},
  {Deil}, {Woillez}, {Conseil}, {Kramer}, {Turner}, {Singer}, {Fox}, {Weaver},
  {Zabalza}, {Edwards}, {Azalee Bostroem}, {Burke}, {Casey}, {Crawford},
  {Dencheva}, {Ely}, {Jenness}, {Labrie}, {Lim}, {Pierfederici}, {Pontzen},
  {Ptak}, {Refsdal}, {Servillat}, \& {Streicher}}]{astropy:2013}
{Astropy Collaboration}, {Robitaille}, T.~P., {Tollerud}, E.~J., {et~al.} 2013,
  \bibinfo{title}{{Astropy: A community Python package for astronomy},} \aap,
  558, A33, \dodoi{10.1051/0004-6361/201322068}

\bibitem[{ {Astropy Collaboration} {et~al.}(2022){Astropy Collaboration},
  {Price-Whelan}, {Lim}, {Earl}, {Starkman}, {Bradley}, {Shupe}, {Patil},
  {Corrales}, {Brasseur}, {N{\"o}the}, {Donath}, {Tollerud}, {Morris},
  {Ginsburg}, {Vaher}, {Weaver}, {Tocknell}, {Jamieson}, {van Kerkwijk},
  {Robitaille}, {Merry}, {Bachetti}, {G{\"u}nther}, {Aldcroft},
  {Alvarado-Montes}, {Archibald}, {B{\'o}di}, {Bapat}, {Barentsen},
  {Baz{\'a}n}, {Biswas}, {Boquien}, {Burke}, {Cara}, {Cara}, {Conroy},
  {Conseil}, {Craig}, {Cross}, {Cruz}, {D'Eugenio}, {Dencheva}, {Devillepoix},
  {Dietrich}, {Eigenbrot}, {Erben}, {Ferreira}, {Foreman-Mackey}, {Fox},
  {Freij}, {Garg}, {Geda}, {Glattly}, {Gondhalekar}, {Gordon}, {Grant},
  {Greenfield}, {Groener}, {Guest}, {Gurovich}, {Handberg}, {Hart},
  {Hatfield-Dodds}, {Homeier}, {Hosseinzadeh}, {Jenness}, {Jones}, {Joseph},
  {Kalmbach}, {Karamehmetoglu}, {Ka{\l}uszy{\'n}ski}, {Kelley}, {Kern},
  {Kerzendorf}, {Koch}, {Kulumani}, {Lee}, {Ly}, {Ma}, {MacBride}, {Maljaars},
  {Muna}, {Murphy}, {Norman}, {O'Steen}, {Oman}, {Pacifici}, {Pascual},
  {Pascual-Granado}, {Patil}, {Perren}, {Pickering}, {Rastogi}, {Roulston},
  {Ryan}, {Rykoff}, {Sabater}, {Sakurikar}, {Salgado}, {Sanghi}, {Saunders},
  {Savchenko}, {Schwardt}, {Seifert-Eckert}, {Shih}, {Jain}, {Shukla}, {Sick},
  {Simpson}, {Singanamalla}, {Singer}, {Singhal}, {Sinha}, {Sip{\H{o}}cz},
  {Spitler}, {Stansby}, {Streicher}, {{\v{S}}umak}, {Swinbank}, {Taranu},
  {Tewary}, {Tremblay}, {de Val-Borro}, {Van Kooten}, {Vasovi{\'c}}, {Verma},
  {de Miranda Cardoso}, {Williams}, {Wilson}, {Winkel}, {Wood-Vasey}, {Xue},
  {Yoachim}, {Zhang}, {Zonca}, \& {Astropy Project
  Contributors}}]{astropy:2022}
{Astropy Collaboration}, {Price-Whelan}, A.~M., {Lim}, P.~L., {et~al.} 2022,
  \bibinfo{title}{{The Astropy Project: Sustaining and Growing a
  Community-oriented Open-source Project and the Latest Major Release (v5.0) of
  the Core Package},} \apj, 935, 167, \dodoi{10.3847/1538-4357/ac7c74}

\bibitem[{R. {Bacon} {et~al.}(2010){Bacon}, {Accardo}, {Adjali}, {Anwand},
  {Bauer}, {Biswas}, {Blaizot}, {Boudon}, {Brau-Nogue}, {Brinchmann},
  {Caillier}, {Capoani}, {Carollo}, {Contini}, {Couderc}, {Daguis{\'e}},
  {Deiries}, {Delabre}, {Dreizler}, {Dubois}, {Dupieux}, {Dupuy}, {Emsellem},
  {Fechner}, {Fleischmann}, {Fran{\c{c}}ois}, {Gallou}, {Gharsa}, {Glindemann},
  {Gojak}, {Guiderdoni}, {Hansali}, {Hahn}, {Jarno}, {Kelz}, {Koehler},
  {Kosmalski}, {Laurent}, {Le Floch}, {Lilly}, {Lizon}, {Loupias}, {Manescau},
  {Monstein}, {Nicklas}, {Olaya}, {Pares}, {Pasquini}, {P{\'e}contal-Rousset},
  {Pell{\'o}}, {Petit}, {Popow}, {Reiss}, {Remillieux}, {Renault}, {Roth},
  {Rupprecht}, {Serre}, {Schaye}, {Soucail}, {Steinmetz}, {Streicher}, {Stuik},
  {Valentin}, {Vernet}, {Weilbacher}, {Wisotzki}, \& {Yerle}}]{Bacon2010MUSE}
{Bacon}, R., {Accardo}, M., {Adjali}, L., {et~al.} 2010, in Society of
  Photo-Optical Instrumentation Engineers (SPIE) Conference Series, Vol. 7735,
  Ground-based and Airborne Instrumentation for Astronomy III, ed. I.~S.
  {McLean}, S.~K. {Ramsay}, \& H.~{Takami}, 773508, \dodoi{10.1117/12.856027}

\bibitem[{M. {Benisty} {et~al.}(2023){Benisty}, {Dominik}, {Follette},
  {Garufi}, {Ginski}, {Hashimoto}, {Keppler}, {Kley}, \&
  {Monnier}}]{Benisty2023}
{Benisty}, M., {Dominik}, C., {Follette}, K., {et~al.} 2023, in Astronomical
  Society of the Pacific Conference Series, Vol. 534, Protostars and Planets
  VII, ed. S.~{Inutsuka}, Y.~{Aikawa}, T.~{Muto}, K.~{Tomida}, \& M.~{Tamura},
  605, \dodoi{10.48550/arXiv.2203.09991}

\bibitem[{S.~K. {Betti} {et~al.}(2022){Betti}, {Follette}, {Jorquera},
  {Duch{\^e}ne}, {Mazoyer}, {Bonnefoy}, {Chauvin}, {P{\'e}rez}, {Boccaletti},
  {Pinte}, {Weinberger}, {Grady}, {Close}, {Defr{\`e}re}, {Downey}, {Hinz},
  {M{\'e}nard}, {Schneider}, {Skemer}, \& {Vaz}}]{Betti2022}
{Betti}, S.~K., {Follette}, K., {Jorquera}, S., {et~al.} 2022,
  \bibinfo{title}{{Detection of Near-infrared Water Ice at the Surface of the
  (Pre)Transitional Disk of AB Aur: Informing Icy Grain Abundance, Composition,
  and Size},} \aj, 163, 145, \dodoi{10.3847/1538-3881/ac4d9b}

\bibitem[{A. {Boccaletti} {et~al.}(2020){Boccaletti}, {Di Folco}, {Pantin},
  {Dutrey}, {Guilloteau}, {Tang}, {Pi{\'e}tu}, {Habart}, {Milli}, {Beck}, \&
  {Maire}}]{Boccaletti2020ABAur}
{Boccaletti}, A., {Di Folco}, E., {Pantin}, E., {et~al.} 2020,
  \bibinfo{title}{{Possible evidence of ongoing planet formation in AB Aurigae.
  A showcase of the SPHERE/ALMA synergy},} \aap, 637, L5,
  \dodoi{10.1051/0004-6361/202038008}

\bibitem[{B.~P. {Bowler} {et~al.}(2025){Bowler}, {Zhou}, {Biddle}, {Yushu
  Jiang}, {Bae}, {Close}, {Follette}, {Franson}, {Kraus}, {Sanghi}, {Tran},
  {Ward-Duong}, {Wu}, \& {Zhu}}]{Bowler2025}
{Bowler}, B.~P., {Zhou}, Y., {Biddle}, L.~I., {et~al.} 2025,
  \bibinfo{title}{{H$\alpha$ Variability of AB Aur b with the Hubble Space
  Telescope: Probing the Nature of a Protoplanet Candidate with Accretion Light
  Echoes},} arXiv e-prints, arXiv:2502.14736.
\newblock \doarXiv{2502.14736}

\bibitem[{C. {Catala} {et~al.}(1999){Catala}, {Donati}, {B{\"o}hm},
  {Landstreet}, {Henrichs}, {Unruh}, {Hao}, {Collier Cameron}, {Johns-Krull},
  {Kaper}, {Simon}, {Foing}, {Cao}, {Ehrenfreund}, {Hatzes}, {Huang}, {de
  Jong}, {Kennelly}, {ten Kulve}, {Mulliss}, {Neff}, {Oliveira}, {Schrijvers},
  {Stempels}, {Telting}, {Walton}, \& {Yang}}]{Catala1999}
{Catala}, C., {Donati}, J.~F., {B{\"o}hm}, T., {et~al.} 1999,
  \bibinfo{title}{{Short-term spectroscopic variability in the pre-main
  sequence Herbig AE star AB Aurigae during the MUSICOS 96 campaign},} \aap,
  345, 884

\bibitem[{G. {Costigan} {et~al.}(2014){Costigan}, {Vink}, {Scholz}, {Ray}, \&
  {Testi}}]{Costigan2014}
{Costigan}, G., {Vink}, J.~S., {Scholz}, A., {Ray}, T., \& {Testi}, L. 2014,
  \bibinfo{title}{{Temperaments of young stars: rapid mass accretion rate
  changes in T Tauri and Herbig Ae stars},} \mnras, 440, 3444,
  \dodoi{10.1093/mnras/stu529}

\bibitem[{T. {Currie}(2024){Currie}}]{Currie2024}
{Currie}, T. 2024, \bibinfo{title}{{Direct Imaging Detection of the Protoplanet
  AB Aur b at Wavelengths Covering Pa{\ensuremath{\beta}}},} Research Notes of
  the American Astronomical Society, 8, 146, \dodoi{10.3847/2515-5172/ad50ce}

\bibitem[{T. {Currie} {et~al.}(2023{\natexlab{a}}){Currie}, {Biller},
  {Lagrange}, {Marois}, {Guyon}, {Nielsen}, {Bonnefoy}, \& {De
  Rosa}}]{Currie2023PPVII}
{Currie}, T., {Biller}, B., {Lagrange}, A., {et~al.} 2023{\natexlab{a}}, in
  Astronomical Society of the Pacific Conference Series, Vol. 534, Astronomical
  Society of the Pacific Conference Series, ed. S.~{Inutsuka}, Y.~{Aikawa},
  T.~{Muto}, K.~{Tomida}, \& M.~{Tamura}, 799,
  \dodoi{10.48550/arXiv.2205.05696}

\bibitem[{T. {Currie} {et~al.}(2015){Currie}, {Cloutier}, {Brittain}, {Grady},
  {Burrows}, {Muto}, {Kenyon}, \& {Kuchner}}]{Currie2015}
{Currie}, T., {Cloutier}, R., {Brittain}, S., {et~al.} 2015,
  \bibinfo{title}{{Resolving the HD 100546 Protoplanetary System with the
  Gemini Planet Imager: Evidence for Multiple Forming, Accreting Planets},}
  \apjl, 814, L27, \dodoi{10.1088/2041-8205/814/2/L27}

\bibitem[{T. {Currie} {et~al.}(2011){Currie}, {Burrows}, {Itoh}, {Matsumura},
  {Fukagawa}, {Apai}, {Madhusudhan}, {Hinz}, {Rodigas}, {Kasper}, {Pyo}, \&
  {Ogino}}]{Currie2011}
{Currie}, T., {Burrows}, A., {Itoh}, Y., {et~al.} 2011, \bibinfo{title}{{A
  Combined Subaru/VLT/MMT 1-5 {\ensuremath{\mu}}m Study of Planets Orbiting HR
  8799: Implications for Atmospheric Properties, Masses, and Formation},} \apj,
  729, 128, \dodoi{10.1088/0004-637X/729/2/128}

\bibitem[{T. {Currie} {et~al.}(2022){Currie}, {Lawson}, {Schneider}, {Lyra},
  {Wisniewski}, {Grady}, {Guyon}, {Tamura}, {Kotani}, {Kawahara}, {Brandt},
  {Uyama}, {Muto}, {Dong}, {Kudo}, {Hashimoto}, {Fukagawa}, {Wagner}, {Lozi},
  {Chilcote}, {Tobin}, {Groff}, {Ward-Duong}, {Januszewski}, {Norris},
  {Tuthill}, {van der Marel}, {Sitko}, {Deo}, {Vievard}, {Jovanovic},
  {Martinache}, \& {Skaf}}]{Currie2022ABAurb}
{Currie}, T., {Lawson}, K., {Schneider}, G., {et~al.} 2022,
  \bibinfo{title}{{Images of embedded Jovian planet formation at a wide
  separation around AB Aurigae},} Nature Astronomy,
  \dodoi{10.1038/s41550-022-01634-x}

\bibitem[{T. {Currie} {et~al.}(2023{\natexlab{b}}){Currie}, {Brandt}, {Brandt},
  {Lacy}, {Burrows}, {Guyon}, {Tamura}, {Liu}, {Sagynbayeva}, {Tobin},
  {Chilcote}, {Groff}, {Marois}, {Thompson}, {Murphy}, {Kuzuhara}, {Lawson},
  {Lozi}, {Deo}, {Vievard}, {Skaf}, {Uyama}, {Jovanovic}, {Martinache},
  {Kasdin}, {Kudo}, {McElwain}, {Janson}, {Wisniewski}, {Hodapp}, {Nishikawa},
  {He{\l}miniak}, {Kwon}, \& {Hayashi}}]{Currie2023a}
{Currie}, T., {Brandt}, G.~M., {Brandt}, T.~D., {et~al.} 2023{\natexlab{b}},
  \bibinfo{title}{{Direct imaging and astrometric detection of a gas giant
  planet orbiting an accelerating star},} Science, 380, 198,
  \dodoi{10.1126/science.abo6192}

\bibitem[{D. {Demars} {et~al.}(2023){Demars}, {Bonnefoy}, {Dougados}, {Aoyama},
  {Thanathibodee}, {Marleau}, {Tremblin}, {Delorme}, {Palma-Bifani}, {Petrus},
  {Bowler}, {Chauvin}, \& {Lagrange}}]{Demars2023}
{Demars}, D., {Bonnefoy}, M., {Dougados}, C., {et~al.} 2023,
  \bibinfo{title}{{Emission line variability of young 10-30 M$_{Jup}$
  companions. I. The case of GQ Lup b and GSC 06214-00210 b},} \aap, 676, A123,
  \dodoi{10.1051/0004-6361/202346221}

\bibitem[{E. {Dykes} {et~al.}(2024){Dykes}, {Currie}, {Lawson}, {Lucas},
  {Kudo}, {Chen}, {Guyon}, {Groff}, {Lozi}, {Chilcote}, {Brandt}, {Vievard},
  {Skaf}, {Deo}, {Morsy}, {Bovie}, {Uyama}, {Grady}, {Sitko}, {Hashimoto},
  {Martinache}, {Jovanovic}, {Tamura}, \& {Kasdin}}]{Dykes2024}
{Dykes}, E., {Currie}, T., {Lawson}, K., {et~al.} 2024,
  \bibinfo{title}{{SCExAO/CHARIS Near-infrared Scattered-light Imaging and
  Integral Field Spectropolarimetry of the AB Aurigae Protoplanetary System},}
  \apj, 977, 172, \dodoi{10.3847/1538-4357/ad8ba0}

\bibitem[{S. {Edwards} {et~al.}(1994){Edwards}, {Hartigan}, {Ghandour}, \&
  {Andrulis}}]{Edwards1994Accretion}
{Edwards}, S., {Hartigan}, P., {Ghandour}, L., \& {Andrulis}, C. 1994,
  \bibinfo{title}{{Spectroscopic Evidence for Magnetospheric Accretion in
  Classical T Tauri Stars},} \aj, 108, 1056, \dodoi{10.1086/117134}

\bibitem[{S.~C. {Eriksson} {et~al.}(2020){Eriksson}, {Asensio Torres},
  {Janson}, {Aoyama}, {Marleau}, {Bonnefoy}, \& {Petrus}}]{Eriksson2020}
{Eriksson}, S.~C., {Asensio Torres}, R., {Janson}, M., {et~al.} 2020,
  \bibinfo{title}{{Strong H{\ensuremath{\alpha}} emission and signs of
  accretion in a circumbinary planetary mass companion from MUSE},} \aap, 638,
  L6, \dodoi{10.1051/0004-6361/202038131}

\bibitem[{D. {Foreman-Mackey} {et~al.}(2013){Foreman-Mackey}, {Hogg}, {Lang},
  \& {Goodman}}]{foreman-mackey+2013}
{Foreman-Mackey}, D., {Hogg}, D.~W., {Lang}, D., \& {Goodman}, J. 2013,
  \bibinfo{title}{{emcee: The MCMC Hammer},} \pasp, 125, 306,
  \dodoi{10.1086/670067}

\bibitem[{L. {Francis} \& N. {van der Marel}(2020){Francis} \& {van der
  Marel}}]{fran2020a}
{Francis}, L., \& {van der Marel}, N. 2020, \bibinfo{title}{{Dust depleted
  inner disks in a large sample of transition disks through long-baseline ALMA
  observations},} arXiv e-prints, arXiv:2003.00079.
\newblock \doarXiv{2003.00079}

\bibitem[{M. {Fukagawa} {et~al.}(2004){Fukagawa}, {Hayashi}, {Tamura}, {Itoh},
  {Hayashi}, {Oasa}, {Takeuchi}, {Morino}, {Murakawa}, {Oya}, {Yamashita},
  {Suto}, {Mayama}, {Naoi}, {Ishii}, {Pyo}, {Nishikawa}, {Takato}, {Usuda},
  {Ando}, {Iye}, {Miyama}, \& {Kaifu}}]{Fukagawa2004ABAur}
{Fukagawa}, M., {Hayashi}, M., {Tamura}, M., {et~al.} 2004,
  \bibinfo{title}{{Spiral Structure in the Circumstellar Disk around AB
  Aurigae},} \apjl, 605, L53, \dodoi{10.1086/420699}

\bibitem[{ {Gaia Collaboration} {et~al.}(2022){Gaia Collaboration},
  {Vallenari}, {Brown}, {Prusti}, {de Bruijne}, {Arenou}, {Babusiaux},
  {Biermann}, {Creevey}, {Ducourant}, {Evans}, {Eyer}, {Guerra}, {Hutton},
  {Jordi}, {Klioner}, {Lammers}, {Lindegren}, {Luri}, {Mignard}, {Panem},
  {Pourbaix}, {Randich}, {Sartoretti}, {Soubiran}, {Tanga}, {Walton},
  {Bailer-Jones}, {Bastian}, {Drimmel}, {Jansen}, {Katz}, {Lattanzi}, {van
  Leeuwen}, {Bakker}, {Cacciari}, {Casta{\~n}eda}, {De Angeli}, {Fabricius},
  {Fouesneau}, {Fr{\'e}mat}, {Galluccio}, {Guerrier}, {Heiter}, {Masana},
  {Messineo}, {Mowlavi}, {Nicolas}, {Nienartowicz}, {Pailler}, {Panuzzo},
  {Riclet}, {Roux}, {Seabroke}, {Sordo{\o}rcit}, {Th{\'e}venin},
  {Gracia-Abril}, {Portell}, {Teyssier}, {Altmann}, {Andrae}, {Audard},
  {Bellas-Velidis}, {Benson}, {Berthier}, {Blomme}, {Burgess}, {Busonero},
  {Busso}, {C{\'a}novas}, {Carry}, {Cellino}, {Cheek}, {Clementini},
  {Damerdji}, {Davidson}, {de Teodoro}, {Nu{\~n}ez Campos}, {Delchambre},
  {Dell'Oro}, {Esquej}, {Fern{\'a}ndez-Hern{\'a}ndez}, {Fraile}, {Garabato},
  {Garc{\'\i}a-Lario}, {Gosset}, {Haigron}, {Halbwachs}, {Hambly}, {Harrison},
  {Hern{\'a}ndez}, {Hestroffer}, {Hodgkin}, {Holl}, {Jan{\ss}en}, {Jevardat de
  Fombelle}, {Jordan}, {Krone-Martins}, {Lanzafame}, {L{\"o}ffler}, {Marchal},
  {Marrese}, {Moitinho}, {Muinonen}, {Osborne}, {Pancino}, {Pauwels},
  {Recio-Blanco}, {Reyl{\'e}}, {Riello}, {Rimoldini}, {Roegiers}, {Rybizki},
  {Sarro}, {Siopis}, {Smith}, {Sozzetti}, {Utrilla}, {van Leeuwen}, {Abbas},
  {{\'A}brah{\'a}m}, {Abreu Aramburu}, {Aerts}, {Aguado}, {Ajaj},
  {Aldea-Montero}, {Altavilla}, {{\'A}lvarez}, {Alves}, {Anders}, {Anderson},
  {Anglada Varela}, {Antoja}, {Baines}, {Baker}, {Balaguer-N{\'u}{\~n}ez},
  {Balbinot}, {Balog}, {Barache}, {Barbato}, {Barros}, {Barstow},
  {Bartolom{\'e}}, {Bassilana}, {Bauchet}, {Becciani}, {Bellazzini},
  {Berihuete}, {Bernet}, {Bertone}, {Bianchi}, {Binnenfeld}, {Blanco-Cuaresma},
  {Blazere}, {Boch}, {Bombrun}, {Bossini}, {Bouquillon}, {Bragaglia},
  {Bramante}, {Breedt}, {Bressan}, {Brouillet}, {Brugaletta}, {Bucciarelli},
  {Burlacu}, {Butkevich}, {Buzzi}, {Caffau}, {Cancelliere}, {Cantat-Gaudin},
  {Carballo}, {Carlucci}, {Carnerero}, {Carrasco}, {Casamiquela}, {Castellani},
  {Castro-Ginard}, {Chaoul}, {Charlot}, {Chemin}, {Chiaramida}, {Chiavassa},
  {Chornay}, {Comoretto}, {Contursi}, {Cooper}, {Cornez}, {Cowell}, {Crifo},
  {Cropper}, {Crosta}, {Crowley}, {Dafonte}, {Dapergolas}, {David}, {David},
  {de Laverny}, {De Luise}, {De March}, {De Ridder}, {de Souza}, {de Torres},
  {del Peloso}, {del Pozo}, {Delbo}, {Delgado}, {Delisle}, {Demouchy},
  {Dharmawardena}, {Di Matteo}, {Diakite}, {Diener}, {Distefano}, {Dolding},
  {Edvardsson}, {Enke}, {Fabre}, {Fabrizio}, {Faigler}, {Fedorets}, {Fernique},
  {Fienga}, {Figueras}, {Fournier}, {Fouron}, {Fragkoudi}, {Gai},
  {Garcia-Gutierrez}, {Garcia-Reinaldos}, {Garc{\'\i}a-Torres}, {Garofalo},
  {Gavel}, {Gavras}, {Gerlach}, {Geyer}, {Giacobbe}, {Gilmore}, {Girona},
  {Giuffrida}, {Gomel}, {Gomez}, {Gonz{\'a}lez-N{\'u}{\~n}ez},
  {Gonz{\'a}lez-Santamar{\'\i}a}, {Gonz{\'a}lez-Vidal}, {Granvik}, {Guillout},
  {Guiraud}, {Guti{\'e}rrez-S{\'a}nchez}, {Guy}, {Hatzidimitriou}, {Hauser},
  {Haywood}, {Helmer}, {Helmi}, {Sarmiento}, {Hidalgo}, {Hilger},
  {H{\l}adczuk}, {Hobbs}, {Holland}, {Huckle}, {Jardine}, {Jasniewicz},
  {Jean-Antoine Piccolo}, {Jim{\'e}nez-Arranz}, {Jorissen}, {Juaristi
  Campillo}, {Julbe}, {Karbevska}, {Kervella}, {Khanna}, {Kontizas},
  {Kordopatis}, {Korn}, {K{\'o}sp{\'a}l}, {Kostrzewa-Rutkowska},
  {Kruszy{\'n}ska}, {Kun}, {Laizeau}, {Lambert}, {Lanza}, {Lasne}, {Le
  Campion}, {Lebreton}, {Lebzelter}, {Leccia}, {Leclerc}, {Lecoeur-Taibi},
  {Liao}, {Licata}, {Lindstr{\o}m}, {Lister}, {Livanou}, {Lobel}, {Lorca},
  {Loup}, {Madrero Pardo}, {Magdaleno Romeo}, {Managau}, {Mann}, {Manteiga},
  {Marchant}, {Marconi}, {Marcos}, {Marcos Santos}, {Mar{\'\i}n Pina},
  {Marinoni}, {Marocco}, {Marshall}, {Polo}, {Mart{\'\i}n-Fleitas}, {Marton},
  {Mary}, {Masip}, {Massari}, {Mastrobuono-Battisti}, {Mazeh}, {McMillan},
  {Messina}, {Michalik}, {Millar}, {Mints}, {Molina}, {Molinaro}, {Moln{\'a}r},
  {Monari}, {Mongui{\'o}}, {Montegriffo}, {Montero}, {Mor}, {Mora},
  {Morbidelli}, {Morel}, {Morris}, {Muraveva}, {Murphy}, {Musella}, {Nagy},
  {Noval}, {Oca{\~n}a}, {Ogden}, {Ordenovic}, {Osinde}, {Pagani}, {Pagano},
  {Palaversa}, {Palicio}, {Pallas-Quintela}, {Panahi}, {Payne-Wardenaar},
  {Pe{\~n}alosa Esteller}, {Penttil{\"a}}, {Pichon}, {Piersimoni}, {Pineau},
  {Plachy}, {Plum}, {Poggio}, {Pr{\v{s}}a}, {Pulone}, {Racero}, {Ragaini},
  {Rainer}, {Raiteri}, {Rambaux}, {Ramos}, {Ramos-Lerate}, {Re Fiorentin},
  {Regibo}, {Richards}, {Rios Diaz}, {Ripepi}, {Riva}, {Rix}, {Rixon},
  {Robichon}, {Robin}, {Robin}, {Roelens}, {Rogues}, {Rohrbasser},
  {Romero-G{\'o}mez}, {Rowell}, {Royer}, {Ruz Mieres}, {Rybicki}, {Sadowski},
  {S{\'a}ez N{\'u}{\~n}ez}, {Sagrist{\`a} Sell{\'e}s}, {Sahlmann}, {Salguero},
  {Samaras}, {Sanchez Gimenez}, {Sanna}, {Santove{\~n}a}, {Sarasso},
  {Schultheis}, {Sciacca}, {Segol}, {Segovia}, {S{\'e}gransan}, {Semeux},
  {Shahaf}, {Siddiqui}, {Siebert}, {Siltala}, {Silvelo}, {Slezak}, {Slezak},
  {Smart}, {Snaith}, {Solano}, {Solitro}, {Souami}, {Souchay}, {Spagna},
  {Spina}, {Spoto}, {Steele}, {Steidelm{\"u}ller}, {Stephenson}, {S{\"u}veges},
  {Surdej}, {Szabados}, {Szegedi-Elek}, {Taris}, {Taylo}, {Teixeira},
  {Tolomei}, {Tonello}, {Torra}, {Torra}, {Torralba Elipe}, {Trabucchi},
  {Tsounis}, {Turon}, {Ulla}, {Unger}, {Vaillant}, {van Dillen}, {van Reeven},
  {Vanel}, {Vecchiato}, {Viala}, {Vicente}, {Voutsinas}, {Weiler}, {Wevers},
  {Wyrzykowski}, {Yoldas}, {Yvard}, {Zhao}, {Zorec}, {Zucker}, \&
  {Zwitter}}]{gaia2022}
{Gaia Collaboration}, {Vallenari}, A., {Brown}, A.~G.~A., {et~al.} 2022,
  \bibinfo{title}{{Gaia Data Release 3: Summary of the content and survey
  properties},} arXiv e-prints, arXiv:2208.00211,
  \dodoi{10.48550/arXiv.2208.00211}

\bibitem[{C. {Ginski} {et~al.}(2016){Ginski}, {Stolker}, {Pinilla}, {Dominik},
  {Boccaletti}, {de Boer}, {Benisty}, {Biller}, {Feldt}, {Garufi}, {Keller},
  {Kenworthy}, {Maire}, {M{\'e}nard}, {Mesa}, {Milli}, {Min}, {Pinte}, {Quanz},
  {van Boekel}, {Bonnefoy}, {Chauvin}, {Desidera}, {Gratton}, {Girard},
  {Keppler}, {Kopytova}, {Lagrange}, {Langlois}, {Rouan}, \&
  {Vigan}}]{Ginski2016}
{Ginski}, C., {Stolker}, T., {Pinilla}, P., {et~al.} 2016,
  \bibinfo{title}{{Direct detection of scattered light gaps in the transitional
  disk around HD 97048 with VLT/SPHERE},} \aap, 595, A112,
  \dodoi{10.1051/0004-6361/201629265}

\bibitem[{S.~Y. {Haffert} {et~al.}(2019){Haffert}, {Bohn}, {de Boer},
  {Snellen}, {Brinchmann}, {Girard}, {Keller}, \& {Bacon}}]{Haffert2019}
{Haffert}, S.~Y., {Bohn}, A.~J., {de Boer}, J., {et~al.} 2019,
  \bibinfo{title}{{Two accreting protoplanets around the young star PDS 70},}
  Nature Astronomy, 329, \dodoi{10.1038/s41550-019-0780-5}

\bibitem[{L. {Hartmann} {et~al.}(2016){Hartmann}, {Herczeg}, \&
  {Calvet}}]{Hartmann2016}
{Hartmann}, L., {Herczeg}, G., \& {Calvet}, N. 2016, \bibinfo{title}{{Accretion
  onto Pre-Main-Sequence Stars},} \araa, 54, 135,
  \dodoi{10.1146/annurev-astro-081915-023347}

\bibitem[{L. {Hartmann} {et~al.}(1994){Hartmann}, {Hewett}, \&
  {Calvet}}]{Hartmann1994Model}
{Hartmann}, L., {Hewett}, R., \& {Calvet}, N. 1994,
  \bibinfo{title}{{Magnetospheric Accretion Models for T Tauri Stars. I. Balmer
  Line Profiles without Rotation},} \apj, 426, 669, \dodoi{10.1086/174104}

\bibitem[{J. {Hashimoto} {et~al.}(2020){Hashimoto}, {Aoyama}, {Konishi},
  {Uyama}, {Takasao}, {Ikoma}, \& {Tanigawa}}]{Hashimoto2020MUSE}
{Hashimoto}, J., {Aoyama}, Y., {Konishi}, M., {et~al.} 2020,
  \bibinfo{title}{{Accretion Properties of PDS 70b with MUSE},} \aj, 159, 222,
  \dodoi{10.3847/1538-3881/ab811e}

\bibitem[{J. {Hashimoto} {et~al.}(2011){Hashimoto}, {Tamura}, {Muto}, {Kudo},
  {Fukagawa}, {Fukue}, {Goto}, {Grady}, {Henning}, {Hodapp}, {Honda},
  {Inutsuka}, {Kokubo}, {Knapp}, {McElwain}, {Momose}, {Ohashi}, {Okamoto},
  {Takami}, {Turner}, {Wisniewski}, {Janson}, {Abe}, {Brandner}, {Carson},
  {Egner}, {Feldt}, {Golota}, {Guyon}, {Hayano}, {Hayashi}, {Hayashi}, {Ishii},
  {Kandori}, {Kusakabe}, {Matsuo}, {Mayama}, {Miyama}, {Morino}, {Moro-Martin},
  {Nishimura}, {Pyo}, {Suto}, {Suzuki}, {Takato}, {Terada}, {Thalmann},
  {Tomono}, {Watanabe}, {Yamada}, {Takami}, \& {Usuda}}]{Hashimoto2011ABAur}
{Hashimoto}, J., {Tamura}, M., {Muto}, T., {et~al.} 2011,
  \bibinfo{title}{{Direct Imaging of Fine Structures in Giant Planet-forming
  Regions of the Protoplanetary Disk Around AB Aurigae},} \apjl, 729, L17,
  \dodoi{10.1088/2041-8205/729/2/L17}

\bibitem[{B. {Hord} {et~al.}(2017){Hord}, {Lyra}, {Flock}, {Turner}, \& {Mac
  Low}}]{Hord2017}
{Hord}, B., {Lyra}, W., {Flock}, M., {Turner}, N.~J., \& {Mac Low}, M.-M. 2017,
  \bibinfo{title}{{On Shocks Driven by High-mass Planets in Radiatively
  Inefficient Disks. III. Observational Signatures in Thermal Emission and
  Scattered Light},} \apj, 849, 164, \dodoi{10.3847/1538-4357/aa8fcf}

\bibitem[{T.-O. {Husser} {et~al.}(2016){Husser}, {Kamann}, {Dreizler}, {Wendt},
  {Wulff}, {Bacon}, {Wisotzki}, {Brinchmann}, {Weilbacher}, {Roth}, \&
  {Monreal-Ibero}}]{Husser2016a}
{Husser}, T.-O., {Kamann}, S., {Dreizler}, S., {et~al.} 2016,
  \bibinfo{title}{{MUSE crowded field 3D spectroscopy of over 12 000 stars in
  the globular cluster NGC 6397. I. The first comprehensive HRD of a globular
  cluster},} \aap, 588, A148, \dodoi{10.1051/0004-6361/201526949}

\bibitem[{Y. {Kasagi} {et~al.}(2022){Kasagi}, {Kotani}, {Kawahara}, {Tajiri},
  {Muto}, {Aizawa}, {Fujii}, {Hattori}, {Masuda}, {Momose}, {Ohsawa}, \&
  {Takita}}]{Kasagi2022}
{Kasagi}, Y., {Kotani}, T., {Kawahara}, H., {et~al.} 2022,
  \bibinfo{title}{{Dippers from TESS Full-frame Images. II. Spectroscopic
  Characterization of Four Young Dippers},} \apjs, 259, 40,
  \dodoi{10.3847/1538-4365/ac4e8b}

\bibitem[{M. {Keppler} {et~al.}(2018){Keppler}, {Benisty}, {M{\"u}ller},
  {Henning}, {van Boekel}, {Cantalloube}, {Ginski}, {van Holstein}, {Maire},
  {Pohl}, {Samland }, {Avenhaus}, {Baudino}, {Boccaletti}, {de Boer},
  {Bonnefoy}, {Chauvin}, {Desidera}, {Langlois}, {Lazzoni}, {Marleau},
  {Mordasini}, {Pawellek}, {Stolker}, {Vigan}, {Zurlo}, {Birnstiel},
  {Brandner}, {Feldt}, {Flock}, {Girard}, {Gratton}, {Hagelberg}, {Isella},
  {Janson}, {Juhasz}, {Kemmer}, {Kral}, {Lagrange}, {Launhardt}, {Matter},
  {M{\'e}nard}, {Milli}, {Molli{\`e}re}, {Olofsson}, {P{\'e}rez}, {Pinilla},
  {Pinte}, {Quanz}, {Schmidt}, {Udry}, {Wahhaj}, {Williams}, {Buenzli},
  {Cudel}, {Dominik}, {Galicher}, {Kasper}, {Lannier}, {Mesa}, {Mouillet},
  {Peretti}, {Perrot}, {Salter}, {Sissa}, {Wildi}, {Abe}, {Antichi},
  {Augereau}, {Baruffolo}, {Baudoz}, {Bazzon}, {Beuzit}, {Blanchard}, {Brems},
  {Buey}, {De Caprio}, {Carbillet}, {Carle}, {Cascone}, {Cheetham}, {Claudi},
  {Costille}, {Delboulb{\'e}}, {Dohlen}, {Fantinel}, {Feautrier}, {Fusco},
  {Giro}, {Gluck}, {Gry}, {Hubin}, {Hugot}, {Jaquet}, {Le Mignant}, {Llored},
  {Madec}, {Magnard}, {Martinez}, {Maurel}, {Meyer}, {M{\"o}ller-Nilsson},
  {Moulin}, {Mugnier}, {Orign{\'e}}, {Pavlov}, {Perret}, {Petit}, {Pragt},
  {Puget}, {Rabou}, {Ramos}, {Rigal}, {Rochat}, {Roelfsema}, {Rousset}, {Roux},
  {Salasnich}, {Sauvage}, {Sevin}, {Soenke}, {Stadler}, {Suarez}, {Turatto}, \&
  {Weber}}]{Keppler2018}
{Keppler}, M., {Benisty}, M., {M{\"u}ller}, A., {et~al.} 2018,
  \bibinfo{title}{{Discovery of a planetary-mass companion within the gap of
  the transition disk around PDS 70},} \aap, 617, A44,
  \dodoi{10.1051/0004-6361/201832957}

\bibitem[{R. {Kurosawa} {et~al.}(2006){Kurosawa}, {Harries}, \&
  {Symington}}]{Kurosawa2006}
{Kurosawa}, R., {Harries}, T.~J., \& {Symington}, N.~H. 2006,
  \bibinfo{title}{{On the formation of H{\ensuremath{\alpha}} line emission
  around classical T Tauri stars},} Monthly Notices of the Royal Astronomical
  Society, 370, 580, \dodoi{10.1111/j.1365-2966.2006.10527.x}

\bibitem[{D. {Mawet} {et~al.}(2014){Mawet}, {Milli}, {Wahhaj}, {Pelat},
  {Absil}, {Delacroix}, {Boccaletti}, {Kasper}, {Kenworthy}, {Marois},
  {Mennesson}, \& {Pueyo}}]{Mawet2014}
{Mawet}, D., {Milli}, J., {Wahhaj}, Z., {et~al.} 2014,
  \bibinfo{title}{{Fundamental Limitations of High Contrast Imaging Set by
  Small Sample Statistics},} \apj, 792, 97, \dodoi{10.1088/0004-637X/792/2/97}

\bibitem[{A. {Modigliani} {et~al.}(2010){Modigliani}, {Goldoni}, {Royer},
  {Haigron}, {Guglielmi}, {Fran{\c{c}}ois}, {Horrobin}, {Bristow}, {Vernet},
  {Moehler}, {Kerber}, {Ballester}, {Mason}, \&
  {Christensen}}]{Modigliani2010X-ShooterPipeline}
{Modigliani}, A., {Goldoni}, P., {Royer}, F., {et~al.} 2010, in Society of
  Photo-Optical Instrumentation Engineers (SPIE) Conference Series, Vol. 7737,
  Observatory Operations: Strategies, Processes, and Systems III, ed. D.~R.
  {Silva}, A.~B. {Peck}, \& B.~T. {Soifer}, 773728, \dodoi{10.1117/12.857211}

\bibitem[{T. {Muto} {et~al.}(2012){Muto}, {Grady}, {Hashimoto}, {Fukagawa},
  {Hornbeck}, {Sitko}, {Russell}, {Werren}, {Cur{\'e}}, {Currie}, {Ohashi},
  {Okamoto}, {Momose}, {Honda}, {Inutsuka}, {Takeuchi}, {Dong}, {Abe},
  {Brandner}, {Brandt}, {Carson}, {Egner}, {Feldt}, {Fukue}, {Goto}, {Guyon},
  {Hayano}, {Hayashi}, {Hayashi}, {Henning}, {Hodapp}, {Ishii}, {Iye},
  {Janson}, {Kandori}, {Knapp}, {Kudo}, {Kusakabe}, {Kuzuhara}, {Matsuo},
  {Mayama}, {McElwain}, {Miyama}, {Morino}, {Moro-Martin}, {Nishimura}, {Pyo},
  {Serabyn}, {Suto}, {Suzuki}, {Takami}, {Takato}, {Terada}, {Thalmann},
  {Tomono}, {Turner}, {Watanabe}, {Wisniewski}, {Yamada}, {Takami}, {Usuda}, \&
  {Tamura}}]{Muto2012}
{Muto}, T., {Grady}, C.~A., {Hashimoto}, J., {et~al.} 2012,
  \bibinfo{title}{{Discovery of Small-scale Spiral Structures in the Disk of
  SAO 206462 (HD 135344B): Implications for the Physical State of the Disk from
  Spiral Density Wave Theory},} \apjl, 748, L22,
  \dodoi{10.1088/2041-8205/748/2/L22}

\bibitem[{J. {Muzerolle} {et~al.}(1998){Muzerolle}, {Calvet}, \&
  {Hartmann}}]{Muzerolle1998Model}
{Muzerolle}, J., {Calvet}, N., \& {Hartmann}, L. 1998,
  \bibinfo{title}{{Magnetospheric Accretion Models for the Hydrogen Emission
  Lines of T Tauri Stars},} \apj, 492, 743, \dodoi{10.1086/305069}

\bibitem[{S. {Paardekooper} {et~al.}(2023){Paardekooper}, {Dong}, {Duffell},
  {Fung}, {Masset}, {Ogilvie}, \& {Tanaka}}]{Paardekooper2023PPVII}
{Paardekooper}, S., {Dong}, R., {Duffell}, P., {et~al.} 2023, in Astronomical
  Society of the Pacific Conference Series, Vol. 534, Astronomical Society of
  the Pacific Conference Series, ed. S.~{Inutsuka}, Y.~{Aikawa}, T.~{Muto},
  K.~{Tomida}, \& M.~{Tamura}, 685

\bibitem[{M.~D. {Perrin} {et~al.}(2009){Perrin}, {Schneider}, {Duchene},
  {Pinte}, {Grady}, {Wisniewski}, \& {Hines}}]{Perrin2009}
{Perrin}, M.~D., {Schneider}, G., {Duchene}, G., {et~al.} 2009,
  \bibinfo{title}{{The Case of AB Aurigae's Disk in Polarized Light: Is there
  Truly a Gap?},} \apjl, 707, L132, \dodoi{10.1088/0004-637X/707/2/L132}

\bibitem[{A.~M. {Price-Whelan} {et~al.}(2018){Price-Whelan}, {Sip{\H{o}}cz},
  {G{\"u}nther}, {Lim}, {Crawford}, {Conseil}, {Shupe}, {Craig}, {Dencheva},
  {Ginsburg}, {VanderPlas}, {Bradley}, {P{\'e}rez-Su{\'a}rez}, {de Val-Borro},
  {Paper Contributors}, {Aldcroft}, {Cruz}, {Robitaille}, {Tollerud},
  {Coordination Committee}, {Ardelean}, {Babej}, {Bach}, {Bachetti}, {Bakanov},
  {Bamford}, {Barentsen}, {Barmby}, {Baumbach}, {Berry}, {Biscani}, {Boquien},
  {Bostroem}, {Bouma}, {Brammer}, {Bray}, {Breytenbach}, {Buddelmeijer},
  {Burke}, {Calderone}, {Cano Rodr{\'\i}guez}, {Cara}, {Cardoso}, {Cheedella},
  {Copin}, {Corrales}, {Crichton}, {D{\textquoteright}Avella}, {Deil},
  {Depagne}, {Dietrich}, {Donath}, {Droettboom}, {Earl}, {Erben}, {Fabbro},
  {Ferreira}, {Finethy}, {Fox}, {Garrison}, {Gibbons}, {Goldstein}, {Gommers},
  {Greco}, {Greenfield}, {Groener}, {Grollier}, {Hagen}, {Hirst}, {Homeier},
  {Horton}, {Hosseinzadeh}, {Hu}, {Hunkeler}, {Ivezi{\'c}}, {Jain}, {Jenness},
  {Kanarek}, {Kendrew}, {Kern}, {Kerzendorf}, {Khvalko}, {King}, {Kirkby},
  {Kulkarni}, {Kumar}, {Lee}, {Lenz}, {Littlefair}, {Ma}, {Macleod},
  {Mastropietro}, {McCully}, {Montagnac}, {Morris}, {Mueller}, {Mumford},
  {Muna}, {Murphy}, {Nelson}, {Nguyen}, {Ninan}, {N{\"o}the}, {Ogaz}, {Oh},
  {Parejko}, {Parley}, {Pascual}, {Patil}, {Patil}, {Plunkett}, {Prochaska},
  {Rastogi}, {Reddy Janga}, {Sabater}, {Sakurikar}, {Seifert}, {Sherbert},
  {Sherwood-Taylor}, {Shih}, {Sick}, {Silbiger}, {Singanamalla}, {Singer},
  {Sladen}, {Sooley}, {Sornarajah}, {Streicher}, {Teuben}, {Thomas},
  {Tremblay}, {Turner}, {Terr{\'o}n}, {van Kerkwijk}, {de la Vega}, {Watkins},
  {Weaver}, {Whitmore}, {Woillez}, {Zabalza}, \& {Contributors}}]{astropy:2018}
{Price-Whelan}, A.~M., {Sip{\H{o}}cz}, B.~M., {G{\"u}nther}, H.~M., {et~al.}
  2018, \bibinfo{title}{{The Astropy Project: Building an Open-science Project
  and Status of the v2.0 Core Package},} \aj, 156, 123,
  \dodoi{10.3847/1538-3881/aabc4f}

\bibitem[{B. {Reipurth} {et~al.}(1996){Reipurth}, {Pedrosa}, \&
  {Lago}}]{Reipurth1996}
{Reipurth}, B., {Pedrosa}, A., \& {Lago}, M.~T.~V.~T. 1996,
  \bibinfo{title}{{H{\ensuremath{\alpha}} emission in pre-main sequence stars.
  I. an atlas of line profiles.},} \aaps, 120, 229

\bibitem[{P.~C. {Schneider} {et~al.}(2018){Schneider}, {Manara}, {Facchini},
  {G{\"u}nther}, {Herczeg}, {Fedele}, \& {Teixeira}}]{Schneider2018}
{Schneider}, P.~C., {Manara}, C.~F., {Facchini}, S., {et~al.} 2018,
  \bibinfo{title}{{Multi-epoch monitoring of the AA Tauri-like star V 354 Mon.
  Indications for a low gas-to-dust ratio in the inner disk warp},} \aap, 614,
  A108, \dodoi{10.1051/0004-6361/201731959}

\bibitem[{J. {Speedie} {et~al.}(2024){Speedie}, {Dong}, {Hall}, {Longarini},
  {Veronesi}, {Paneque-Carre{\~n}o}, {Lodato}, {Tang}, {Teague}, \&
  {Hashimoto}}]{Speedie2024}
{Speedie}, J., {Dong}, R., {Hall}, C., {et~al.} 2024,
  \bibinfo{title}{{Gravitational instability in a planet-forming disk},} \nat,
  633, 58, \dodoi{10.1038/s41586-024-07877-0}

\bibitem[{S. {Str{\"o}bele} {et~al.}(2012){Str{\"o}bele}, {La Penna},
  {Arsenault}, {Conzelmann}, {Delabre}, {Duchateau}, {Dorn}, {Fedrigo},
  {Hubin}, {Quentin}, {Jolley}, {Kiekebusch}, {Kirchbauer}, {Klein}, {Kolb},
  {Kuntschner}, {Le Louarn}, {Lizon}, {Madec}, {Pettazzi}, {Soenke}, {Tordo},
  {Vernet}, \& {Muradore}}]{Strobele2012MUSE-AO}
{Str{\"o}bele}, S., {La Penna}, P., {Arsenault}, R., {et~al.} 2012, in Society
  of Photo-Optical Instrumentation Engineers (SPIE) Conference Series, Vol.
  8447, Adaptive Optics Systems III, ed. B.~L. {Ellerbroek}, E.~{Marchetti}, \&
  J.-P. {V{\'e}ran}, 844737, \dodoi{10.1117/12.926110}

\bibitem[{Y.~W. {Tang} {et~al.}(2012){Tang}, {Guilloteau}, {Pi{\'e}tu},
  {Dutrey}, {Ohashi}, \& {Ho}}]{Tang2012}
{Tang}, Y.~W., {Guilloteau}, S., {Pi{\'e}tu}, V., {et~al.} 2012,
  \bibinfo{title}{{The circumstellar disk of AB Aurigae: evidence for envelope
  accretion at late stages of star formation?},} \aap, 547, A84,
  \dodoi{10.1051/0004-6361/201219414}

\bibitem[{Y.-W. {Tang} {et~al.}(2017){Tang}, {Guilloteau}, {Dutrey}, {Muto},
  {Shen}, {Gu}, {Inutsuka}, {Momose}, {Pietu}, {Fukagawa}, {Chapillon}, {Ho},
  {di Folco}, {Corder}, {Ohashi}, \& {Hashimoto}}]{Tang2017}
{Tang}, Y.-W., {Guilloteau}, S., {Dutrey}, A., {et~al.} 2017,
  \bibinfo{title}{{Planet Formation in AB Aurigae: Imaging of the Inner Gaseous
  Spirals Observed inside the Dust Cavity},} \apj, 840, 32,
  \dodoi{10.3847/1538-4357/aa6af7}

\bibitem[{T. {Thanathibodee} {et~al.}(2019{\natexlab{a}}){Thanathibodee},
  {Calvet}, {Bae}, {Muzerolle}, \& {Hern{\'a}ndez}}]{Thanathibodee2019PDS70}
{Thanathibodee}, T., {Calvet}, N., {Bae}, J., {Muzerolle}, J., \&
  {Hern{\'a}ndez}, R.~F. 2019{\natexlab{a}}, \bibinfo{title}{{Magnetospheric
  Accretion as a Source of H{\ensuremath{\alpha}} Emission from Protoplanets
  around PDS 70},} \apj, 885, 94, \dodoi{10.3847/1538-4357/ab44c1}

\bibitem[{T. {Thanathibodee} {et~al.}(2019{\natexlab{b}}){Thanathibodee},
  {Calvet}, {Muzerolle}, {Brice{\~n}o}, {Hern{\'a}ndez}, \&
  {Mauc{\'o}}}]{Thanathibodee2019acc}
{Thanathibodee}, T., {Calvet}, N., {Muzerolle}, J., {et~al.}
  2019{\natexlab{b}}, \bibinfo{title}{{Complex Magnetospheric Accretion Flows
  in the Low Accretor CVSO 1335},} \apj, 884, 86,
  \dodoi{10.3847/1538-4357/ab4127}

\bibitem[{T. {Uyama} {et~al.}(2021){Uyama}, {Xie}, {Aoyama}, {Beichman},
  {Hashimoto}, {Dong}, {Hasegawa}, {Ikoma}, {Mawet}, {McElwain}, {Ruffio},
  {Wagner}, {Wang}, \& {Zhou}}]{Uyama2021}
{Uyama}, T., {Xie}, C., {Aoyama}, Y., {et~al.} 2021,
  \bibinfo{title}{{Keck/OSIRIS Pa{\ensuremath{\beta}} High-contrast Imaging and
  Updated Constraints on PDS 70b},} \aj, 162, 214,
  \dodoi{10.3847/1538-3881/ac2739}

\bibitem[{S. {van der Walt} {et~al.}(2011){van der Walt}, {Colbert}, \&
  {Varoquaux}}]{VanDerWalt2011}
{van der Walt}, S., {Colbert}, S.~C., \& {Varoquaux}, G. 2011,
  \bibinfo{title}{{The NumPy Array: A Structure for Efficient Numerical
  Computation},} Computing in Science and Engineering, 13, 22,
  \dodoi{10.1109/MCSE.2011.37}

\bibitem[{J.~J. {Wang} {et~al.}(2015){Wang}, {Ruffio}, {De Rosa}, {Aguilar},
  {Wolff}, \& {Pueyo}}]{Wang2015pyKLIP}
{Wang}, J.~J., {Ruffio}, J.-B., {De Rosa}, R.~J., {et~al.} 2015,
  \bibinfo{title}{{pyKLIP: PSF Subtraction for Exoplanets and Disks},}
  \doeprint{1506.001}

\bibitem[{P.~M. {Weilbacher} {et~al.}(2020){Weilbacher}, {Palsa}, {Streicher},
  {Bacon}, {Urrutia}, {Wisotzki}, {Conseil}, {Husemann}, {Jarno}, {Kelz},
  {P{\'e}contal-Rousset}, {Richard}, {Roth}, {Selman}, \&
  {Vernet}}]{Weilbacher2020MUSEpipeline}
{Weilbacher}, P.~M., {Palsa}, R., {Streicher}, O., {et~al.} 2020,
  \bibinfo{title}{{The data processing pipeline for the MUSE instrument},}
  \aap, 641, A28, \dodoi{10.1051/0004-6361/202037855}

\bibitem[{T.~J.~G. {Wilson} {et~al.}(2022){Wilson}, {Matt}, {Harries}, \&
  {Herczeg}}]{Wilson2022Model}
{Wilson}, T.~J.~G., {Matt}, S., {Harries}, T.~J., \& {Herczeg}, G.~J. 2022,
  \bibinfo{title}{{Hydrogen emission from accretion and outflow in T Tauri
  stars},} \mnras, 514, 2162, \dodoi{10.1093/mnras/stac1397}

\bibitem[{C. {Xie} {et~al.}(2020){Xie}, {Haffert}, {de Boer}, {Kenworthy},
  {Brinchmann}, {Girard}, {Snellen}, \& {Keller}}]{Xie2020MUSE}
{Xie}, C., {Haffert}, S.~Y., {de Boer}, J., {et~al.} 2020,
  \bibinfo{title}{{Searching for proto-planets with MUSE},} \aap, 644, A149,
  \dodoi{10.1051/0004-6361/202038242}

\bibitem[{Y. {Zhou} {et~al.}(2022){Zhou}, {Sanghi}, {Bowler}, {Wu}, {Close},
  {Long}, {Ward-Duong}, {Zhu}, {Kraus}, {Follette}, \& {Bae}}]{Zhou2022ABAurb}
{Zhou}, Y., {Sanghi}, A., {Bowler}, B.~P., {et~al.} 2022,
  \bibinfo{title}{{HST/WFC3 H{\ensuremath{\alpha}} Direct-imaging Detection of
  a Pointlike Source in the Disk Cavity of AB Aur},} \apjl, 934, L13,
  \dodoi{10.3847/2041-8213/ac7fef}

\bibitem[{Y. {Zhou} {et~al.}(2023){Zhou}, {Bowler}, {Yang}, {Sanghi},
  {Herczeg}, {Kraus}, {Bae}, {Long}, {Follette}, {Ward-Duong}, {Zhu}, {Biddle},
  {Close}, {Yushu Jiang}, \& {Wu}}]{Zhou2023ABAurb}
{Zhou}, Y., {Bowler}, B.~P., {Yang}, H., {et~al.} 2023,
  \bibinfo{title}{{UV-Optical Emission of AB Aur b is Consistent with Scattered
  Stellar Light},} arXiv e-prints, arXiv:2308.16223,
  \dodoi{10.48550/arXiv.2308.16223}

\bibitem[{A. {Zurlo} {et~al.}(2016){Zurlo}, {Vigan}, {Galicher}, {Maire},
  {Mesa}, {Gratton}, {Chauvin}, {Kasper}, {Moutou}, {Bonnefoy}, {Desidera},
  {Abe}, {Apai}, {Baruffolo}, {Baudoz}, {Baudrand}, {Beuzit}, {Blancard},
  {Boccaletti}, {Cantalloube}, {Carle}, {Cascone}, {Charton}, {Claudi},
  {Costille}, {de Caprio}, {Dohlen}, {Dominik}, {Fantinel}, {Feautrier},
  {Feldt}, {Fusco}, {Gigan}, {Girard}, {Gisler}, {Gluck}, {Gry}, {Henning},
  {Hugot}, {Janson}, {Jaquet}, {Lagrange}, {Langlois}, {Llored}, {Madec},
  {Magnard}, {Martinez}, {Maurel}, {Mawet}, {Meyer}, {Milli},
  {Moeller-Nilsson}, {Mouillet}, {Orign{\'e}}, {Pavlov}, {Petit}, {Puget},
  {Quanz}, {Rabou}, {Ramos}, {Rousset}, {Roux}, {Salasnich}, {Salter},
  {Sauvage}, {Schmid}, {Soenke}, {Stadler}, {Suarez}, {Turatto}, {Udry},
  {Vakili}, {Wahhaj}, {Wildi}, \& {Antichi}}]{Zurlo2016}
{Zurlo}, A., {Vigan}, A., {Galicher}, R., {et~al.} 2016, \bibinfo{title}{{First
  light of the VLT planet finder SPHERE. III. New spectrophotometry and
  astrometry of the HR 8799 exoplanetary system},} \aap, 587, A57,
  \dodoi{10.1051/0004-6361/201526835}

\end{thebibliography}
\bibliographystyle{aasjournal}

\appendix 
\section{Raw Data Halo Intensity Profile and Spectrum}
\begin{figure*}[!htbp]
\centering
    \includegraphics[width=0.45\linewidth]{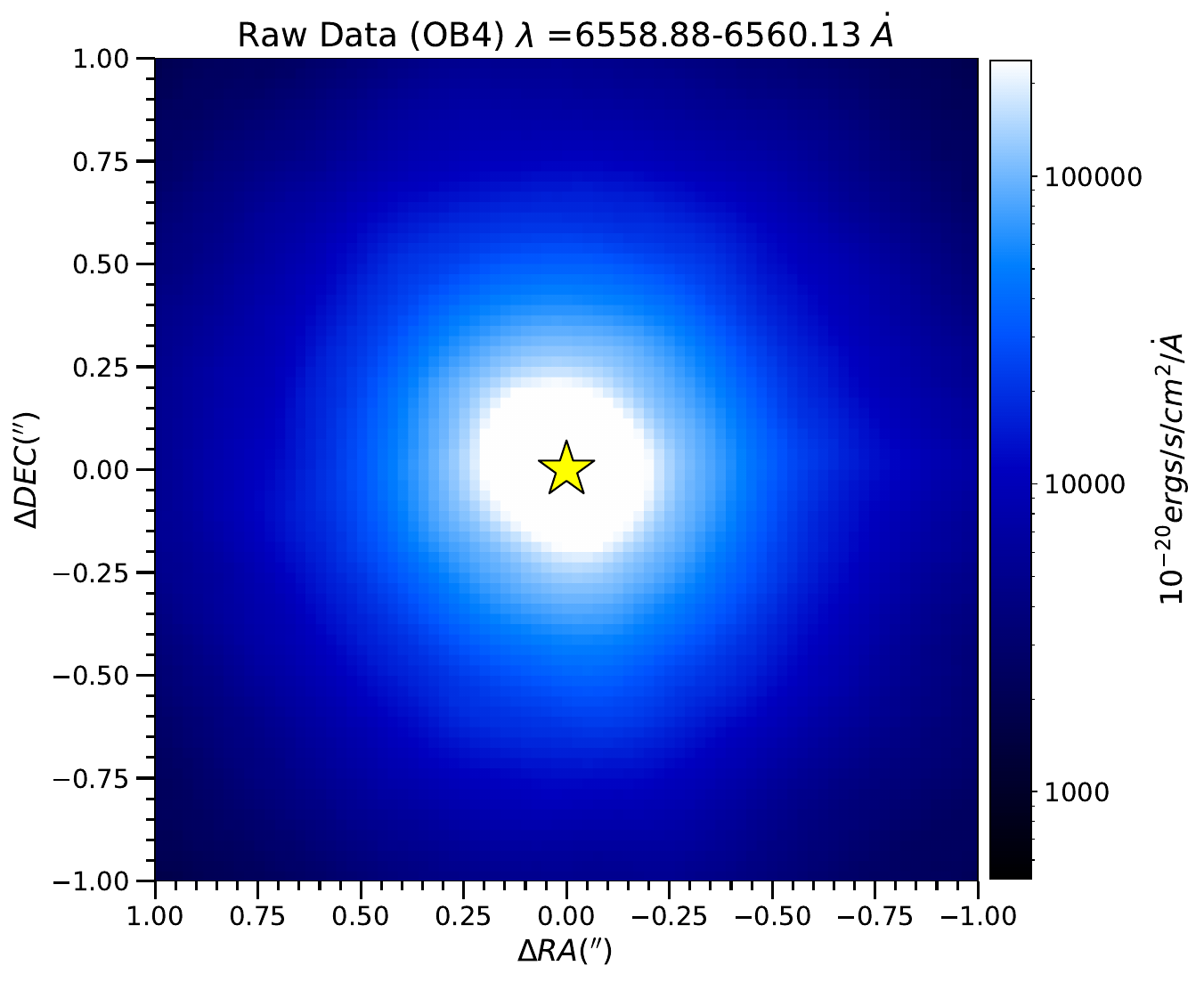}
    \includegraphics[width=0.45\linewidth]{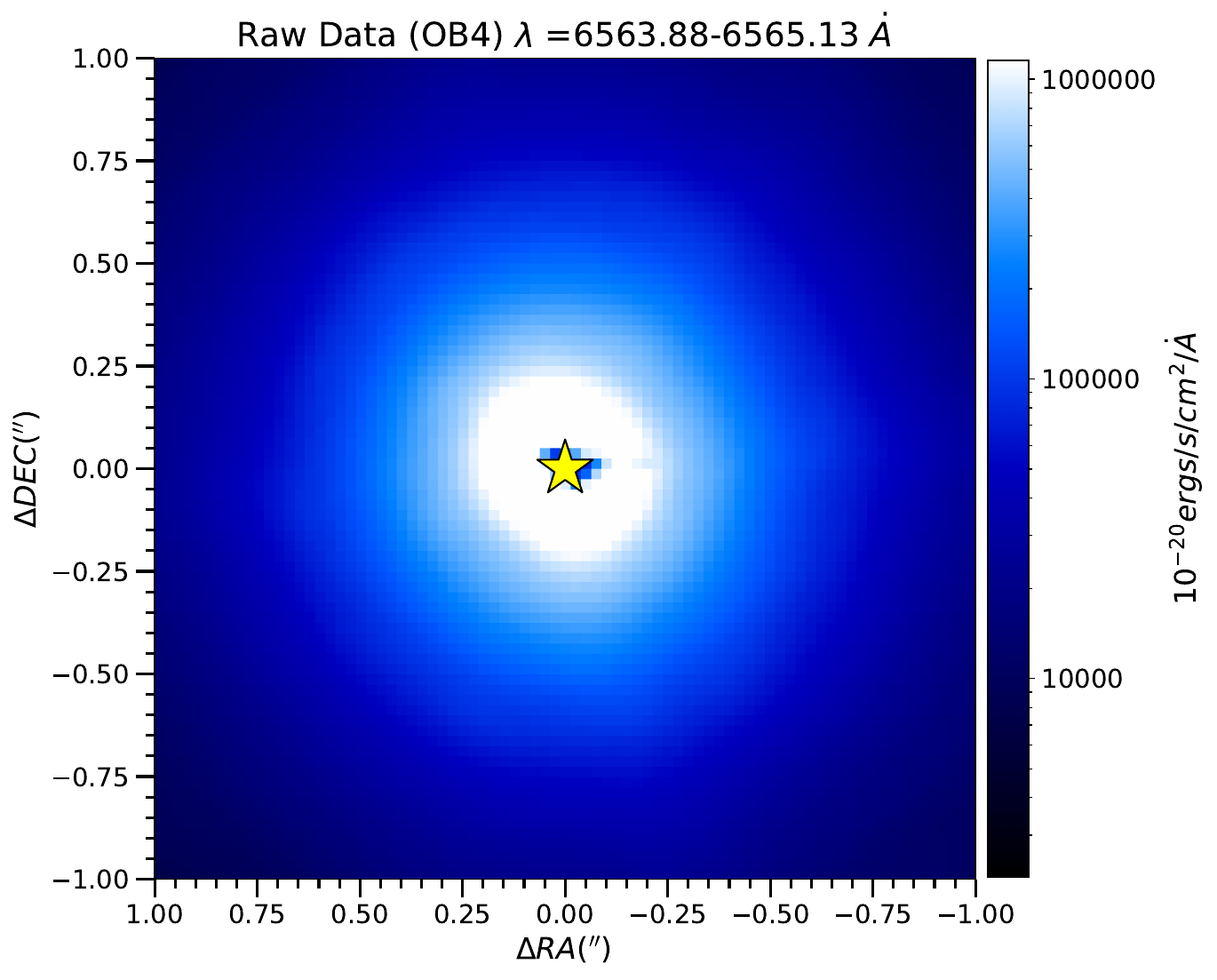}\\
    \includegraphics[width=0.9\linewidth]{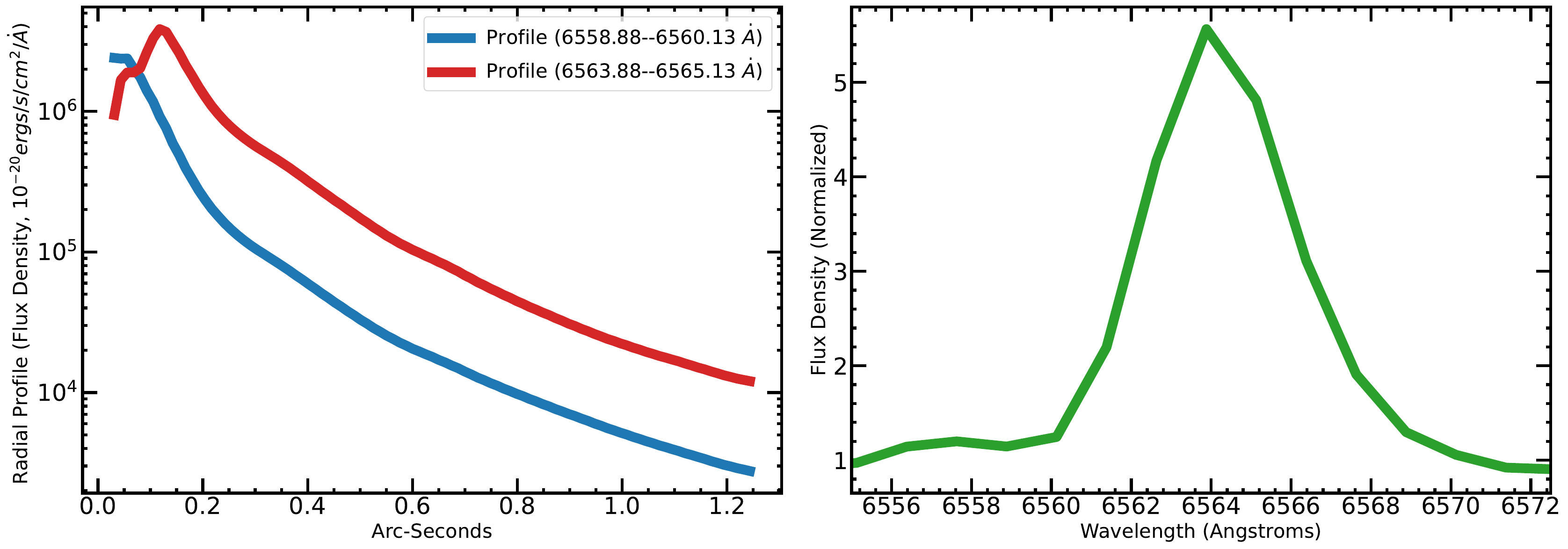}
    \caption{(top panels) Raw (pre-PSF subtraction) OB4 MUSE image at blue shifted and red-shifted wavelengths.  (bottom-left) Azimuthally-averaged radial intensity profile for the blue and red-shifted channels.  (bottom-right) Raw spectrum at the location of AB Aur b in OB4 normalized to the median intensity value across the entire wavelength range.  
    }\label{fig:abaur_psfsub_rawdata} 
\end{figure*}

Figure \ref{fig:abaur_psfsub_rawdata} analyses the raw MUSE data prior to PSF subtraction.   For channels displaced from the $H_{\rm \alpha}$ line center ($\approx$ 6563--6566 \AA), the inner $\approx$ 4--5 pixels are saturated or are non linear but the rest of the image plane is unaffected.  For channels covering the line center, the core is saturated/non-linear out to $\approx$0\farcs{}3 and the image plane is affected by horizontal banding 0\farcs{}3 above and below the star's x position, which is difficult to remove at the smallest ($\rho$ $\lesssim$ 0\farcs{}3) separations.  

At the position of AB Aur b, the halo is a factor of $\approx$6 fainter than at 0\farcs{}3 in both blue shifted and red-shifted channels.  AB Aur b's signal is about 5--9\% as bright as the halo intensity: detecting sources lying below the halo at this level is common \citep[e.g.][]{Currie2015}.  The raw spectrum of the halo at AB Aur b's location likewise is peaked at $\approx$6564 \AA.
 \newpage
 \section{Results Using Alternate Assumptions for Reference Spectrum Fitting}

\begin{figure*}[!htbp]
\centering
    \includegraphics[width=0.45\linewidth]{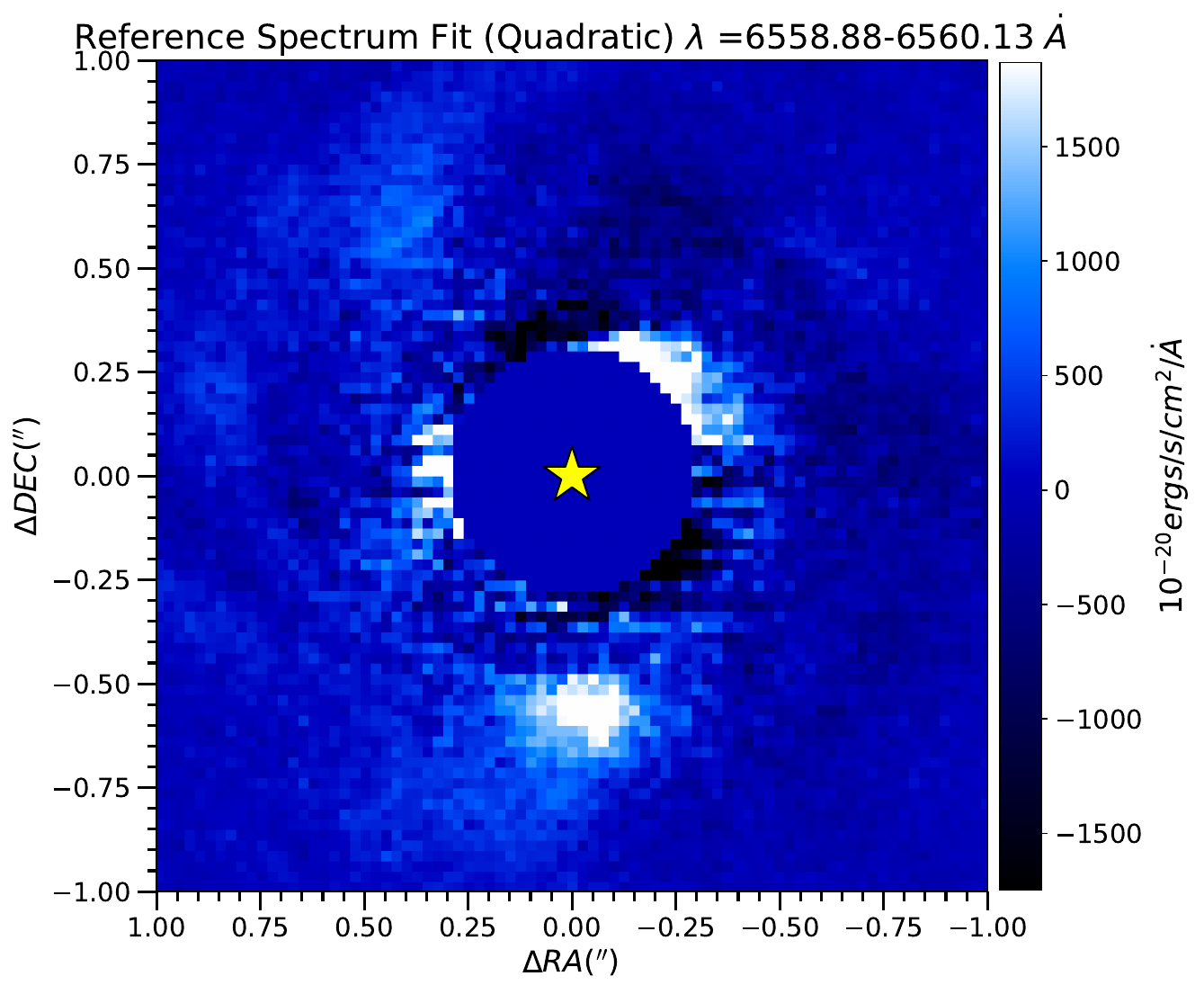}
    \includegraphics[width=0.45\linewidth]{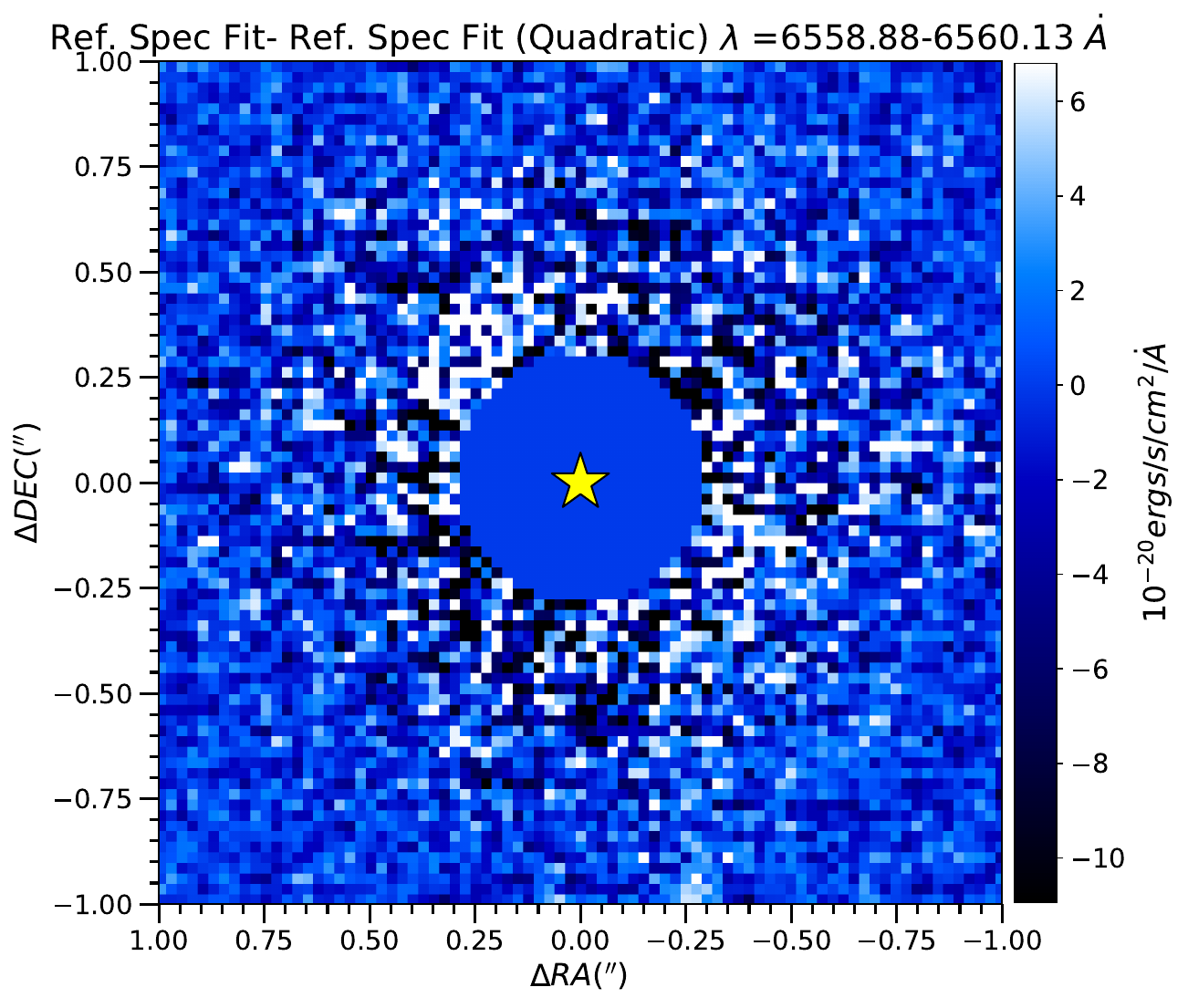}\\
        \includegraphics[width=0.45\linewidth]{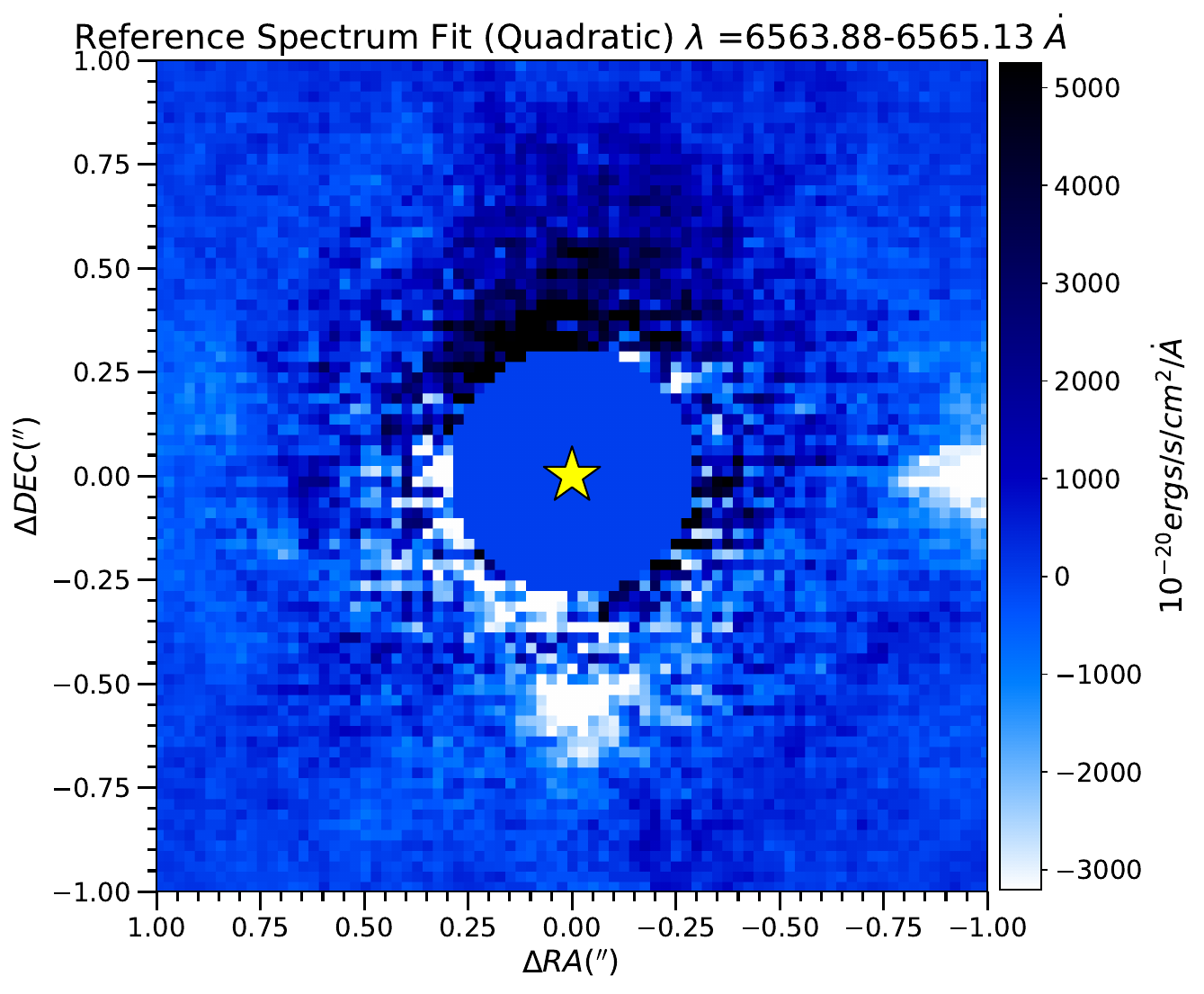}
    \includegraphics[width=0.45\linewidth]{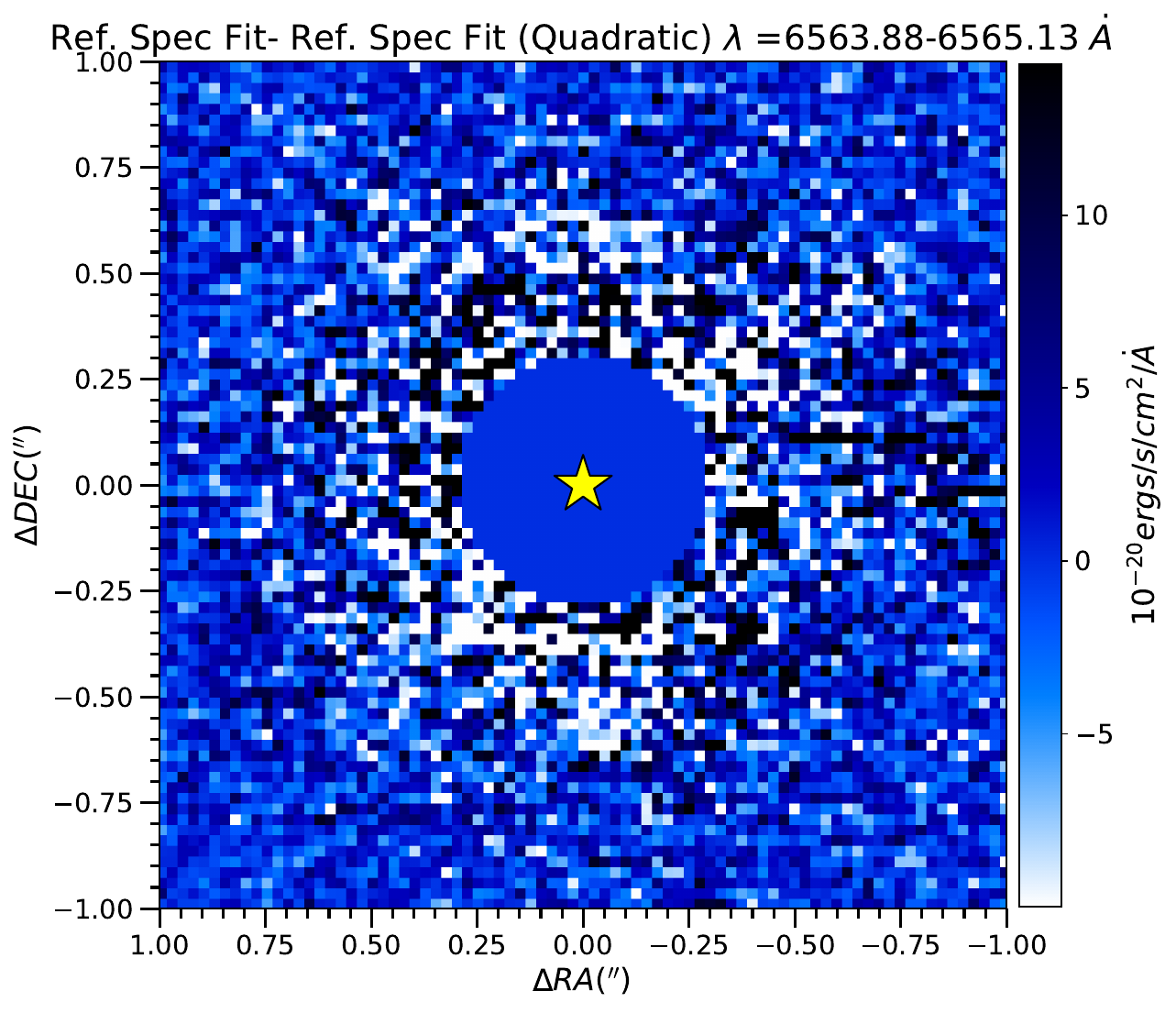}
    \caption{(top-left) PSF-subtracted MUSE image of AB Aurigae constructed from the average of the 6558.88 \AA\ and 6560.13 \AA\ slices in OB4 using instead a 2nd-degree polynomial fit. (top-right) The difference between this image and the one presented in Figure \ref{fig:abaur_blueshift1}. (Bottom panels) The results shown for red-shifted wavelengths showing AB Aur b in absorption.  Note the difference in the intensity scale for the difference images (right-hand panels).
    }\label{fig:abaur_psfsub_alt} 
\end{figure*}

\begin{figure*}[!htbp]
\centering
    \includegraphics[width=0.45\linewidth]{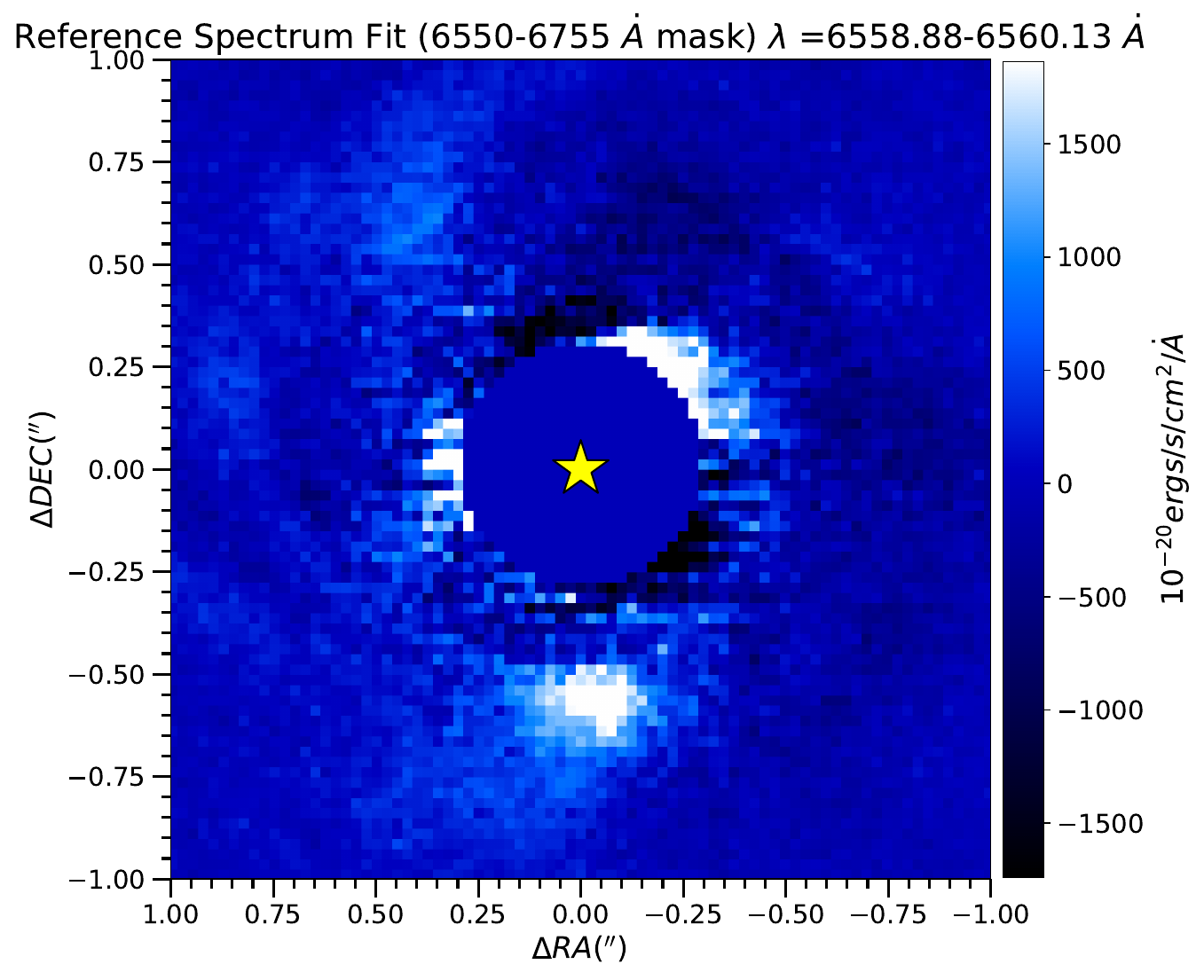}
    \includegraphics[width=0.45\linewidth]{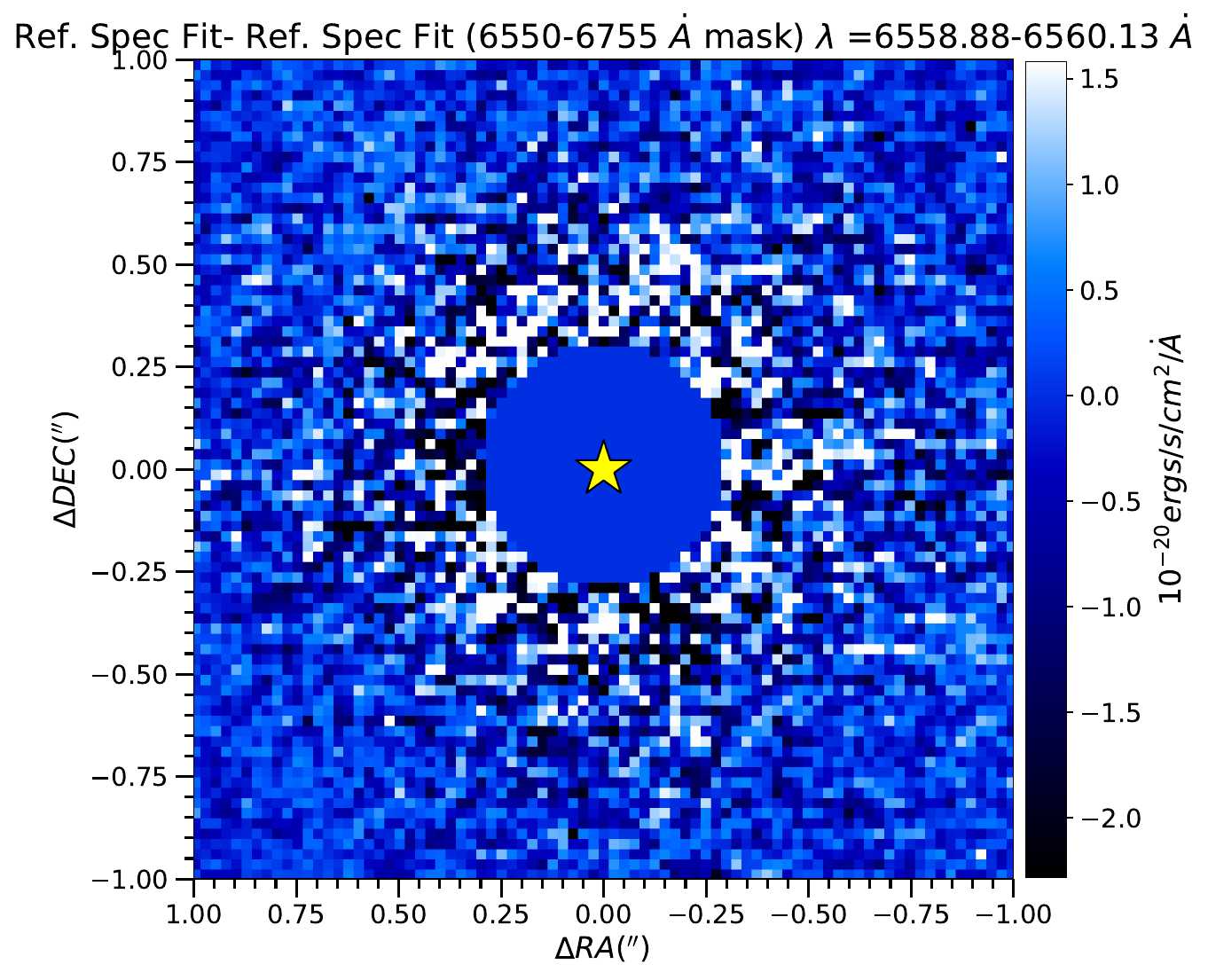}\\
        \includegraphics[width=0.45\linewidth]{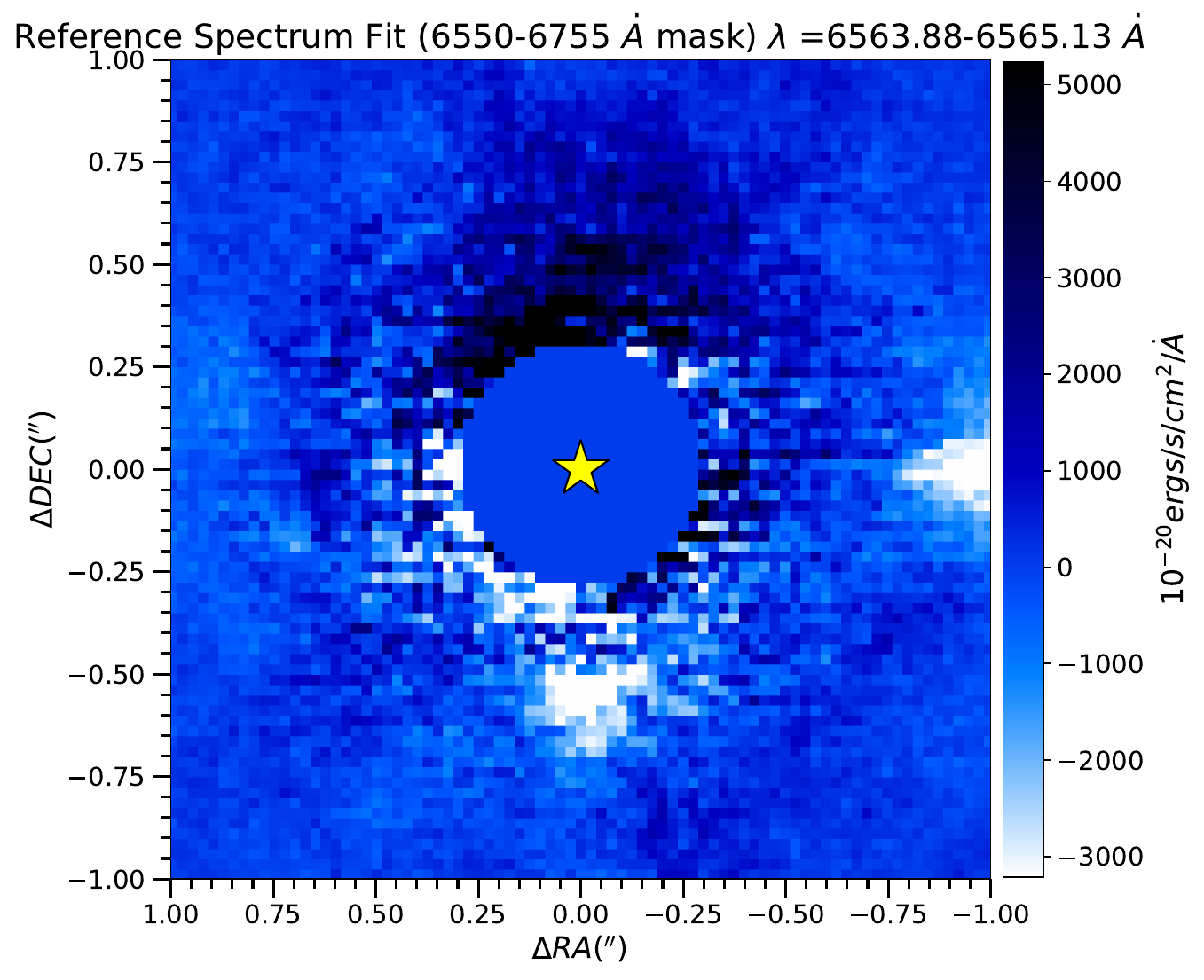}
    \includegraphics[width=0.45\linewidth]{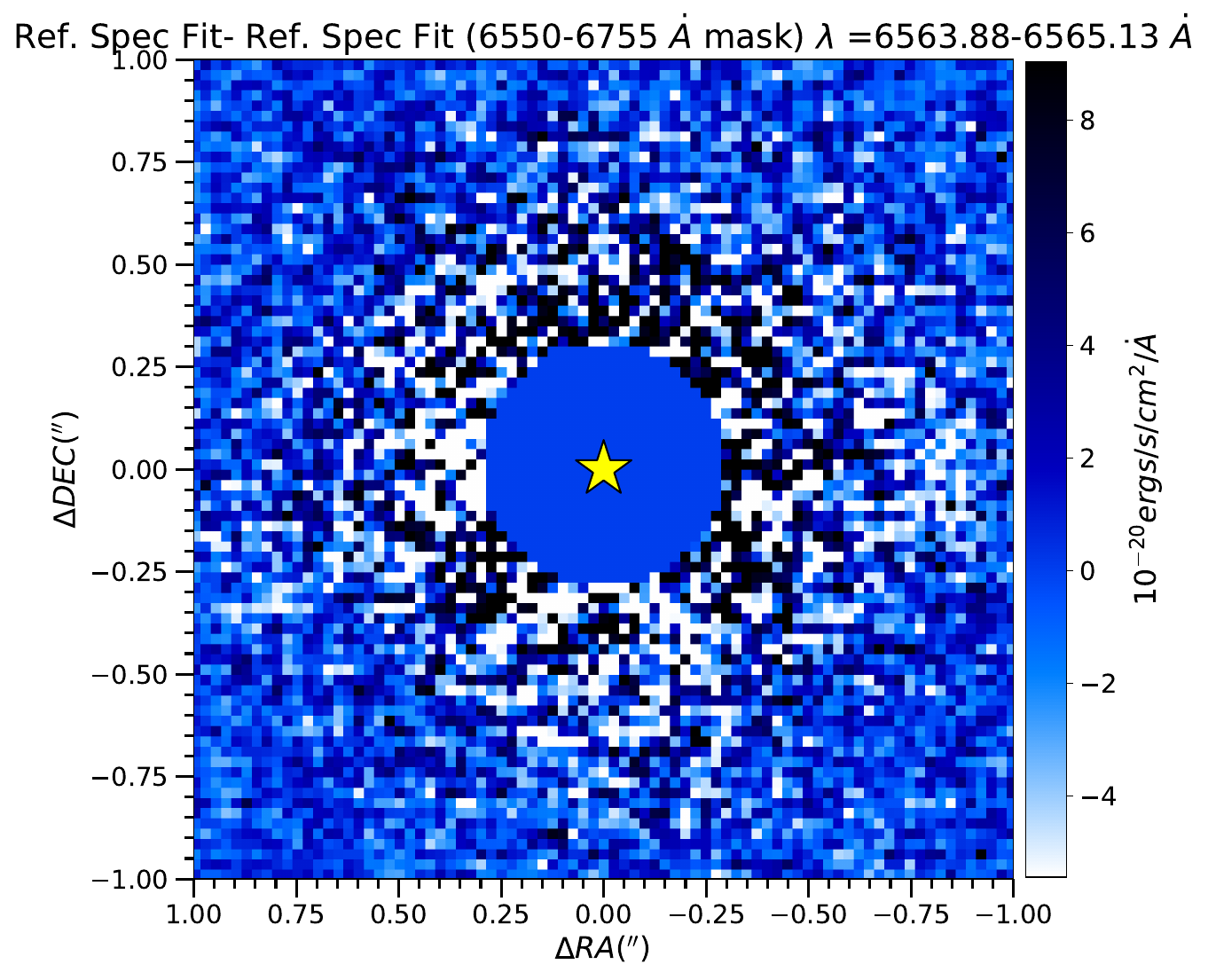}
    \caption{The same as Figure \ref{fig:abaur_psfsub_alt} except for differences in the range of wavelengths masked when constructing the reference spectrum.
    }\label{fig:abaur_psfsub_alt2} 
\end{figure*}

Figures \ref{fig:abaur_psfsub_alt} and \ref{fig:abaur_psfsub_alt2} show reductions using a second-order polynomial fit or a wide masked wavelength range for reference spectrum fitting.  In all cases, AB Aur b is detected at both blue shifted and red-shifted channels: its intensity differs by less than 1\% in all channels.

\newpage

\section{Data Reduction for PDS 70 and V345 Mon}

\subsection{PDS 70}
We reprocessed MUSE data first reported in \citet{Haffert2019} (ESO Program ID 60.A-9100(K)).  To analyze PDS 70 data, we followed the same basic reduction and PSF subtraction procedures outlined in Sections 2 and 3 of this work.  Like with AB Aurigae, we built a reference spectrum for each spaxel within a 1 pixel wide annulus using a third-order polynomial fit to residuals of the flux-normalized spaxels smoothed by a Savitz-Golay filter with a window size of 12.6 \AA, masking the channels covering $H_{\rm \alpha}$ (6550--6570 \AA). 

Unlike AB Aur b, PDS 70 b and c do not lie embedded in a protoplanetary disk and thus the $H_{\rm \alpha}$ signal at their positions is simply the sum of scattered light from the star and accretion-induced emission.  Thus, we used simple circular annuli, not those aligned to the disk's geometry.  We determined a FWHM of 0\farcs{}068 at $H_{\rm \alpha}$, leading to an aperture correction of $\approx$ 9.84 between a 0\farcs{}034 aperture radius and an infinite aperture. 

\begin{figure*}[!htbp]
\centering
        \includegraphics[width=0.4\linewidth]{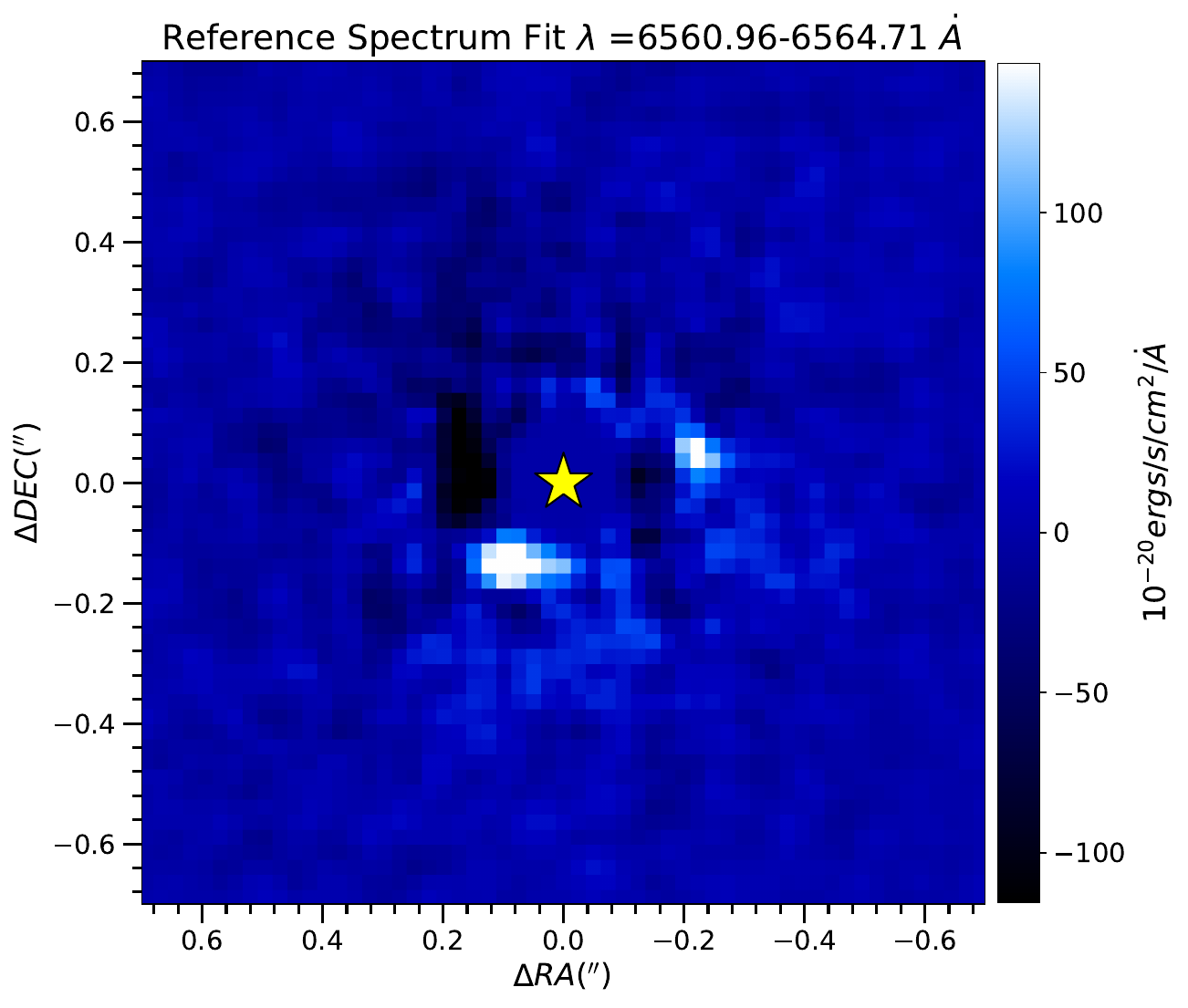}
      \includegraphics[width=0.475\linewidth]{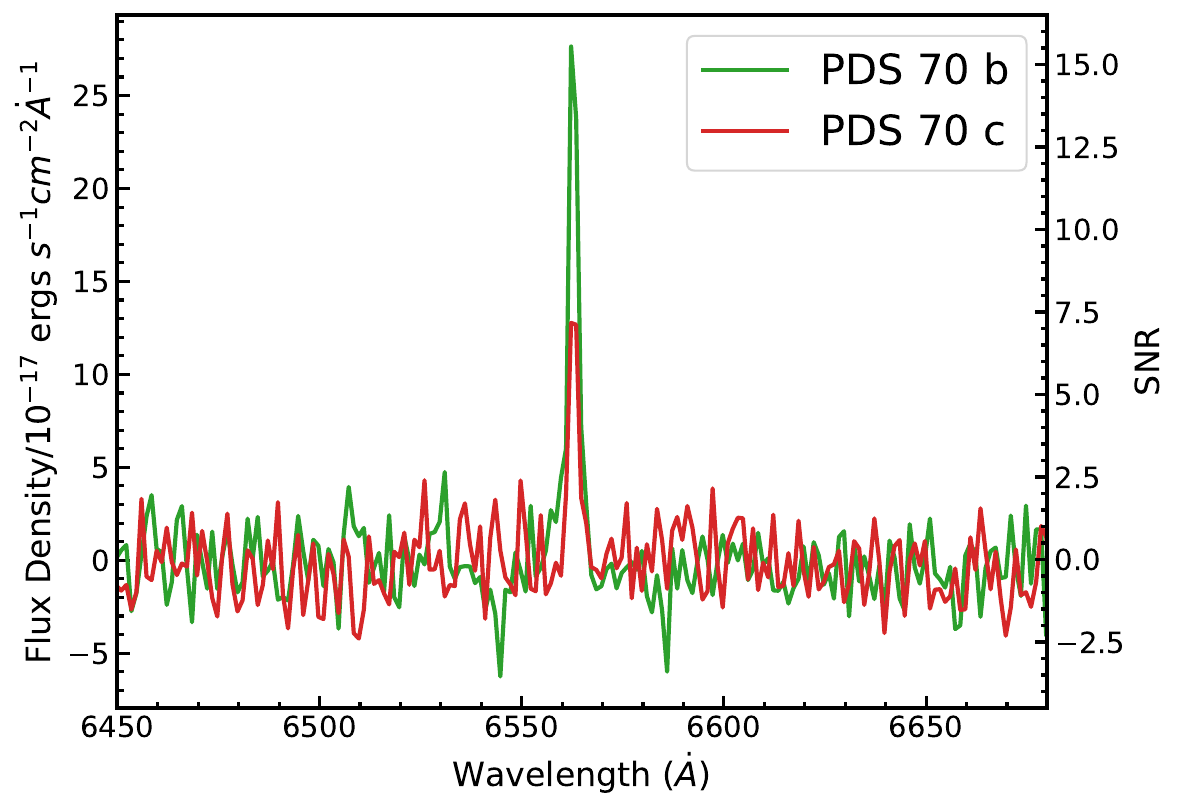}  
    \caption{VLT/MUSE $H_{\rm \alpha}$ image (left) and extracted spectra (right) for for PDS 70 bc from data originally published in \citet{Haffert2019} and also reduced in \citet{Hashimoto2020MUSE}.   
    }\label{fig:pds70} 
\end{figure*}

Figure \ref{fig:pds70} displays our results.  PDS 70 b and c are detected at SNR $\sim$ 18 and 9.7 (peak channel SNR of $\sim$ 15 and 7), respectively, intermediate between values from \citet{Haffert2019} and \citet{Hashimoto2020MUSE}.  The estimated line fluxes for the b and c protoplanets are 8.07 $\pm$ 0.45 $\times$ 10$^{-16}$ ergs s$^{-1}$ cm$^{-2}$ $\dot{A}^{-1}$ and 3.18 $\pm$ 0.33 $\times$ 10$^{-16}$ ergs s$^{-1}$ cm$^{-2}$ $\dot{A}^{-1}$.  These values agree with those published by \citet{Hashimoto2020MUSE} using a more aggressive PSF subtraction approach, which required modeling to correct for throughput, whereas the throughput is $\sim$100\% with the method adopted in this paper since the $H_{\rm \alpha}$ channels themselves are masked when building the reference spectrum.

\subsection{V345 Mon}
 X-Shooter data for V345 Mon draws from Program 084.C-1095(A) obtained on 2010 Jan 18 (PI. G. Herczeg).   The data underwent calibration through the X-Shooter data processing pipeline v2.7.0 \citep{Modigliani2010X-ShooterPipeline}. The calibration process included basic calibration of the raw data such as bias- and dark subtraction, correction of bad- and hot pixeles, flatfielding, optimal extraction, correction for the instrumental flexures, wavelength calibration, and sky subtraction, and correction for telescope and instrumental throughput. Additional flux calibration due to slit losses was not performed because we compare only the H$_\alpha$ line shape in this work. 

 \newpage

\label{appendix:mcmc}
\section{MCMC Corner Plots Modeling the AB Aur b Morphology in Blue-Shifted Emission and Red-Shifted Absorption}

\begin{figure*}[!htbp]
\centering
    \includegraphics[width=1.\linewidth]{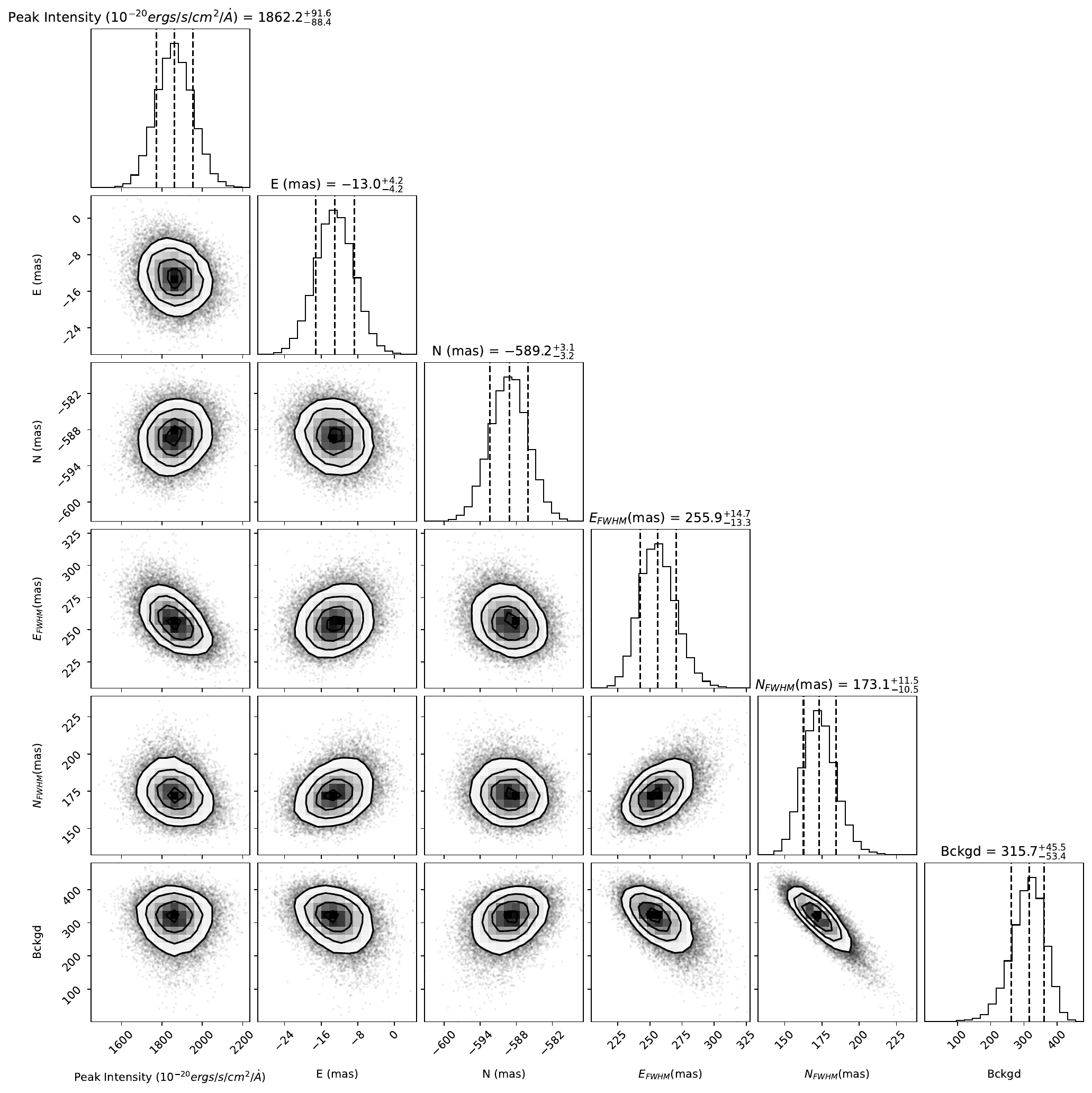}
    \caption{Corner plot showing the posterior distributions estimated from \texttt{emcee} for AB Aur b in blue-shifted $H_{\rm \alpha}$.  
    }\label{fig:abaur_corner_halphablue} 
\end{figure*}

\begin{figure*}[!htbp]
\centering
    \includegraphics[width=1.\linewidth]{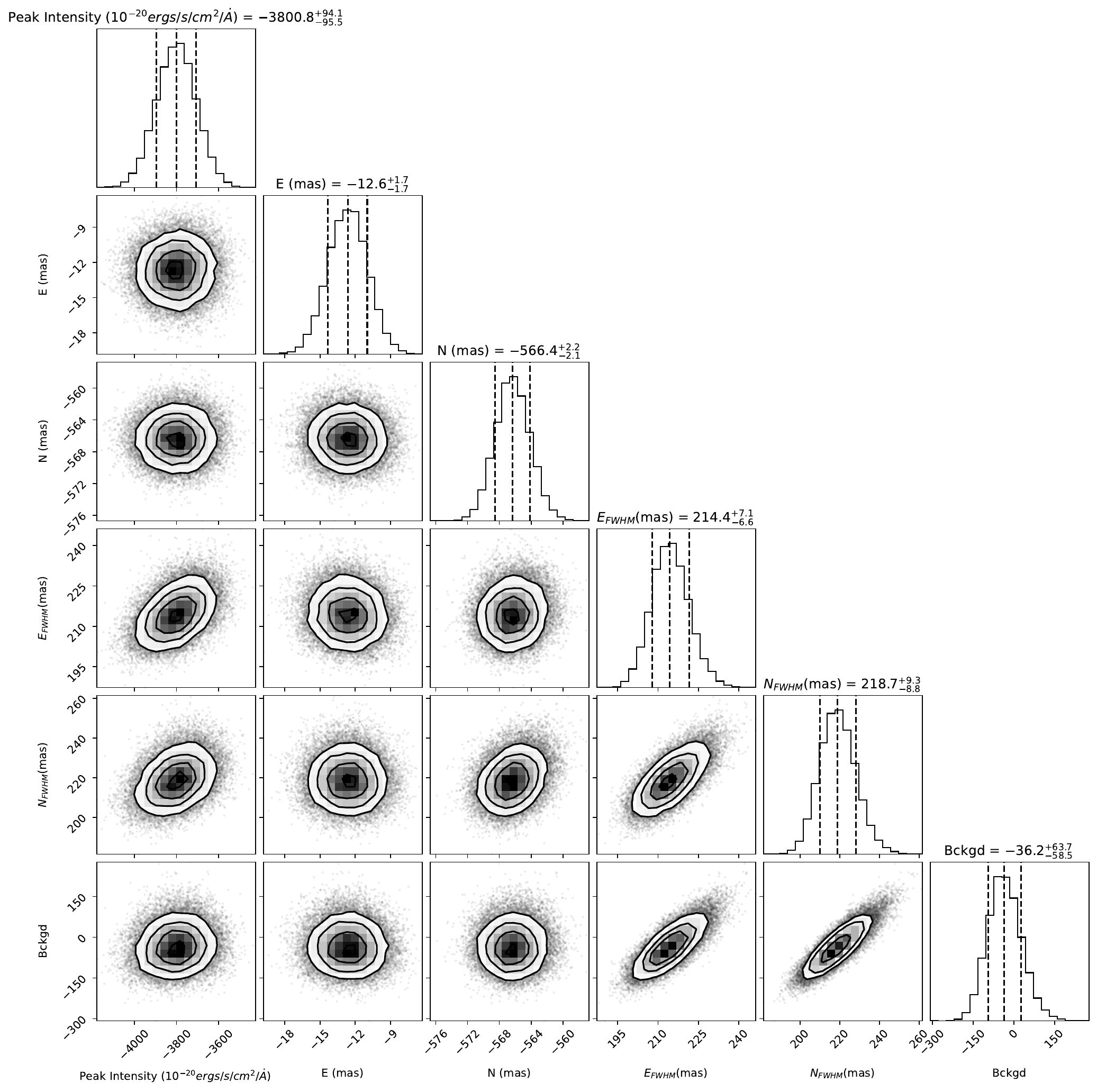}
    \caption{Corner plot showing the posterior distributions estimated from \texttt{emcee} for AB Aur b in red-shifted $H_{\rm \alpha}$.  
    }\label{fig:abaur_corner_halphared} 
\end{figure*}

Figures \ref{fig:abaur_corner_halphablue} and \ref{fig:abaur_corner_halphared} show posterior distributions to model AB Aur b's peak intensity, location, and morphology at blue shifted and red-shifted $H_{\rm \alpha}$.

\newpage
\section{Co-Location of the AB Aur b Signal in MUSE and SCExAO/CHARIS}
\begin{figure*}[!htbp]
\centering
    \includegraphics[width=0.45\linewidth]{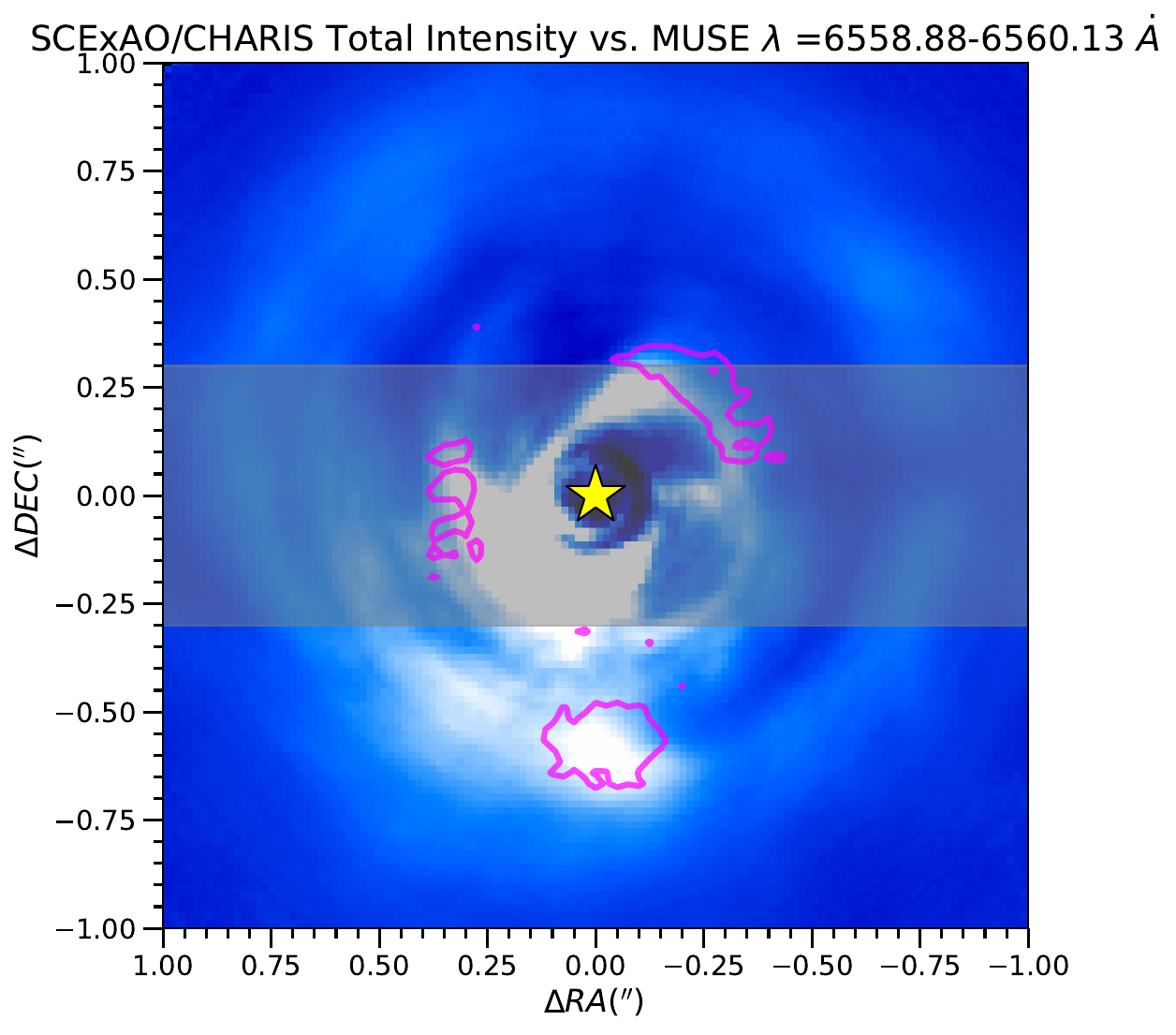}
    \includegraphics[width=0.45\linewidth]{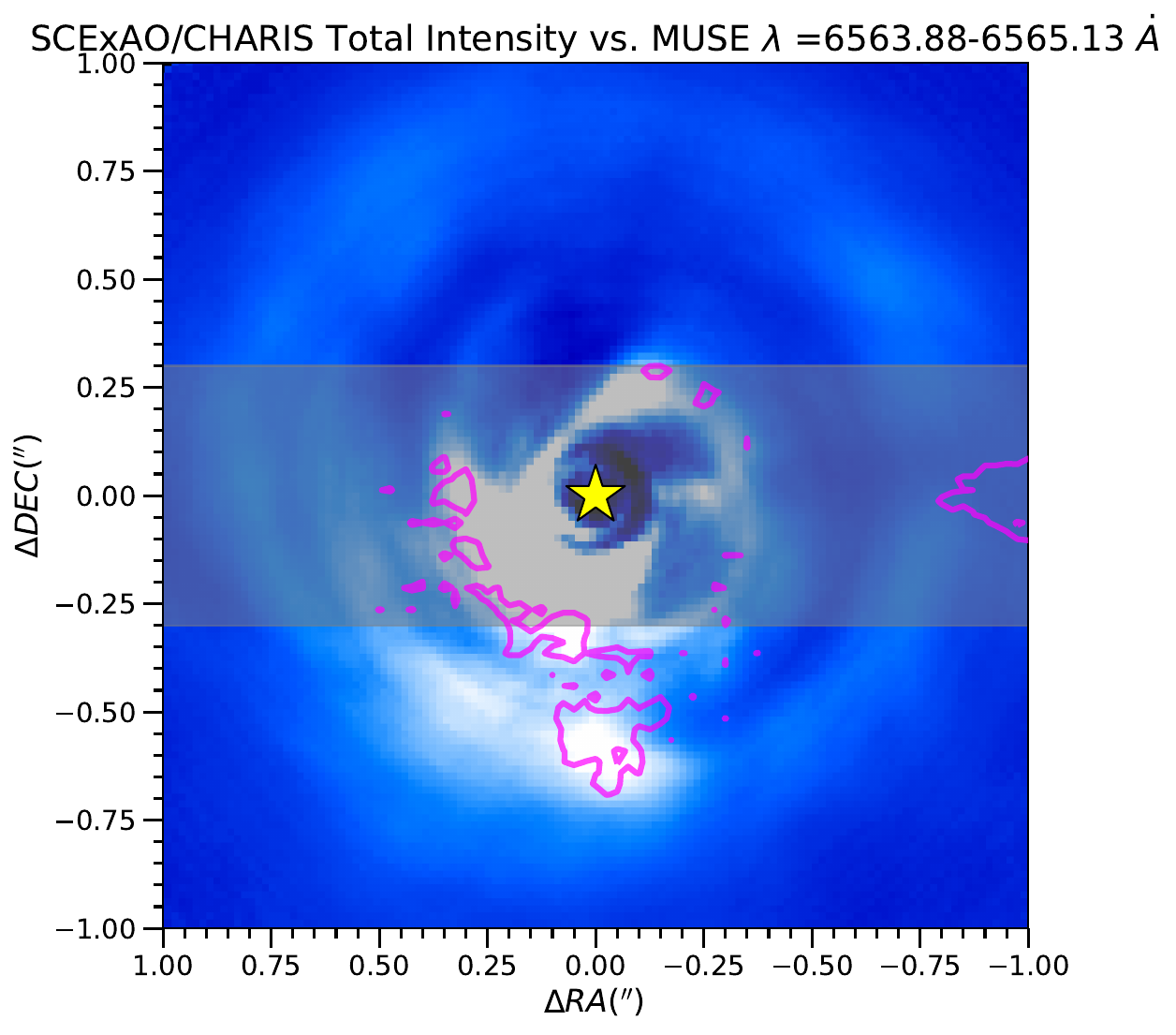}\\
    \includegraphics[width=0.45\linewidth]{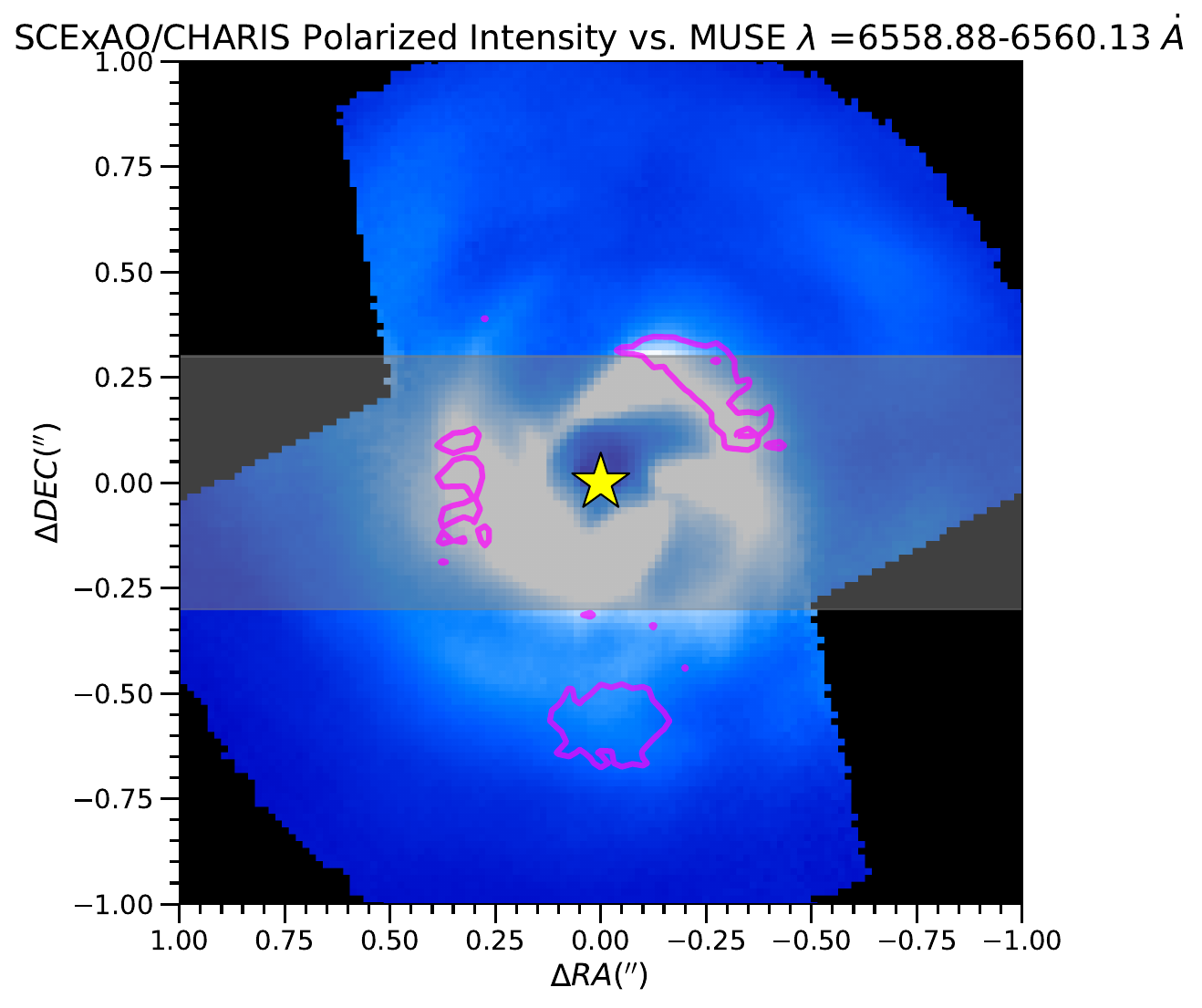}
    \includegraphics[width=0.45\linewidth]{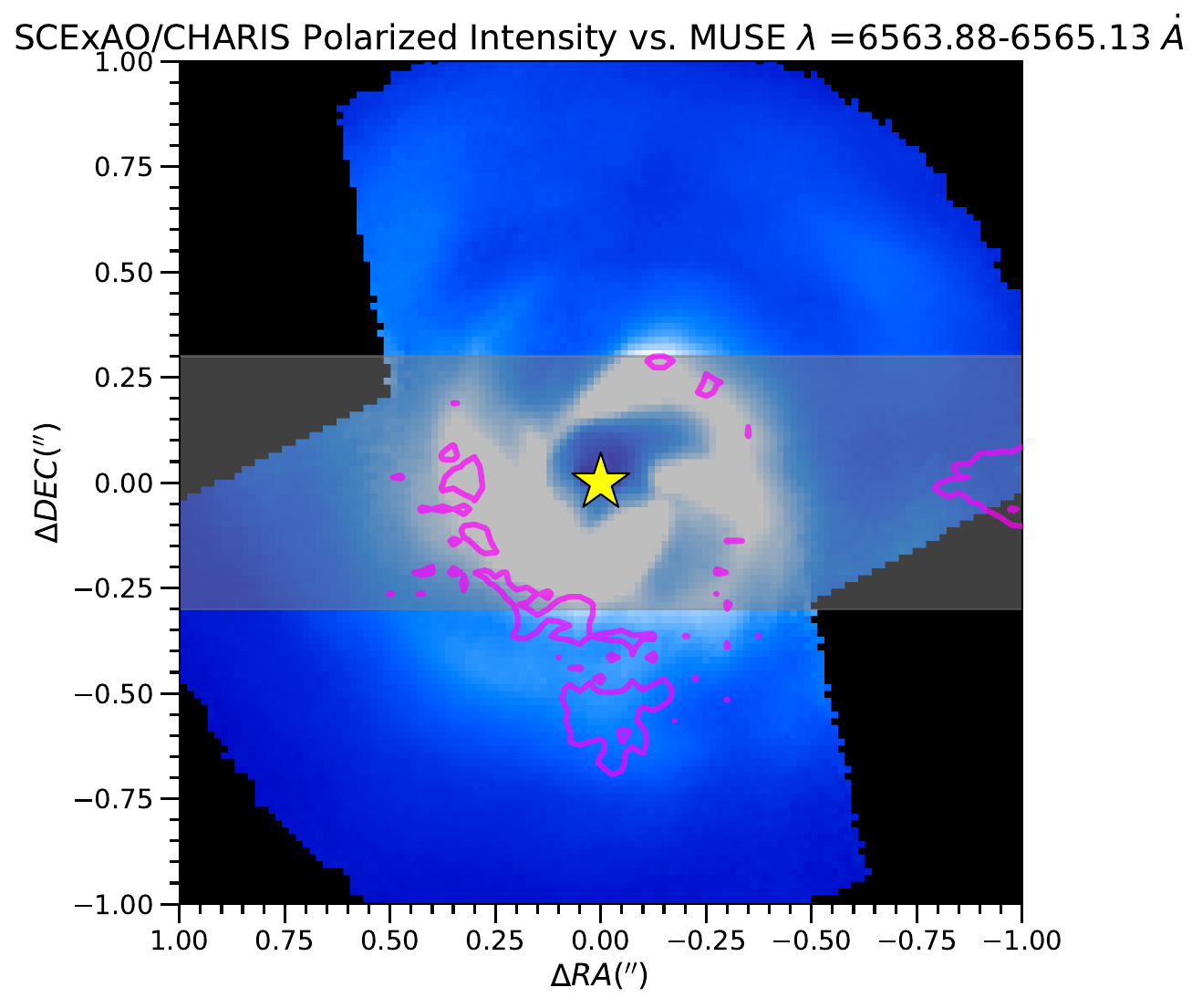}
    \caption{CHARIS total intensity wavelength-collapsed image \citep{Currie2022ABAurb,Currie2024} and polarized intensity image \citep{Dykes2024} with MUSE contours (magenta) overplotted at intensities of 1250 and -2500 $\times$10$^{-20}$ ergs s$^{-1}$ cm$^{-2}$ \AA\ for the blue shifted and red-shifted channels, respectively.  The horizontal shaded region identifies where MUSE data are affected by horizontal striping artifacts and (for $\rho$ $\lesssim$ 0\farcs{}3) non linearity and saturation. AB Aur b is the only signal clearly detected with MUSE at SNR $>$ 5 and co-located with a feature seen with CHARIS.  Bright regions at smaller separations ($\rho$ $\lesssim$ 0\farcs{}3) are affected by detector non-linearity and residual striping: they do not clearly trace any extended feature and the significance of their detections using standard high-contrast imaging metrics described in the main text \citep{Currie2011,Mawet2014} is SNR $\lesssim$2.
    }\label{fig:abaur_charis_muse} 
\end{figure*}

Figure \ref{fig:abaur_charis_muse} compares the MUSE intensity contours at blue shifted and red-shifted channels to CHARIS total and polarized intensity images \cite{Currie2022ABAurb,Currie2024,Dykes2024}.  As noted in the main text, AB Aur b's positions in the MUSE data agree with published estimates from CHARIS and STIS to within 0.2–2$\sigma$.   The main tension between the MUSE astrometry and CHARIS's is that AB Aur b's angular separation is $\approx$10-20 mas smaller with MUSE ($\sim$ 0.4--0.8 MUSE pixels).   Agreement is better with the STIS astrometry ($\rho$ $\sim$ 0\farcs{}575), which like MUSE is also obtained at optical wavelengths.  

\end{document}